\documentclass[onefignum,onetabnum]{siamart171218}

\usepackage{lipsum}
\usepackage{amsfonts}
\usepackage{graphicx}
\usepackage{epstopdf}
\usepackage{algorithm}
\usepackage{algorithmic}
\usepackage{xfrac,subfigure}
\ifpdf
  \DeclareGraphicsExtensions{.eps,.pdf,.png,.jpg}
\else
  \DeclareGraphicsExtensions{.eps}
\fi

\newsiamremark{remark}{Remark}
\newsiamremark{hypothesis}{Hypothesis}
\crefname{hypothesis}{Hypothesis}{Hypotheses}
\newsiamthm{claim}{Claim}

\headers{Mathieu Functions: Historical Perspective}{C.~Brimacombe, R.~M.~Corless, and M.~Zamir}

\title{COMPUTATION AND APPLICATIONS OF MATHIEU FUNCTIONS: A HISTORICAL PERSPECTIVE\thanks{Submitted to the editors DATE.
\funding{This work was supported by NSERC under grants numbered RGPIN-2020-06438 (RMC) and RGPIN-2019-04749 (MZ); and, while RMC was visiting the Isaac Newton Institute
during the programme
Complex Analysis: Tools, techniques, and applications, by EPSRC Grant \# EP/R014604/1.}}}

\author{Chris Brimacombe\thanks{University of Toronto, Toronto, \textsc{Canada}
  (\email{chris.brimacombe@mail.utoronto.ca }).}
\and Robert M.~Corless\thanks{School of Mathematical and Statistical Sciences, Western University, London, \textsc{Canada}
  (\email{rcorless@uwo.ca}, \email{zamir@uwo.ca}).}
\and Mair Zamir\footnotemark[3]}

\usepackage{amsopn}

\newcommand{\mat}[1]{\mathbf{#1}}

\newcommand{\Ce}{\ensuremath{\mathrm{Ce}}}
\newcommand{\ce}{\ensuremath{\mathrm{ce}}}
\newcommand{\Se}{\ensuremath{\mathrm{Se}}}
\newcommand{\se}{\ensuremath{\mathrm{se}}}

\makeatletter
\newcommand*{\addFileDependency}[1]{
  \typeout{(#1)}
  \@addtofilelist{#1}
  \IfFileExists{#1}{}{\typeout{No file #1.}}
}
\makeatother

\ifpdf
\hypersetup{
  pdftitle={Computation and {Mathieu} functions},
  pdfauthor={C.~Brimacombe, R.~M.~Corless, and M.~Zamir}
}
\fi

\renewcommand{\emph}[1]{\textsl{#1}}
\begin{document}

\maketitle

\begin{abstract}
Mathieu functions of period $\pi$ or $2\pi$, also called elliptic
cylinder functions, were introduced in 1868 by \'Emile Mathieu together with so-called modified Mathieu functions, in order to help understand the vibrations of an elastic membrane set in a fixed elliptical hoop. These functions still occur frequently in applications today: our interest, for instance, was stimulated by a problem of pulsatile blood flow in a blood vessel compressed into an elliptical cross-section.
This paper surveys and recapitulates the historical development of the theory and methods of computation for Mathieu functions and modified Mathieu functions and identifies some gaps in current software capability, particularly to do with double eigenvalues of the Mathieu equation.
We demonstrate how to compute Puiseux expansions of the Mathieu eigenvalues about such double eigenvalues, and give methods to compute the generalized eigenfunctions that arise there.
In examining Mathieu's original contribution, we bring out that his use of anti-secularity predates that of Lindstedt.
For interest, we also provide short biographies of some of the major mathematical researchers involved in the history of the Mathieu functions: \'Emile Mathieu, Sir Edmund Whittaker, Edward Ince, and Gertrude Blanch.
\end{abstract}



\section{Introduction}
What is the sound of an elliptic drum?
In a memoir presented at the Sorbonne in 1868, \'Emile Mathieu showed the way to find the answer,
when he described the solution of a
mechanical vibration problem characterized by an elliptic boundary.
The memoir was groundbreaking in that it introduced a new differential equation whose
eigenvalues and corresponding periodic solutions
led to the definition of a new class of functions.
In 1912 Whittaker named these new functions in
honour of their discoverer: the differential equation is now known as the
Mathieu equation and the $\pi$ and $2\pi$ periodic solutions (and \emph{only} these periodic solutions) are known as the
Mathieu functions.
While work on theoretical and analytical aspects of Mathieu functions has continued since the
introduction of these functions in 1868, the focus has shifted in recent years to work on
numerical and computational aspects. This development, driven by steep advances in digital
technology, has given rise to heavy reliance today on ``packaged software'' for the
evaluation of Mathieu functions.
This practice comes with the risk of concealing as yet
unresolved (or at least not \emph{completely} resolved) analytical and computational issues
involved in the use of Mathieu functions.

Modern approaches to solving such vibrational problems are more likely to be direct.  See~\cite{gander2019class} for an exemplar of this approach.  Spectral expansions using special functions do remain of interest for approximation, however; see~\cite{Townsend2015} for an exemplar of this approach.  But regardless of application, the Mathieu functions themselves remain of interest because the equation is simple enough to occur in a wide variety of contexts, and there is a need to compute them directly.

We review methods of computation of periodic Mathieu functions and note that all the methods we review fall short at some ``exceptional'' or ``double'' points, as noted both by~\cite{blanch1969double} and by~\cite{hunter1981eigenvalues}. Although those objections were made more than fifty and forty years ago respectively, and theoretical work on these was completed in~\cite{meixner}, we believe that there is still no fully satisfactory code available, as we will detail.
We will discuss what to do in practice at these double points, where the eigenvalues merge and we lose one independent eigenfunction and thus lose completeness of our set of eigenfunctions.

While we survey many results here, we cannot cover everything.  Instead we list several references that contain many details (and indeed many important results) that we omit.
The first and most important reference is to Chapter 20 of the classic~\cite{abramowitz}, written by Gertrude Blanch, and its successor, Chapter 28 of the Digital Library of Mathematical Functions (henceforth DLMF, \hyperlink{https://dlmf.nist.gov/28.1}{https://dlmf.nist.gov/28.1}) written by G.~Wolf. The DLMF is intended to be an updated replacement for~\cite{abramowitz} in this computer age. Both are (perfectly legally) free online.
The DLMF gives pointers to many of the alternative notations used elsewhere in the rather substantial literature. The original Chapter 20 of~\cite{abramowitz} has yet more references.  We will mention others as we go.

One of our favourite reference works for methods of computation of special functions, namely the very thorough and carefully laid out book~\cite{Gil2007}, unfortunately does not include Mathieu functions as examples (indeed, there may be a reason for that, which we will discuss).  However, many of the techniques that are described in that book are potentially applicable for Mathieu functions, and we will see variants of some of them used in this present paper.

\subsection{Organization of the paper}

In section~\ref{sec:applications} we briefly outline three applications of Mathieu functions, including the one that motivated us to perform this study.
We then recapitulate in section~\ref{sec:history} some of the historical development of Mathieu functions. We are not historians of mathematics or science, but we have done our best.
In section~\ref{sec:double} we look at a method to compute double points and the corresponding eigenvalues, and Puiseux series, about those points, for the eigenvalues.  In section~\ref{sec:alg} we look at algorithms for computing solutions of the Mathieu equations, including Mathieu functions (eigenfunctions).  We finish that section with a discussion of how to compute \emph{generalized} eigenfunctions for double eigenvalues, which are needed for completeness.  We provide concluding remarks in section~\ref{sec:concl}.  In appendix~\ref{app:pulsatiledetails} we provide more details of the application that motivated us to undertake this work.  In appendix~\ref{app:Bessel2Mathieu} we discuss confocal ellipses, and give a singular perturbation argument relating Mathieu functions to Bessel functions in the limit as ellipses become circles. In appendix~\ref{app:MathieuPerturbForm} we compare Mathieu's perturbative solution to the Mathieu equation with that produced by a computer algebra system, and apart from minor errors and typos in his paper confirm his results.  Finally, in appendix~\ref{app:DoubleEigTable} we provide, as an \emph{homage} to the great scientific table-makers, a table of Puiseux expansions about double points.  Of course, the computer-readable version at \href{https://github.com/rcorless/Puiseux-series-Mathieu-double-points}{https://github.com/rcorless/Puiseux-series-Mathieu-double-points} (together with the Maple code used to generate it) is much more useful nowadays; still, there's something to be said for looking at a collection of related numbers.  We remark that printing that table nicely with \LaTeX---at least, nicely enough compared to old-style high-quality production---made for an interesting challenge, and some quirks remain in the table as ``Easter Eggs'' for any that are willing to hunt for them.

\vfill
\pagebreak
\section{Applications\label{sec:applications}}
\subsection{Columns and Strings Under Periodically Varying Forces}
One of the more surprising physics experiments in the undergraduate curriculum is the stabilization of a vertical hinged column by up-and-down vibration at the right frequency.  There are many YouTube videos demonstrating this, and, after reading this paper, the reader may choose to search some of those videos out: a good set of keywords to search with is ``stability of the inverted pendulum''.  This phenomenon is explained using topological terms in~\cite{levi1988stability}.  The underlying mathematics has been known at least since 1928~\cite{vanderPol:1943:stability}, and is concerned directly with the Mathieu equation, if the periodic force is a simple sine or cosine.

There are similar, though more complicated, problems which still need the Mathieu equation. We will briefly describe the model studied in~\cite{lubkin1943stability}, namely the stability of a column under periodically-varying compression, or of a string under tension with a periodically-varying tension force.  These problems have infinitely many degrees of freedom, as opposed to the simple inverted pendulum above.  Consider for instance the case of a column under compression.  If the force $F$ is greater than the Euler load, then it is well known that the column can buckle; the question at issue here is if $F = P + H\cos \omega t$, which consists of a steady part plus a periodically-varying part, can one choose $H$ and $\omega$ so that, even if $P$ is larger than the Euler load, the column remains stable?

\begin{figure}
  \centering
  \includegraphics[width=10cm]{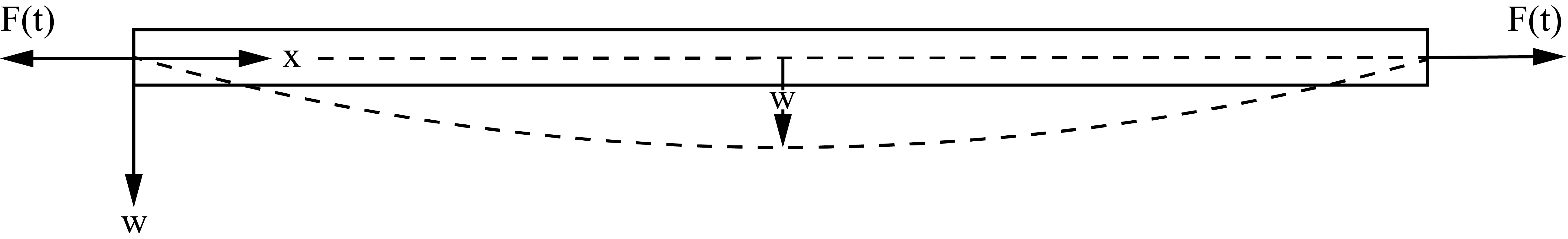}
  \caption{A flexible string with a time-varying tension $F(t)$. The figure indicates that the deformation $w(t)$ is confined to the plane.  We model this figure on Figure~1 of~\cite{lubkin1943stability}.}\label{fig:TensionString}
\end{figure}

The answer is a qualified yes; one qualification is that, at least part of the time, the force must be less than the Euler load, which makes sense.  More, the qualifications require the study of the \emph{stability} of the solutions to the Mathieu equation, and therefore require a good knowledge of the periodic solutions of the Mathieu equation.
The paper~\cite{lubkin1943stability} also studies the motion of a string with a periodically-forced tension with a similar model; see figure~\ref{fig:TensionString}.  The equations of motion they derive are
\[
EI \frac{\partial^4 w}{\partial x^4} - F(t) \frac{\partial^2 w}{\partial x^2} + m \frac{\partial^2 w}{\partial t^2} = 0 \>,
\]
where they take $F(t) = P + H\cos \omega t$ and the familiar Young's modulus $E$ characterizes the relationship between stress and strain, and $I$ is the moment of inertia of the body's cross-section.

``In all of these problems the Mathieu equation (more properly, a \emph{sequence} of Mathieu equations in the continuous systems) plays a central r\^ole, since the decision as to stability depends on the character of the solutions to such equations." [From the introduction in~\cite{lubkin1943stability}.]

In fact the study of the stability of these systems requires more than we are going to cover in this paper: it needs the Floquet theory, which we only lightly touch on.

\subsection{Pulsatile Flow in Tubes of Elliptic Cross Sections}
Our motivation for this present paper originated from a problem in pulsatile blood flow. Under normal circumstances, blood flow occurs in vessels of circular cross sections, but under a number of important pathological conditions the vessels are deformed by external forces to the effect that their cross sections are no longer circular. In a separate study (currently in progress) we are examining this phenomenon using an elliptic cross section as a simple, mathematically tractable, departure from a circular cross section, with the main focus being on the hemodynamic consequences of that departure. In the present paper our focus is on the mathematical and computational consequences.

\begin{figure}
  \centering
  \includegraphics[width=6cm]{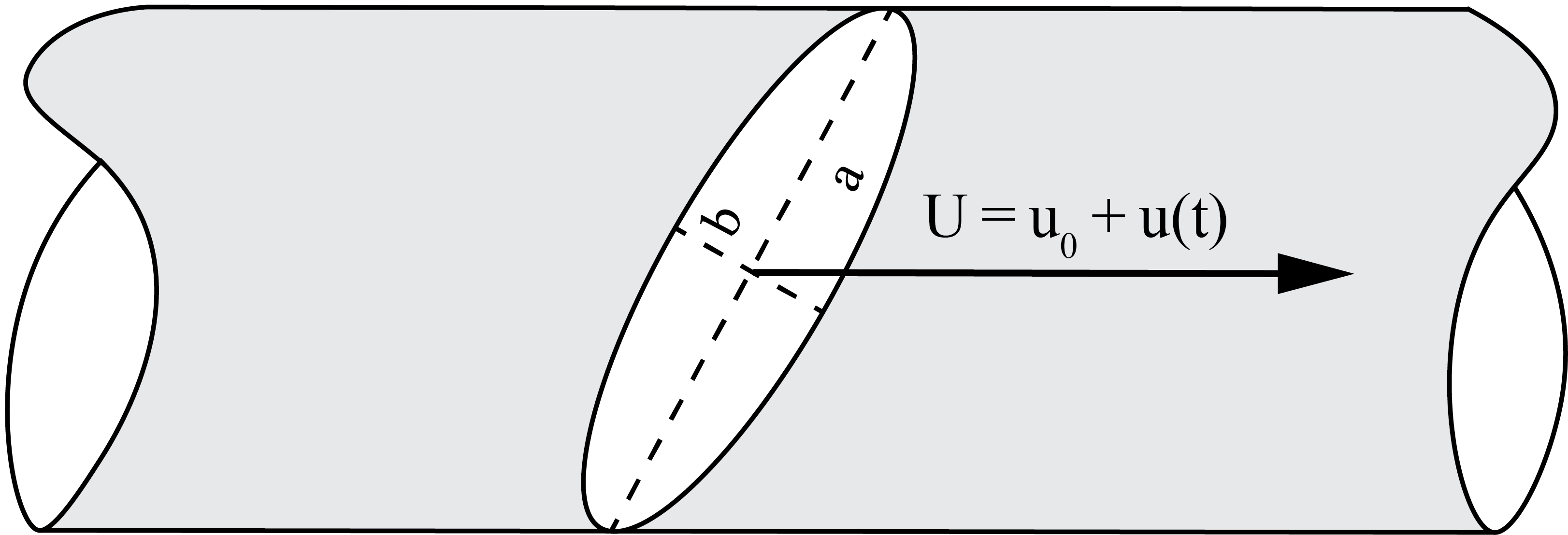}
  \caption{Flow in a tube of elliptic cross-section.  The elliptic cross-section is meant to model some  pathologies where the blood vessels are deformed so that they are no longer circular.  This figure is modelled after Figure 3.1, page 45, of~\cite{zamir}.}\label{fig:FlowTube}
\end{figure}

Fluid flow in a tube (see figure~\ref{fig:FlowTube}) is in general governed by a simplified form of the Navier-Stokes equations~\cite{zamir}. If the tube is straight and of uniform cross section, the flow can be described in terms of a single velocity component $U$ along the axis of the tube. If the flow is \textit{pulsatile}, as in the cardiovascular system, the velocity $U$ consists of a steady part $u_0$ plus an oscillatory part $u(t)$ such that

\begin{equation}
U=u_0+u(t)\>,
\end{equation}
where $t$ is time. Only the oscillatory component of velocity, $u(t)$, is relevant to the present discussion since the transition from Bessel to Mathieu functions occurs in the governing equations of this component of the flow as the cross section of the tube changes from circular to elliptic. We give a brief sketch here, with more details of our motivating application in appendix~\ref{app:pulsatiledetails}. In that appendix the focus is on the transition from Bessel to Mathieu equations in the case of pulsatile flow in a tube as the cross section of the tube transitions from circular to elliptic geometry.

While pulsatile flow in tubes of circular cross sections is governed by Bessel equations, with solutions in terms of Bessel functions, the corresponding situation in tubes of elliptic cross sections involves the Mathieu equation and Mathieu functions. From a mathematical perspective, this transition is not only a matter of curiosity but also a matter of practical importance because of the comparative difficulties involved in the numerical computation of these two kinds of functions. The difficulties and pitfalls involved in the computation of Mathieu functions are the focus of much of the present paper.

\subsection{Vibrating membrane bounded by an ellipse}
\begin{figure}
    \centering
    \includegraphics[width=10cm]{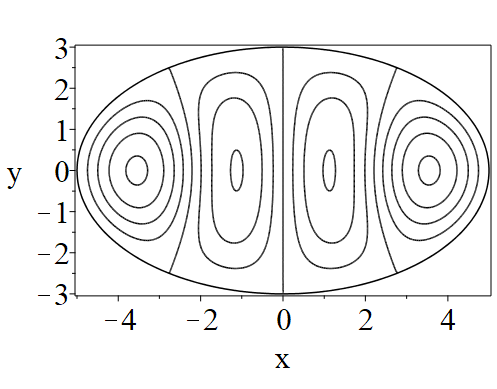}
    \caption{Contours and nodal lines of a possible pure vibration mode of an elliptic drum.  The aspect ratio of the ellipse is $3:5$.   At the time of this snapshot, the rightmost loops in this mode have positive values for the contours, and all adjacent cells have opposite signs.  Notice the hyperbolic nodal lines where the membrane does not move, separating the cells. We have suppressed here the details of which Mathieu functions were used to produce this figure (we give them in section~\ref{sec:modifiedMathieuFunctions}), but we followed the method outlined by Mathieu in his 1868 paper~\cite{mathieu1868memoire}.  See~\cite{corless2020pure} for complete details. }
    \label{fig:contourcemode3}
\end{figure}
One of the simplest physical problems whose solutions involve the Mathieu equation and the Mathieu functions is the sound made by an elliptical drum.  This in fact was the problem Mathieu himself studied, and in section~\ref{sec:Mathieu} we will look in detail at how he solved it.  The physical problem being modelled is, in fact, remarkably easy to define: imagine an elliptical hoop, fixed immovably, and a thin homogeneous membrane stretched tight across the hoop, making a drum.  What are the natural modes of vibration of this drum, and how could we describe mathematically its motion once struck?

One possible natural mode for one particular drum is pictured in figure~\ref{fig:contourcemode3}.  There we see contours of vibration, including contours where there is no motion (the so-called ``nodal lines'').  This figure was drawn using our own software to compute the relevant Mathieu functions, but many software packages exist which could do this.

\section{Historical overview, introducing notions and notation\label{sec:history}}
We introduce the Mathieu equation and the Mathieu functions in historical order, by discussing the contributions of several of the main researchers involved.  The result is a tour of several aspects of late nineteenth-century and early twentieth-century mathematics.  We also give some biographical details of these main figures.  For concreteness in the discussion to follow,
here is the Mathieu equation in one common modern notation:
\begin{equation}
    \frac{d^2 y}{dx^2} + \left[a -2q\cos(2x)\right]y =0\>. \label{eq:mathieueq}
\end{equation}
The parameter~$q$ is given by the physics or the geometry of the specific problem at hand; the eigenvalue $a$ must be calculated in order to ensure periodicity of $y$, given~$q$ and a desired order.
The so-called \textsl{modified} Mathieu equation is related to equation~\eqref{eq:mathieueq} by the transformation $z = \pm i x$ (the sign makes no difference):
\begin{equation}
    \frac{d^2 y}{dz^2} - \left[a -2q\cosh(2z)\right]y =0\>. \label{eq:modmathieueq}
\end{equation}
The \emph{even} solutions are conventionally written as $\Ce_{g}(z,q) = \ce_{g}(\pm i z,q)$. The eigenvalues for even solutions are conventionally written $a_g(q)$. The \emph{odd} solutions are written similarly, as $\se_{g}(z,q)$ and $\Se_{g}(z,q)$, and the odd eigenvalues are conventionally written $b_g(q)$.  Here $g$ is a nonnegative integer, and the solutions split into further classes if $g$ is itself even or odd, as we will see.  The use of the letter $g$ for an integer contradicts the usual $I-N$ convention in use nowadays, from \textsc{Fortran}; but Mathieu used the letter $g$ in this way and we find it convenient when the parity of $g$ is unspecified.

If we write the general solution of the Mathieu equation (with no initial or boundary conditions applied) with arbitrary constants $\alpha$ and $\beta$ as
\begin{align}
    y(x) = \alpha w_I(x; a, q ) + \beta w_{II}(x; a, q )
\end{align}
using the fundamental pair of solutions satisfying $w_I(0;a,q) = 1$ with $w_I'(0;a,q)=0$ and $w_{II}(0;a,q)=0$ with $w_{II}'(0;a,q)=1$, (using the notation of the DLMF and where $'$ denotes $d/dx$) then the general solution of the  Modified Mathieu equation can be written
\begin{align}
    y(z) = \alpha w_I(\pm i z; a, q ) + \beta w_{II}(\pm iz; a, q )\>.
\end{align}
Some software packages denote these functions $C$ and $S$ respectively.
When $a(q)$ or $b(q)$ is an eigenvalue, the periodic Mathieu functions must satisfy (now allowing $x$ to be complex and renaming it $z$)
\begin{align}
    \ce_{m}(z,q) &= c_m w_I(z; a_{m}(q), q ) \\
    \se_m(z,q) &= s_m w_{II}(z; b_m(q), q )\>,
\end{align}
for some normalization constants $c_m$ and $s_m$.  \textbf{In this paper we take those normalization constants to be $1$.}

In theory, the use of Mathieu
functions and modified Mathieu functions in the solution of the aforementioned physical problems is attractive because the functions are analogous to harmonic functions, and expansions in terms of them can be efficient in comparison with direct numerical solution of the PDE model. In practice, there are annoying difficulties: the available software might be restricted to real arguments, or use a different normalization than the one desired (the authors of~\cite{erricolo} state that there are at least three normalizations in common use; they themselves use the same normalization that we do here), or the software may fail to be accurate for ``difficult" values of the problem parameters, say for large values.
A more serious problem is the approximation properties of the expansion itself, followed by the numerical stability of the expansion.  Approximation properties are explored in~\cite{shen2009spectral} for real~$q$, with all the power of the Sturm--Liouville theory (which indeed Mathieu himself used in his 1868 paper).
Numerical stability, on the other hand, has received less attention.

Now that we have sketched where we want to go with Mathieu functions, let us recapitulate their development.

\subsection{\'Emile L\'eonard Mathieu (1835--1890)\label{sec:Mathieu}}
Mathieu began his 1868 discussion~\cite{mathieu1868memoire} of the
vibrations of an elastic membrane held fixed by a hoop in the shape of an ellipse by first considering the simpler problem when the hoop is, in fact, circular\footnote{This memoir was translated from its nineteenth century French for us by Dr.~Robert H.~C.~Moir, and the translation---which we believe may be of interest on its own---has been made available~\cite{moir2021memoir}.}.  Mathieu's discussion of the circular case starts with the PDE
\begin{displaymath}
\frac{\partial^2 w}{\partial t^2} = m^2\left( \frac{\partial^2 w}{\partial x^2} + \frac{\partial^2 w}{\partial y^2}\right)
\end{displaymath}
(Mathieu used $d$ and not our modern Russian $\partial$ for partial derivatives)
and then transformed $x=r\cos\alpha$, $y=r\sin\alpha$ to polar coordinates to get
\begin{displaymath}
\frac{\partial^2 w}{\partial t^2} = m^2\left( \frac{\partial^2 w}{\partial r^2} +\frac{1}{r}\frac{\partial w}{\partial r}+ \frac{1}{r^2}\frac{\partial^2 w}{\partial\alpha^2}\right)\>.
\end{displaymath}
Thereafter Mathieu used what is now the standard method of separation of variables for a pure oscillation $w = \sin(2\lambda m t) u(r,\alpha)$ and $u(r,\alpha) = P(\alpha)Q(r)$ to give a harmonic equation for $P(\alpha)$ and is equivalent to what we now call Bessel's equation for $Q(r)$:
\begin{displaymath}
{r}^{2}{\frac {d^{2}}{d{r}^{2}}}Q \left( r \right) +r{
\frac d{dr}}Q \left( r \right) - \left({n}^{2} -4\,{\lambda}^{2}{
r}^{2} \right) Q \left( r \right) \>.
\end{displaymath}
[Bessel's equation in standard form is $z^2 d^2y/dz^2 + z dy/dz + (z^2-\nu^2)y = 0$; see DLMF 10.2.1.]
One can then write the solution to the original vibration problem as a linear combination of products of these eigenfunctions, and determine the unknown coefficients by matching to the boundary conditions using orthogonality.

Although Bessel (1784--1846) did his work first (and although it was actually Daniel Bernoulli who first identified this equation, even earlier) Mathieu did not call this ``Bessel's equation'', or give it a name at all, but merely solved it in series in what must have been common practice at the time.
He then demonstrated that one must find the \textsl{zeros} of the various series for the Bessel functions, and cited Bourget's Memoirs~\cite{Bourget1866} for a method for doing so, in order to identify the eigenvalues and eigenmodes of the vibration of the membrane in the circular case.

Following the same strategy in the elliptic case, but using a confocal elliptic transformation\footnote{Mathieu initially spelled the hyperbolic functions out explicitly, and later used ``Pour simplifier" the notation $E(\beta) = (e^\beta + e^{-\beta})/2$ for $\cosh(\beta)$ and $\mathcal{E}(\beta) = (e^\beta -e^{-\beta})/2$ for $\sinh(\beta)$. Johann Heinrich Lambert had already in the 18th century introduced the notation which we use today; it is interesting that it was not universally used in the 19th.} $x = c \cosh(\beta)\cos(\alpha)$ and
$y = c\sinh(\beta)\sin(\alpha)$
where $2c$ is the distance between the foci of the ellipse\footnote{Notice that this coordinate transformation is singular if $c=0$. Therefore the connection of Mathieu functions to Bessel functions, while present, requires art to tease out. See Appendix~\ref{app:Bessel2Mathieu}.}, Mathieu arrived at an equation that on separation gives two equations equivalent to those now known as
the \emph{Mathieu equation} and the \emph{modified} Mathieu equation:
\begin{displaymath}
-\frac{1}{P}\frac{d^2 P}{d\alpha^2} + 4\lambda^2 c^2\cos^2(\alpha) = +\frac{1}{Q}\frac{d^2Q}{d\beta^2} + 4\lambda^2 c^2 \cosh^2(\beta)\>.
\end{displaymath}
``Comme le premier membre ne peut renfermer que $\alpha$, et le second que $\beta$, ils sont \'egaux a une m\^eme constante $N$.'' Readers may be pleased to learn that separation of variables reads the same in the 21st century as it did in the 19th,  \textsl{mutatis mutandis}.

There is one remaining difference to the modern notation, and that is the use of $\cos^2\alpha$ and $\cosh^2\beta$.  Using double-angle identities, Mathieu later in this same paper transformed these to something that we write in modern notation as

\begin{align}
    \frac{d^2 P}{d\alpha^2} + (a - 2q\cos 2\alpha)P &= 0 \label{eq:etaeq} \\
    \frac{d^2 Q}{d\beta^2} - (a + 2q \cosh 2\beta)Q &=0\>. \label{eq:xieq}
\end{align}
Notice that these two equations can be transformed into each other by the change of variable $\beta = i\alpha$.
Here $q = \lambda^2c^2$ is a relabeling of the parameter that contained the physics, in Mathieu's case the elastic constant, as well as the focal distance $c$, and $a$ (which Mathieu called $R$) is the separation constant, adjusted from Mathieu's earlier variable $N$ by changing from $\cos^2\alpha = (1+\cos2\alpha)/2$ and $\cosh^2\beta = (1+\cosh2\beta)/2$ to the double-angle forms, for a reason that will become apparent.
Mathieu used $h^2$ where we have~$q$ here, and that notation is still occasionally used.

The boundary conditions of the original problem reflect the elliptic geometry. The angular coordinate $\alpha$ runs from $0$ to $2\pi$ (or, of course, from $-\pi$ to $\pi$), requiring periodicity.  Often the problem is taken to have $\pi$-symmetry, to reflect the symmetry between the top and the bottom of the ellipse.

\subsubsection{Basic properties: orthogonality, eigenvalues}

Mathieu noted that the theory of Jacques Charles Fran\c{c}ois Sturm\footnote{Sturm was the successor of Poisson in the chair of mechanics in the Facult\'e de Sciences, Paris. This fact has a certain poignancy when read together with the obituary of Mathieu~\cite{duhem1892emile}, where we learn that Mathieu had desired that chair.} (1803--1855) implied that, for real~$q$,  what we now call the Sturm--Liouville form of the Mathieu equation could be written as
\begin{align}
\mathcal{L}(y) :=    \frac{d}{dx}\left( L(x)\frac{dy}{dx}\right) - G(q; x)y = N r(x) y(x)\>,
\end{align}
(here, trivially, $L(x)=1$, $r(x)=1$, and $G(q;x) = 2q\cos2x$, while the eigenvalue $N$ has been moved to the right hand side)
and that several useful properties naturally followed. See Chapter 39 of~\cite{korner1989fourier}, and in particular Lemma 39.4 there; note that a \emph{necessary} element of the proof of the properties described in that lemma is that $G(q;x)$ be \emph{real}.  See also~\cite{pryce1993numerical} for numerical treatment of Sturm--Liouville problems.

Strictly speaking, Mathieu had to extend the now-classical theory to the case of periodic boundary conditions in order to establish what he needed.  Indeed, he also showed how to compute the eigenvalues by an interesting perturbative argument which we will take up in section~\ref{sec:Mathieuperturb}.  Before that, he established an inequality around each neighbourhood of $g^2$, the square of an integer; to us, this seems a convincing argument for the existence of each eigenvalue (and thus of the infinite collection) for small real~$q$.

\begin{figure}
\centering     
\subfigure[Some $\ce_{2k}(\alpha)$ for $q=1.5$]{\label{fig:evenpi}\includegraphics[width=60mm]{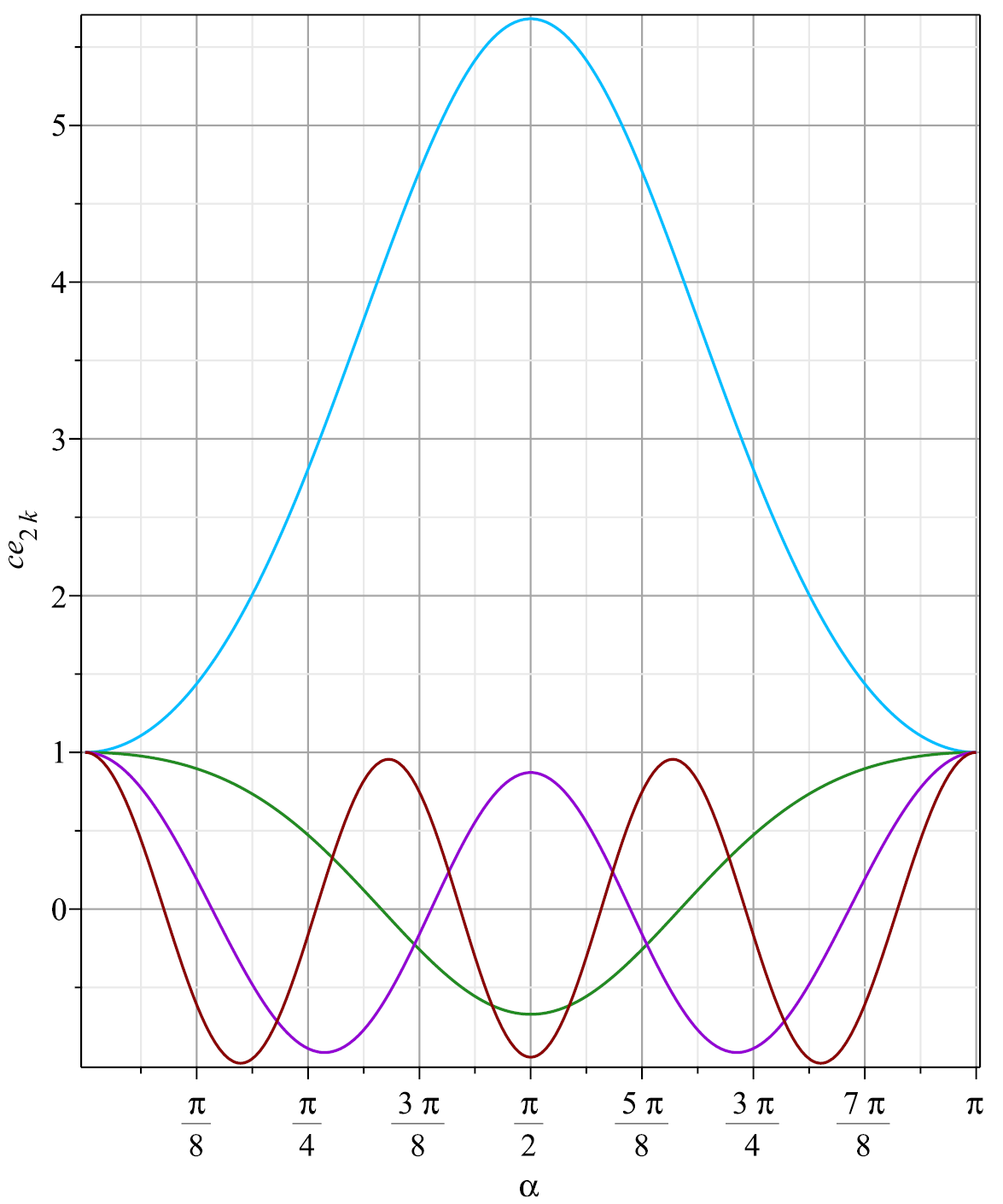}}
\subfigure[Some $\se_{2k}(\alpha)$ for $q=1.5$]{\label{fig:oddpi}\includegraphics[width=60mm]{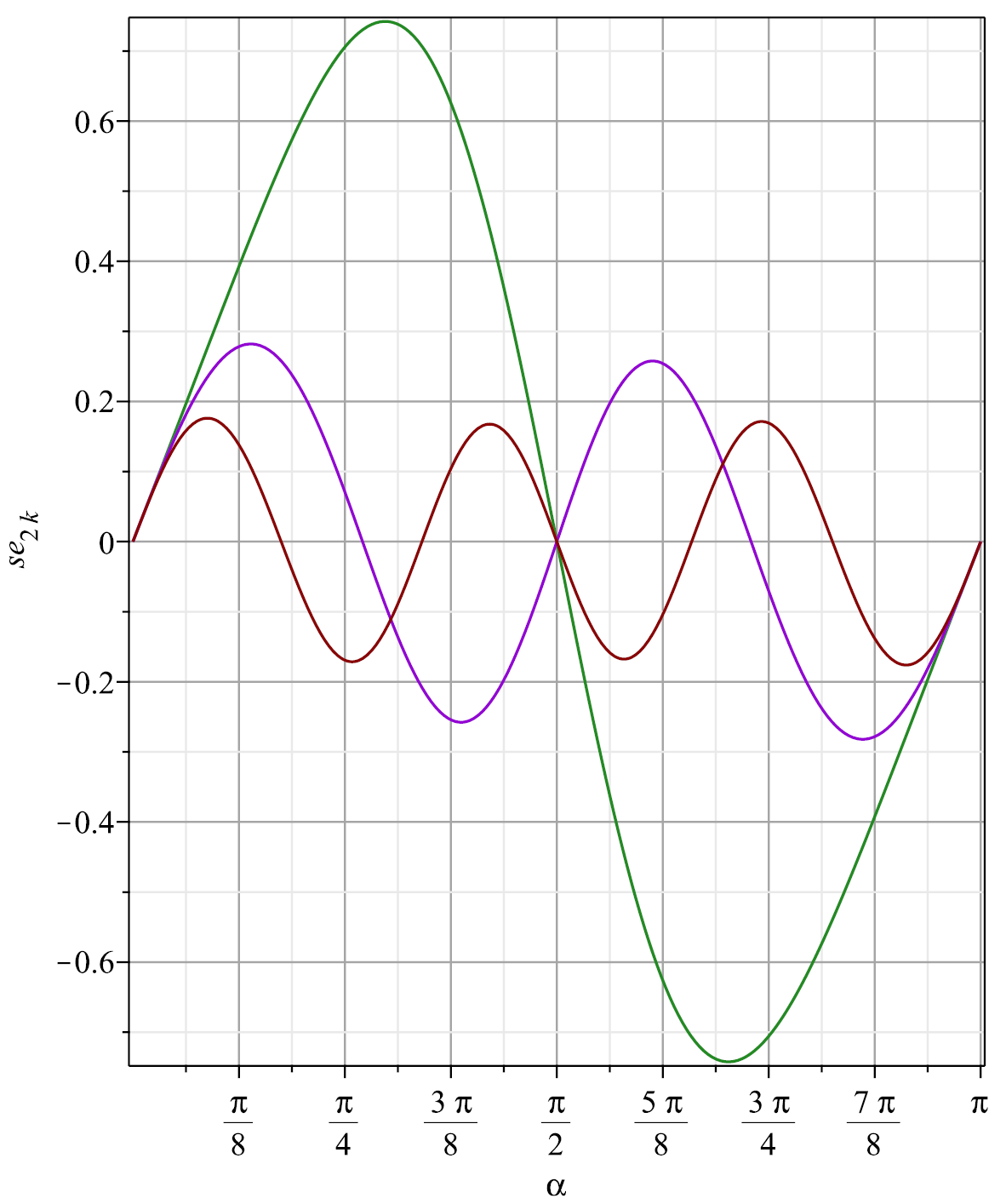}}
\newline
\subfigure[Some $\ce_{2k+1}(\alpha)$ for $q=1.5$]{\label{fig:eventwopi}\includegraphics[width=60mm]{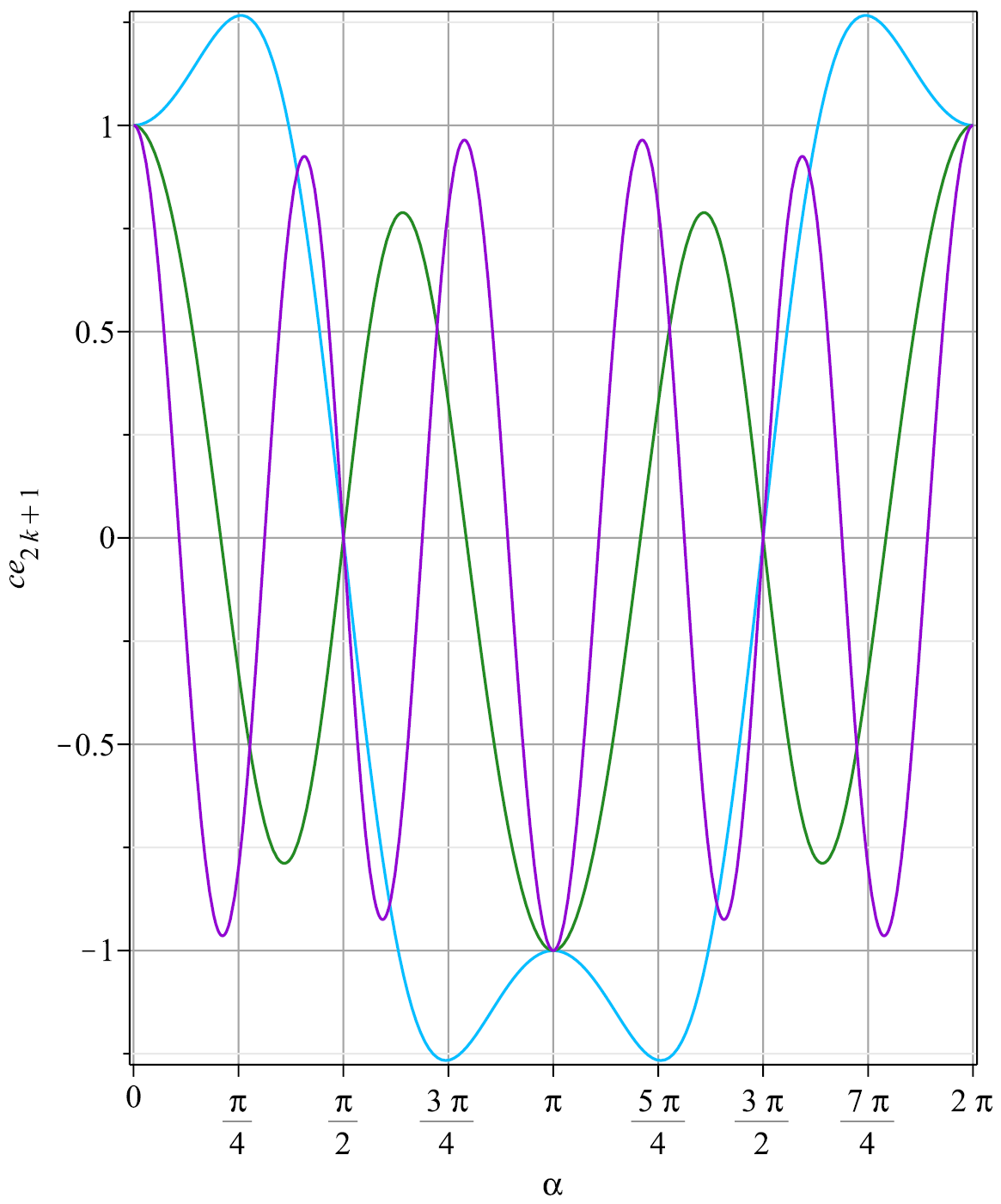}}
\subfigure[Some $\se_{2k+1}(\alpha)$ for $q=1.5$]{\label{fig:oddtwopi}\includegraphics[width=60mm]{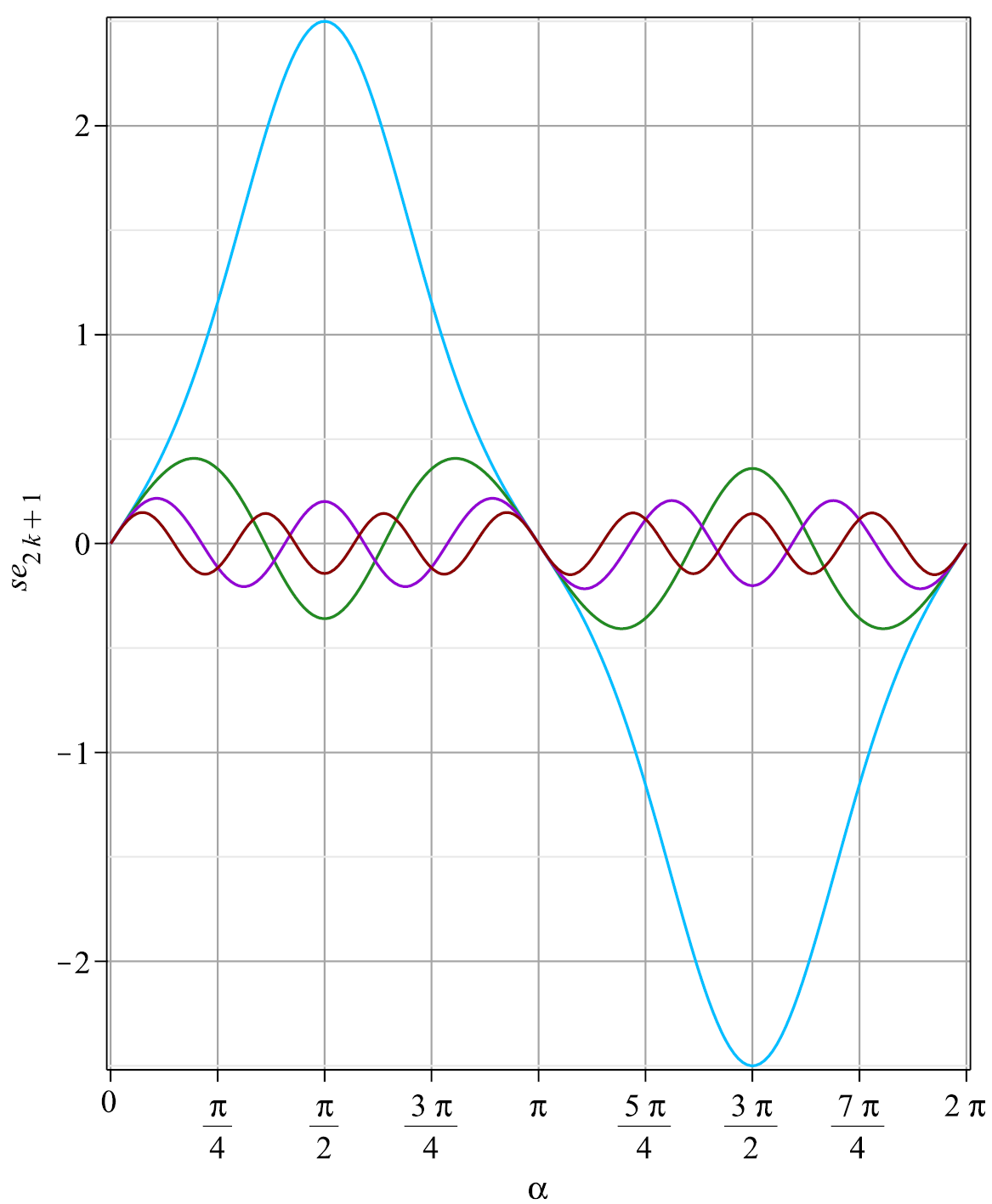}}
\caption{The first few Mathieu functions with $q=1.5$. The normalization shown here has $y(0)=1$ in the case of the even functions and $y'(0)=1$ in the case of the odd functions. Period $\pi$ functions are shown in the top row, period $2\pi$ functions in the bottom row.\label{fig:Mathieufnsq2}}
\end{figure}

The Sturm--Liouville theory shows that given a real value of $q$ there are a countable number of eigenvalues $N$ of the Mathieu equation, conventionally written either $N= -a_k(q)$ or $N = -b_k(q)$ depending on the type, which if $q > 0$ can be arranged in the sequence $a_0(q) < b_1(q) < a_1(q) < b_2(q) < a_2(q) < \cdots$, with the eigenvalues tending to infinity.
Next, to each eigenvalue there corresponds an eigenfunction, unique up to normalization. These eigenfunctions are now called the Mathieu functions, and they come in four classes. If they are \emph{even} and of period $\pi$ they are denoted by $\ce_{2k}(x)$, for $k\ge0$. If they are even and of period $2\pi$ they are denoted by $\ce_{2k+1}(x)$, for $k\ge0$.  If they are \emph{odd} and of period $\pi$ they are denoted by $\se_{2k}(x)$ for $k\ge1$. If they are odd and of period $2\pi$ they are denoted by $\se_{2k+1}(x)$ for $k\ge 1$.  See figure~\ref{fig:Mathieufnsq2} for a representative graph of a few low-frequency Mathieu functions, with $q=1.5$.
Mathieu established that these eigenfunctions are orthogonal with respect to the bilinear form defined by
\begin{align}
    \left\langle y_k, y_\ell \right\rangle := \int_0^{2\pi} y_k(x) y_\ell(x)\,dx = \mathrm{const}\cdot[\mathcal{C}\ {\bf\mathrm{and}}\  k=\ell]\>.
    \label{eq:orthogonality}
\end{align}
We prove this in detail in appendix~\ref{app:orthogonality}.
By the proposition $\mathcal{C}$ we mean here a proposition that is true if $y_k$ and $y_\ell$ are in the same class; that is, both are even and of period $\pi$, both are odd and of period $\pi$, both are even and of period $2\pi$, or both are odd and of period $2\pi$.
We have used here the Iverson convention $[\cdot ]$ to mean $1$ if the proposition inside the brackets is true and $0$ otherwise, because we prefer its generality over the more restrictive Kroenecker delta symbol, and do not think it will be confused with citations~\cite{Knuth1992}.
In other words, eigenfunctions of one class are orthogonal to eigenfunctions of any other class using this bilinear form, which is an inner product if $q$ and $a$ are real.
If appropriate, one could integrate to $\pi$ instead of $2\pi$.
The exact method used for normalization is a matter of convention. Finally, these classes of orthogonal eigenfunctions are \textsl{complete}: every reasonable\footnote{As with Fourier series, one can prove convergence for a quite wide class of functions; however, for utility and \emph{rapid} convergence, one needs a large number of continuous derivatives, and the more smoothness the better.} even/odd period-$p$ function $f(x)$
can be expanded in a convergent series
\begin{align}
    f(x) = \sum_{k\ge0} \alpha_k y_k(x)\>.
\end{align}
Periodic functions which are neither even nor odd can of course be written
as a sum of an even function and an odd function
\[
f(x) = \frac{f(x)+f(-x)}{2} + \frac{f(x)-f(-x)}{2}
\]
and thus will use both classes of the given period in its expansion.
Since the coefficients $\alpha_k$ can be determined by orthogonality, this series is expected to be practical and to enable us to find computationally useful solutions to the original PDE by matching the boundary condition at the edge of the ellipse.

We emphasize that Mathieu only established this for real~$q$. The case of complex~$q$ is a different matter, as noted by~\cite{meixner} and by~\cite{blanch1969double}.
The Sturmian theory fails there (see appendix~\ref{app:orthogonality})
because there may be (and in fact are) \textsl{double} eigenvalues, and in that case Blanch points out that for the eigenfunction associated with the double eigenvalue $\left<y_k, y_k\right> = 0$ and thus normalization by making the bilinear form equal to a nonzero constant is not possible.  We will return to this later.
\subsubsection{Normalizations}
Mathieu normalized the solutions of his equation in a way that might seem curious to modern eyes.  First, he noted that it is easy to see that the solution $y(x)$ of any linear second order ordinary differential equation (ODE) may be written as $y = P_1 + P_2$, where $P_1$ is either a maximum or a minimum at $x=a$  while $P_2$ is zero at $x=a$ ($a$ is arbitrary).  A moment's reflection shows that Mathieu was correct, and that this is true for any $a$ in the domain of definition of $y(x)$.  What is curious is that this separates a second order equation into two parts, each of which imposes only \emph{one} condition: $P_1'(a) = 0$, or $P_2(a)=0$. Clearly any multiple of $P_1$ or of $P_2$ will satisfy the same condition, but only one combination of such functions will equal $y(x)$.  This left Mathieu free to normalize his functions in any way convenient to him, and he took great advantage of it, in particular in his perturbative solutions.  To normalize his functions, he chose to make the coefficient of $\cos g\alpha$ in its series expansion to be unity (and similarly the coefficient of $\sin g\alpha$ in the case of odd eigenfunctions); see appendix~\ref{app:MathieuPerturbForm} for a comparison with one modern normalization.  As Ince pointed out later in~\cite{ince1927mathieu}, this is \emph{not} always possible because for some values of~$q$, even for some real values of $q$, the coefficient of $\cos g\alpha$ is actually zero; but it is at least \emph{almost} always possible in the modern sense; that is, except on a set of measure zero in parameter space.

Other normalizations are in use today: we use the universally-possible normalization discussed at the end of section 20.5 in~\cite{abramowitz}, namely, $y(x) = a \ce_g(x) + b\se_g(x)$ and we specify that $\ce_g(x)$ satisfy not only $\ce_g'(0)=0$ but also $\ce_g(0) = 1$, and similarly $\se_g(0)=0$ and $\se_g'(0)=1$.  As Blanch states in the aforementioned reference, conversion between normalizations is ``rather easy,'' but we wanted one that would always work.

However, in many published papers and codes the ``norm'' using the bilinear form~\eqref{eq:orthogonality} is nominally enforced to be $\pi$ (or $2\pi$), which can (again) \emph{almost} always be done, but not universally: at double eigenvalues, the ``norm'' must vanish.  This is a little better than Mathieu's normalization in that all such exceptional values of~$q$ must be complex: the norm will always be nonzero for real~$q$.  But for some complex~$q$, which we look at carefully in this present paper, the ``norm'' does vanish.  This means that in modern terms the norm is an ``indefinite'' norm~\cite{arbenz2004jacobi}, and requires some care in handling. To emphasize, in this paper we do not normalize by the bilinear form, but instead choose to enforce initial conditions as above; this is also done elsewhere in the literature, but not commonly.

\subsubsection{Perturbative solution: anticipating Lindstedt\label{sec:Mathieuperturb}}
In the 1868 paper under discussion, Mathieu developed series solutions for the first few eigenvalues,  $a_{k}(q)$ and $b_k(q)$ in modern notation; in some cases to sixth order in~$q$ (twelfth order in $h$).
It is interesting to note that to do so he essentially used what we now know as \emph{anti-secularity}: he chose series coefficients in the eigenvalue expansion in order to eliminate secular terms in the expansion for the eigenfunction and thereby enforced periodicity of the solution.  This notion is typically introduced nowadays as the Lindstedt--Poincar\'e method. There are alternatives, now, too: one can instead use the method of multiple scales, or in an even more modern way, use \emph{renormalization}.  See for instance~\cite{corless2019backward} and the references therein for more details of those methods.

When Mathieu published his memoir in 1868, Anders Lindstedt was in his early teens and his work on perturbation~\cite{Lindstedt1882} was fourteen years in the future.  Mathieu might have good grounds for a claim to priority, even though (perhaps) Lindstedt's work was somewhat more general\footnote{Lindstedt's method applies to \emph{weakly nonlinear} equations, which are linear if the small parameter is set to zero.  Lindstedt suggested simultaneous expansion of the eigenvalue.  This generates a sequence of linear equations to solve for subsequent terms, and the overall process is not much different from what Mathieu did.}.  Mathieu's use of anti-secularity is clear, however, once one tries to retrace his steps; it seems very natural, although Mathieu does not comment on it explicitly.  Indeed, his section 11 which details the perturbation solution reads more like an informal summary of notes of how to proceed, with many details left out.  Nonetheless, using anti-secularity to enforce periodicity is exactly what he did.  He also made several elegant uses of his freedom to normalize in the problem in order to reduce the labour involved.  We have implemented his solution in a computer algebra system, to retrace his steps and fill in the details; some of our computations are compared to his in Appendix~\ref{app:MathieuPerturbForm}.

Mathieu's first computation was to find the even period $2\pi$ solution of equation~\eqref{eq:etaeq} when $q=h^2$ was small and the eigenvalue $a$ approached $g^2$, the square of an unspecified integer.  The solution in his notation and with his normalization\footnote{Ince pointed out in~\cite{ince1933xxii} that this fails for some $q$; we discuss this later.} and to fewer terms than he calculated to is:
\begin{align}\label{eq:Mathieugenericg}
  \ce_g(\alpha) = & \cos g\alpha + \left( {\frac {\cos
 \left( g-2 \right)\alpha  }{4(g-1)}}-{\frac {\cos
 \left( g+2 \right)\alpha  }{4(g+1)}} \right) {h}^{2} \nonumber \\
 & + \left( {
\frac {\cos \left( g-4 \right)\alpha  }{ 32(
g-2)  \left( g-1 \right) }}+{\frac {\cos
 \left( g+4 \right)\alpha   }{  32(g+2)  \left( g+1
 \right) }} \right) {h}^{4} \nonumber\\
 &  + \left( {\frac {\cos
 \left( g-6 \right)\alpha   }{  384(g-4)  \left( g-2
 \right)  \left( g-1 \right) }}+{\frac { \left( {g}^{2}-4\,g+7
 \right) \cos  \left( g-2 \right)\alpha   }{
128(g-2)  \left( g+1 \right)  \left( g-1 \right) ^{3}}}\right.\nonumber\\
& \left.-{
\frac { \left( {g}^{2}+4\,g+7 \right) \cos   \left( g+2
 \right)\alpha   }{  128(g+2)  \left( g+1 \right) ^{3
} \left( g-1 \right) }}-{\frac {\cos  \left( g+6
 \right)\alpha  }{ 384(g+2)  \left( g+3 \right)
 \left( g+1 \right) }} \right) {h}^{6} + O(h^8)
\end{align}
As Mathieu noted, this series is valid only for large enough integers $g$.  He also correctly computed the corresponding eigenvalue (he called it $R$ in this part of his paper) as
\begin{equation}\label{eq:Mathieueigeng}
  a = {g}^{2}+{\frac {{h}^{4}}{  2(g-1)  \left( g+1 \right) }}
+{\frac { \left( 5\,{g}^{2}+7 \right) {h}^{8}}{ 32\,\left( g-2
 \right)  \left( g+2 \right)  \left( g-1 \right) ^{3} \left( g+1
 \right) ^{3}}} + \cdots\>.
\end{equation}
Mathieu then goes on to show how to compute perturbation solutions for specific, smaller, frequencies $g$.  See appendix~\ref{app:MathieuPerturbForm} for details.

The idea of a series expression for the Mathieu functions was, of course, natural for the time.  Whether the idea of enforcing periodicity by expanding the eigenvalue in series was original to Mathieu, we do not know; but its presence in his paper certainly predates Lindstedt's work.

For Mathieu, $q=h^2$ was real, and small (if the interfocal distance $2c$ was small).  In many modern applications, $q$ might be complex, or large, or both.
It took many years of further research by others to go beyond these series.

\subsubsection{D-finite, or `holonomic', formulation}
After finding these perturbation expansions, Mathieu took a different approach: he changed variables, first with $\nu = \cos\alpha$, whereupon the Mathieu equation becomes (equation 28.2.3 in the DLMF)
\begin{equation}\label{eq:Dfinite1}
  (1-\nu^2)\frac{d^2P}{d\nu^2} - \nu \frac{dP}{d\nu} + (a + 2q(1-2\nu^2))P = 0
\end{equation}
and alternatively by $\mu = \sin\alpha$, (Mathieu used $\nu'$ rather than $\mu$, which we originally kept; but a referee pointed out that this was confusing, and on second thought we agreed) whereupon the Mathieu equation becomes
\begin{equation}\label{eq:Dfinite2}
 (1-\mu^2)\frac{d^2P}{d\mu^2} - \mu \frac{dP}{d\mu} + (a - 2q(1-2{\mu}^2))P = 0\>.
\end{equation}
The DLMF gives yet another algebraic form, using the change of variables $\zeta=\sin^2\alpha$.  These are interesting for several reasons, and we will mention in section~\ref{sec:Whittaker} some of the further properties that can be deduced from these equations.  What Mathieu used them for first was to generate recurrence relations for their Taylor series expansion, which can be used about any point.  This can also be done for the original formulation, of course, but an important difference is seen between the two forms: in the original formulation, the Taylor series recurrence depends on \emph{all} previously computed terms; for the algebraic formulations, the recurrence relation depends only on a \emph{finite number} of the previous terms.  We give the recurrence relation for the original formulation in equation~\eqref{eq:TaylorRecurrence} in section~\ref{sec:alg}, while the recurrence relations for the algebraic formulations are already given in~\cite{mathieu1868memoire}. For instance, for equation~\eqref{eq:Dfinite1} if
\[
P(\nu) = \sum_{k\ge0} \rho_k (\nu-\nu_0)^k
\]
then (after the first few terms which have to be separately investigated),
\begin{align*}
\rho_{n+4} =& \frac{1}{(\nu_0^2-1)(n+3)(n+4)}\Bigg( 4q\rho_n +8\nu_0q\rho_{n+1} \\
&  + \left(
4{\nu_0}^{2}q-{n}^{2}+a-4n-2q-4 \right) \rho_{n+2} -
 \left( 2n+5 \right)\left( n+3
 \right) \nu_0\rho_{n+3}
\Bigg)\>.
\end{align*}
We computed this recurrence relation automatically from equation~\eqref{eq:Dfinite1} by using the \texttt{gfun} package in Maple, specifically its \texttt{diffeqtorec} command~\cite{salvy1994gfun}.  Mathieu gave the simpler form at $\nu_0 = 0$, and separated out the even and odd series so that each recurrence relation involved only \emph{two} prior terms.
This means that the Mathieu functions are what is now called \emph{$D$-finite} or \emph{holonomic}, and can therefore be computed to high precision with an asymptotically fast algorithm.  See~\cite{benoit2017rigorous,mezzarobba2010numgfun,mezzarobba2012note,van2001fast}.

Mathieu then considered properties of the functions that could be deduced from these power series, which could also be interpreted as series in powers of $\cos\alpha$ or of $\sin\alpha$.  In particular, he used them to count real roots.
\subsubsection{Modified Mathieu Functions\label{sec:modifiedMathieuFunctions}}
\begin{figure}
  \centering
  \subfigure[$\Ce_0(\beta;2)=\ce_0(i\beta;2)$]{\label{fig:modce0}\includegraphics[width=6cm]{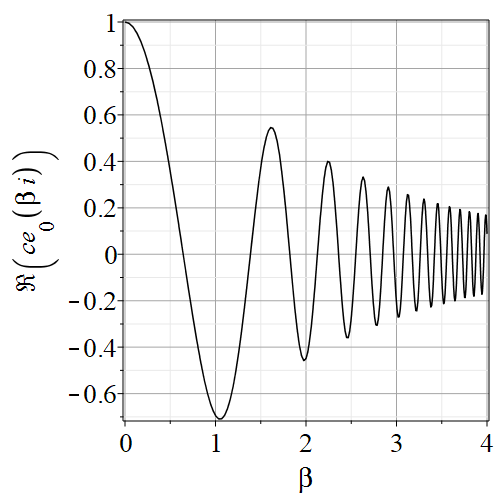}}
  \subfigure[$\Se_1(\beta;2)=\Im(\se_1(i\beta;2))$]{\label{fig:modse1}\includegraphics[width=6cm]{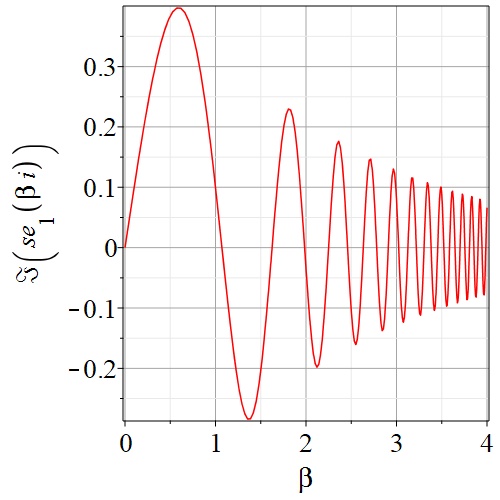}}
  \caption{(Left) A graph of $\Ce_0(q,\beta)$ when $q=2$. As the argument $\beta$ increases, the function becomes increasingly oscillatory.  This value of~$q$ could be used for an elliptical drum whose vertical dimension was such as to coincide with a zero of this function (units depending on the locations of the foci at $\pm c$.)
  (Right) A graph of $\Se_1(\beta;q)=\Im(\se_1(i\beta;q))$ when $q=2$. }\label{fig:modifiedMathieu0q2}
\end{figure}

In order to solve the vibrating drum problem, Mathieu had also to solve equation~\eqref{eq:xieq}.  He chose to do this in a way analogous to the series solution for Bessel's equation that he gave in his introduction, and discussed how to find the real roots thereof, which are necessary for matching the fixed boundary condition at the elliptical rim of the membrane.  In our terms, the modified Mathieu functions are simply the Mathieu functions with purely imaginary argument: $\Ce_g(x;q) = \ce_g(ix;q)$ and $\Se_g(x;q) = -i\se_g(ix;q)$.  We show two such functions in figure~\ref{fig:modifiedMathieu0q2}.

We may now discuss the details of figure~\ref{fig:contourcemode3}.  We chose a drum shape with aspect ratio $5:3$.
We also chose to look at an even mode corresponding to $a_3(q)$, so this means our pure tone will be described by $\Ce_3(\beta;q)\ce_3(\alpha;q)$.  We decided that it should be, for the introduction, a simple picture with no elliptical nodal lines, only hyperbolic; this meant that we were looking for the first zero of $\Ce_3(\beta;q)$. Since the coordinates are $x = c\cosh\beta \cos\alpha$, $y = c\sinh\beta \sin\alpha$ (alternatively, $x+iy = \cos(\alpha - i\beta)$) we will want $c\cosh\beta = 5$ and $c\sinh\beta = 3$; this gives $\beta = \ln 2$ and $c=4$. Now we want the value of~$q$ so that $\Ce_3(\ln2;q) = 0$.  By zerofinding on~$q$, we find that $q \approx 8.5676$. We used simple bisection, because we had not at that time implemented differentiation with respect to~$q$.
The physics of the membrane would then give the frequency of oscillation $\sin 2\lambda m t$ via $q = \lambda^2 c^2$, with the membrane parameter $m$. This value of~$q$ gives the eigenvalue $a_3 \approx 14.6695$.  The  hyperbolic nodal lines are at approximately $\alpha=\pm 0.9857$, $\pm \pi/2 $, and $\pm 2.156$. The contours plotted were at levels $\pm[0,11,22,33,44]/40$, remembering that our normalization is so that $\ce_3(0;q) = 1 = \Ce_3(0;q)$.  More details and more figures can be found in~\cite{corless2020pure}.

\subsubsection{A short biography of Mathieu}
\'Emile L\'eonard Mathieu (15 May 1835---19 October 1890) attended the \'Ecole Polytechnique de Paris, taking the entrance examination in 1854.  He defended his doctoral thesis in pure mathematics in 1859, before a committee consisting of Lam\'e, Liouville, and Serret. He was well-regarded by the community at the time for some of his work, and indeed is still known today for what are called ``Mathieu groups''. Apparently, though, he did not receive the positions that he truly wanted; he then turned to applied mathematics (specifically, Mathematical Physics) to see if that would ``more engage the interest of scientific men''~\cite{duhem1892emile}.  In spite of this change, and in spite of winning a Gold Medal in 1867, he was repeatedly passed over, and in 1869 he left for a position at Besan\c{c}on, becoming Chair of Pure Mathematics there in 1871. He remained there until 1873 when he took up the Chair in Pure Mathematics at Nancy, where he remained until his death in 1890. His obituary is a very interesting read.  In it Duhem~\cite{duhem1892emile} praises Mathieu's achievements, calling him the natural successor of Poisson, and blaming ``fashion'' and ``politics'' for passing him over (thus suggesting that the fashion and politics in science and mathematics were as alive and well then as they are today!).

His being passed over may have been a result of changing fashion, as Duhem contends, or may simply have been an artifact of the Golden Age (for mathematics) that he lived in.  For instance, one of the positions that he wanted, a Chair at Sorbonne, was awarded to Picard instead, who was Hermite's son-in-law and Mathieu thought this was a scandal; to be fair, it could be argued that Mathieu's record was superior at the time. But in other cases it is clearer now. Mathieu complained that he came second to another favourite of Hermite for another Chair, in this case to Hermite's student Henri Poincar\'e. The modern view must be different from Mathieu's: It would be very difficult today to imagine choosing Mathieu over Poincar\'e for any Chair.

We find in~\cite{bolmont:2015:Mathieu} (which contains an interesting view of the tension between Paris and the provinces, and passages from Mathieu's correspondence) still other reasons why Mathieu was perpetually not chosen, and it seems that the judgement of the Establishment that others were more worthy was, in the end, justifiable.  Even more, it was a turbulent time in France generally: the coup when Napoleon III took power happened in 1851, and the Franco-Prussian war in which Napoleon III was captured ended when the Prussians took Paris in 1870, just as one example of how the larger world may have intruded on academic life.  It is quite believable that in these turbulent social circumstances many deserving people did not receive all the recognition that they had earned.

In spite of all the difficulties of the times, however, Mathieu left a very significant body of mathematical advances for posterity.  We have surveyed only a small corner of his work in this present paper, and even from just this it is clear today that he was one of the best mathematicians of his age, which included some of the greatest ever known.

\subsection{Sir Edmund Taylor Whittaker (1873--1956)\label{sec:Whittaker}}
\subsubsection{Whittaker \& Watson on Mathieu functions}
It was Whittaker, another of the giants of nineteenth- and early twentieth-century mathematics, who bestowed the name \emph{the Mathieu equation} on equation~\eqref{eq:mathieueq} and the name \emph{the Mathieu functions} on the \emph{even} and \emph{odd} periodic solutions of the Mathieu equation, ``and these only''~\cite{whittaker1927course}. According to Whittaker's obituary~\cite{temple1956edmund}, the name was given in the paper in the fifth ICM~\cite{whittaker1912functions}.  Attributing a person's name to an equation or a function is a significant event in mathematics because people (even mathematicians) are social animals, and we simply \emph{pay more attention} when a person's name is involved.  Such namings often get it wrong, of course: ``Stigler's Law of Eponymy'' states that no scientific discovery is named after its original discoverer.  For instance, Puiseux series are named after Victor Puiseux (1820--1883) but were in fact discovered by Newton, and for another, Young's modulus---written about in 1807 by Thomas Young---was described 25 years prior to that by Ricatti, but in any case was also discovered by Newton.  In this case however we think Whittaker got it right, and Mathieu deserves the credit.  The more descriptive ``elliptic cylinder functions'' is also used on occasion, and was also used by Whittaker.  Nowadays Mathieu functions and Modified Mathieu functions are also called Angular Mathieu functions and Radial Mathieu functions, in accordance with their roles in the elliptic coordinate system: $\beta$ is more like a radius, and $\alpha$ more like an angle.

A full chapter of ``A Course of Modern Analysis'' by Whittaker and Watson~\cite{whittaker1927course} is devoted entirely to Mathieu functions.  Any serious study of these functions should begin with that book.  For some reason, the authors rescaled the equation and have $16q$ there instead of the $2q$ (or $2h^2$) that Mathieu had.  This is of no real consequence.  More interestingly, they use the same normalization convention that Mathieu did in his perturbation series computations: they take the coefficient of $\cos g z$ in the expansion of $\ce_g(z)$ to be unity, and similarly the coefficient of $\sin gz$ in the expansion of $\se_g(z)$ to be unity.  This differs from modern practice.

Several important theorems are established in that chapter: the orthogonality under the bilinear form of equation~\eqref{eq:orthogonality} is proved (by appeal to results from an earlier chapter), several integral equations are established, and the Floquet theory of the solutions to periodically-forced ODEs\footnote{Achille Marie Gaston Floquet (15 December 1847, \'Epinal – 7 October 1920, Nancy)} is applied to the \emph{non}-periodic solutions of the Mathieu equation, and more generally to Hill's equation.  The Floquet theory shows that solutions must exist in the form $\exp(\mu z)\phi(z)$ where $\phi(z)$ is periodic with period $\pi$ (because the periodic forcing of the Mathieu equation has that period) and $\mu$ (actually, in modern works starting in~\cite{abramowitz}, $\nu$ where $\mu = i\pi \nu$) is called the \emph{characteristic exponent}.  Regions where $\Re(\mu)>0$ indicate that the solution is unstable.  The Mathieu equation is an important but special example for Floquet theory, because the Mathieu equation is also \emph{even} and therefore $\exp(-\mu z)\phi(-z)$ is also a solution; this implies that for \emph{both} solutions to be stable, we must have $\Re(\mu)=0$; moreover the regions in $(a,q)$ parameter space where the solutions are stable are bounded by \emph{characteristic lines} containing periodic solutions.  For more information about Floquet theory, an excellent introduction can be found in~\cite{teschl2012ordinary}.

Below is one of the integral equations established in~\cite{whittaker1927course}: if $G(\eta)$ is an even Mathieu function, then there is a characteristic number $\lambda$ for which (translating from the $16q$ convention of Whittaker to the $2q$ convention used in this paper)
\begin{equation}\label{eq:WhittakerIntegralEquation}
  G(\eta) = \lambda \int_{-\pi}^{\pi} e^{2\sqrt{q}\cos\eta\cos\theta}G(\theta)\,d\theta\>.
\end{equation}
We have not tried using this as a method of computing even Mathieu functions, although Whittaker and Watson claim that it ``affords a simple manner of constructing'' them.

Whittaker and Watson attribute  to Lindemann the change of variable $\zeta = \cos^2 z$ (the DLMF has $\sin^2 z$ but this is the same) which turns the Mathieu equation into an \emph{algebraic differential equation} or ADE, even though Mathieu had done the same using $\nu = \cos z$ and $\mu = \sin z$, as previously discussed.  An ADE is, loosely speaking, an ordinary differential equation with rational functions---usually polynomials---as coefficients; this is not the same as a \emph{differential algebraic equation} or DAE, which notion is more familiar to numerical analysts.
Whittaker and Watson make further references to works of Abel, Stieltjes, Sylvester and others in the study that results from this observation.
Whittaker and Watson make an asymptotic connection of Mathieu functions to Bessel functions using this form of the equation: in this formulation it is more natural, but still a bit involved.

Whittaker and Watson also show that the Fourier series for Mathieu functions are well-defined, at least for small enough~$q$, by exhibiting convergent power series for the coefficients.  This marks an important step.

\subsubsection{A short biography of Whittaker}
We take the following material from the remarkably well-written and well-informed Wikipedia article on Whittaker, supplemented by the mathematical obituary written by G.~Temple~\cite{temple1956edmund}.
Sir Edmund Taylor Whittaker (24 October 1873--24 March 1956) studied mathematics and physics at Trinity College, Cambridge, starting in 1892.  He was elected Fellow of the College in 1896 and continued there until 1906.  He was elected Fellow of the Royal Society of London in 1905. He then became Royal Astronomer of Ireland, and Andrews Professor of Astronomy at Trinity College Dublin. In 1911 he joined the University of Edinburgh and in 1912 was elected Fellow of the Royal Society of Edinburgh, later becoming President.  He was knighted by George VI in 1945  ``for service to mathematics''.

His mathematical obituary previously cited runs $22$ pages, plus a facing portrait and a further four pages listing Whittaker's complete works.  It is a mathematical biography well worth reading.  The topic headings are 1.---Algebra 2.---Interpolation (exhibiting significant work in statistics and in numerical analysis, which the anonymous Wikipedia author also takes particular note of) 3.---Automorphic functions 4.---Astronomy 5.---Potential theory and special functions [It is here that we learn that Whittaker's 1912 paper established that the Mathieu equation is ``the simplest linear differential
equation which is not reducible to a particular or degenerate case of the hypergeometric differential equation''; and it is also here that we learn that E.~L.~Ince was a research student working with Whittaker in Edinburgh and it was from this period that Ince gained his interest in Mathieu functions.] 6.---Dynamics 7.---Relativity and Electromagnetic Theory 8.---Quantum Theory 9.---Scientific Books and Monographs and 10.---Historical and philosophical writings. The very final section, ``11.---Influence'' ends on a slightly ironic note: Temple claims that Whittaker's influence was felt not just by his works, but also by his coinage of mathematical terms, some of which are listed in the final paragraph.  Sadly, of all those terms listed, namely `isometric circle', `adelphic integral', `cotabular functions',
`cardinal function', `congruent hypergeometric function', `Mathieu function', and
`calamoids', few apart from the Mathieu functions are widely known today.  We confess to curiosity as to what he meant by a ``calamoid": the term seems to have survived only in botany, and has to do with palm leaves.

But the mountain of scientific achievement that Whittaker built still stands on its own.

\bigskip\goodbreak
\subsection{Edward Lindsay Ince (1891--1941)}
\subsubsection{Fourier series recurrence relations\label{sec:Ince}}
Mathieu's perturbative solution of his equation and the resulting series in powers of $\cos\alpha$ or $\sin\alpha$ already suggest that it is natural to think of using Fourier series to represent these periodic functions.
It seems, however, that it was Ince who first made practical use of Fourier series for Mathieu functions.

The basic idea is simple: we expand our proposed periodic solution $y(x)$ in Fourier series:
\begin{align}
    y(x) = \sum_{k\ge 0} A_k \cos kx + \sum_{k\ge 1} B_k \sin kx \>.
\end{align}
Inserting this series into the Mathieu equation~\eqref{eq:mathieueq}, using the multiplication identities
\begin{align*}
    \cos(2x)\cos(kx) &= \tfrac{1}{2}\Bigl( \cos(k+2)x + \cos(k-2)x \Bigr) \\
    \cos(2x)\sin(kx) &= \tfrac{1}{2}\Bigl( \sin(k+2)x + \sin(k-2)x \Bigr)\>,
\end{align*}
and then equating coefficients of $\cos kx$ and of $\sin kx$,
gives us a collection of recurrence relations.  By circumstance (which Mathieu made simpler by converting from $\cos^2 x$ to the double-angle form), the $A_k$ coefficients only involve other $A_k$ coefficients, and moreover only those that differ by $2$ in index; similarly for the $B_k$ coefficients.  The edge conditions (those recurrences specialize when $k=0$ and $k=1$ to slightly different forms) produce a set of equations that can be written as follows:
\begin{align}
    a A_0\phantom{XXX} - q &A_2  &= 0 \label{eq:recurrencebase}\\
    -2q A_0 + (a-4) &A_2\phantom{XXX} - q A_4  &= 0
\end{align}
and, for all $k\ge2$,
\begin{align}
    -qA_{2k-2} + (a-(2k)^2)A_{2k} - q A_{2k+2} = 0\>.
\end{align}
The odd cosine coefficients must instead satisfy
\begin{align}
    (a-1-q) A_1 - q &A_3  &= 0
\end{align}
and, for all $k\ge 1$,
\begin{align}
    -qA_{2k-1} + (a-(2k+1)^2)A_{2k} - q A_{2k+3} = 0\>.
\end{align}
The even sine coefficients give
\begin{align}
    (a-4)& B_2 - q B_4  &= 0
\end{align}
and, for all $k\ge2$,
\begin{align}
    -qB_{2k-2} + (a-(2k)^2)B_{2k} - q B_{2k+2} = 0\>.
\end{align}
Finally, the odd sine coefficients give
\begin{align}
    (a-1+q) B_1 - q &B_3  &= 0
\end{align}
and, for all $k\ge1$,
\begin{align}
    -qB_{2k-1} + (a-(2k+1)^2)B_{2k+1} - q B_{2k+3} = 0\>. \label{eq:Brecurrence}
\end{align}
All four of these sets of recurrence relations give rise to infinite tridiagonal matrices, each one symmetric except the first.  By an artifice, namely replacing $A_0$ by $A_0\sqrt{2}$, we may symmetrize even the first one.

\medskip\par
The infinite eigenvalue problem then becomes
\begin{equation}
 \left[ \begin {array}{cccccc} 0&\sqrt {2}q&0&0&0&\cdots\\ \noalign{\medskip}
\sqrt {2}q&4&q&0&0&\cdots\\ \noalign{\medskip}0&q&16&q&0&\cdots\\ \noalign{\medskip}0
&0&q&36&q &\cdots\\
\noalign{\medskip}0
&0&0&q&64 &\ddots\\
\noalign{\medskip}\vdots&\vdots&\vdots&\vdots&\ddots&\ddots\end {array} \right]
\left[ \begin{array}{c}
\sqrt{2}A_0\\
\noalign{\medskip}
A_2\\
\noalign{\medskip}
A_4\\
\noalign{\medskip}
A_6\\
\noalign{\medskip}
A_8\\
\noalign{\medskip}
\vdots
\end{array}
\right]
= a\left[ \begin{array}{c}
\sqrt{2}A_0\\
\noalign{\medskip}
A_2\\
\noalign{\medskip}
A_4\\
\noalign{\medskip}
A_6\\
\noalign{\medskip}
A_8\\
\noalign{\medskip}
\vdots
\end{array}
\right]\>.
\label{eq:tridceeven}
\end{equation}
The eigenvalues of this matrix are denoted $a_0(q)$, $a_2(q)$, $a_4(q)$, $\ldots$ and indeed for real~$q$ these occur in increasing order: $a_0(q) < a_2(q) < a_4(q) < \cdots$.  For complex~$q$ the situation is more complicated, but the set of eigenvalues remains discrete and countable, essentially because the diagonal of the matrix above contains entries that grow rapidly enough with the row index.

The other three symmetric tridiagonal matrices are constructed analogously and have analogous properties.  For completeness, we include them here:
\begin{equation}
 \left[ \begin {array}{ccccc} 1+q&q&0&0&\cdots\\ \noalign{\medskip}
q&9&q&0&\cdots\\ \noalign{\medskip}0&q&25&q&\cdots\\ \noalign{\medskip}0
&0&q&49&\ddots\\ \noalign{\medskip}\vdots&\vdots&\vdots&\ddots&\ddots\end {array} \right]
\left[ \begin{array}{c}
A_1\\
\noalign{\medskip}
A_3\\
\noalign{\medskip}
A_5\\
\noalign{\medskip}
A_7\\
\noalign{\medskip}
\vdots
\end{array}
\right]
= a\left[ \begin{array}{c}
A_1\\
\noalign{\medskip}
A_3\\
\noalign{\medskip}
A_5\\
\noalign{\medskip}
A_7\\
\noalign{\medskip}
\vdots
\end{array}
\right]
\label{eq:tridceodd}
\end{equation}
The eigenvalues of equation~\eqref{eq:tridceodd} are denoted $a_{2k+1}(q)$.

\begin{equation}
 \left[ \begin {array}{ccccc}
4&q&0&0&\cdots \\ \noalign{\medskip}q&16&q&0&\cdots\\ \noalign{\medskip}
0&q&36&q&\ddots\\ \noalign{\medskip}
0&0&q&64&\ddots\\
\noalign{\medskip}\vdots&\vdots&\vdots&\ddots&\ddots\end {array} \right]
\left[ \begin{array}{c}
B_2\\
\noalign{\medskip}
B_4\\
\noalign{\medskip}
B_6\\
\noalign{\medskip}
B_8\\
\noalign{\medskip}
\vdots
\end{array}
\right]
= a\left[ \begin{array}{c}
B_2\\
\noalign{\medskip}
B_4\\
\noalign{\medskip}
B_6\\
\noalign{\medskip}
B_8\\
\noalign{\medskip}
\vdots
\end{array}
\right]
\label{eq:tridseeven}
\end{equation}
The eigenvalues of equation~\eqref{eq:tridseeven} are denoted $b_{2k}(q)$.

\begin{equation}
 \left[ \begin {array}{ccccc} 1-q&q&0&0&\cdots\\ \noalign{\medskip}
q&9&q&0&\cdots\\ \noalign{\medskip}0&q&25&q&\cdots\\ \noalign{\medskip}0
&0&q&49&\ddots\\ \noalign{\medskip}\vdots&\vdots&\vdots&\ddots&\ddots\end {array} \right]
\left[ \begin{array}{c}
B_1\\
\noalign{\medskip}
B_3\\
\noalign{\medskip}
B_5\\
\noalign{\medskip}
B_7\\
\noalign{\medskip}
\vdots
\end{array}
\right]
= a\left[ \begin{array}{c}
B_1\\
\noalign{\medskip}
B_3\\
\noalign{\medskip}
B_5\\
\noalign{\medskip}
B_7\\
\noalign{\medskip}
\vdots
\end{array}
\right]
\label{eq:tridseodd}
\end{equation}
The eigenvalues of equation~\eqref{eq:tridseodd} are denoted $b_{2k+1}(q)$.

\medskip\par\noindent
\textbf{Remark}. If~$q$ is real, then these are real symmetric eigenvalue problems, which have special properties connected to orthogonal polynomials~\cite{wilf1962}.  If~$q$ is complex, then the matrices are \textsl{complex} symmetric, not Hermitian.  This has some important consequences that we will pursue in section~\ref{sec:double}.

Since matrices are nearly universally familiar nowadays, it may be simplest to think of these infinite complex symmetric tridiagonal matrix eigenproblems as determining the eigenvalues and corresponding eigenfunctions of the Mathieu equation. Truncating these infinite matrices causes no problems, as previously stated.  This, or at least in terms of determinants, seems to have been the way that Ince thought of the problem~\cite{ince1927mathieu}, although he did not have direct numerical methods to solve the eigenproblem and so he went on to solve the problems perturbatively, recovering and extending Mathieu's power series in~$q$ for the first few eigenvalues and for the coefficients $A_k$ and $B_k$ of the corresponding eigenfunctions. These series and their computation remain of interest, with (to quote Blanch~\cite{blanch1966numerical}) ``an algorithm for generating the successive [series coefficients], suitable for computers" being published in~\cite{rubin1964anecdote}  \footnote{This last paper is interesting: the author, a researcher at ``TRG Incorporated in Melville, NY" appeared to be disappointed at their main result, which was a set of recurrence relations which could be used numerically to generate the numerical series coefficients of both eigenfunctions and eigenvalues: it seemed that they were really searching for a good test problem for early symbolic computation programs!
We were intrigued, but unable to find out much about TRG Incorporated except that staff there carried out research on computational fluid simulations and lasers in the 1960's.  As a further remark on this quotation, we are not sure if Blanch is referring to \textsl{human} computers or to machines, although we incline to the latter interpretation because she does use two kinds of IBM computers for that article.}.
Subsequent papers give similar recurrence relations, and, if the order is large enough and not too many terms are needed, explicit symbolic formulae for the perturbative coefficients.  Such a formula is termed ``generic" in~\cite{frenkel2001algebraic}.

Ince also derived a continued fraction\footnote{The authors of~\cite{chaos2002mathieu} attribute the derivation of the continued fraction form to Heine in~\cite{heine1878handbuch} but it seems likely that Ince's derivation was independent.} from the above recurrence relations. We will continue to use tridiagonal matrices for the moment, as is done in many numerical methods today such as~\cite{ziener}.  The paper~\cite{chaos2002mathieu} shows especially good results, and the authors advocate strongly for the conceptual and computational use of these matrices.

\begin{figure}
  \centering
  \includegraphics[width=8cm]{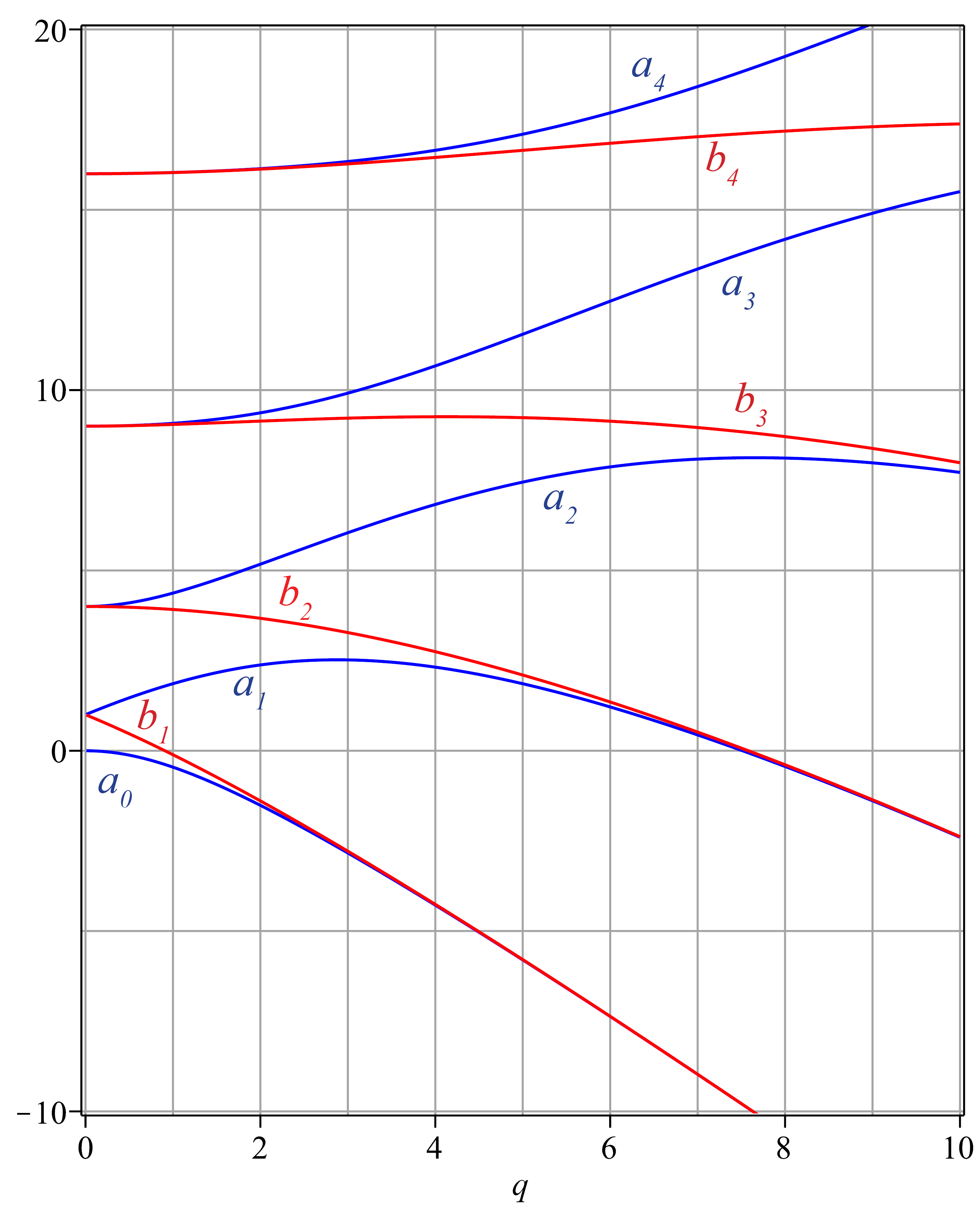}
  \caption{The first few eigenvalues $a_g(q)$ and $b_g(q)$, that is, the eigenvalues of the infinite tridiagonal matrices in equations~\eqref{eq:tridceeven}--\eqref{eq:tridseodd}, as functions of the real variable~$q$. This graph is modelled after the similar one in Fig 28.1 of the DLMF.}\label{fig:eigenvaluesLikeDLMF}
\end{figure}

When $q=0$ the matrix eigenvalues are simply the squares of the whole numbers: $0$, $1$, $4$, $9$, and so on.  They are, technically, double eigenvalues (taking even and odd functions together), but in this case they retain two independent eigenfunctions $\cos kx $ and $\sin kx$, except if $k=0$. For real~$q$ the eigenvalues can be sorted in increasing order.  See Figure~\ref{fig:eigenvaluesLikeDLMF}.

The evenness of the Mathieu equation, and its invariance under the transformation $z \to z\pm \pi/2$ and $q \to -q$, mean that the eigenvalues have the following symmetries:  $a_{2n}(-q) = a_{2n}(q)$, $a_{2n+1}(q) = b_{2n+1}(q)$, and $b_{2n+2}(-q) = b_{2n+2}(q)$. These are equations 28.2.26--28.2.28 in the DLMF.  There is also the conjugate symmetry $a(\overline{q}) = \overline{a(q)}$ and similarly for $b(q)$.

Obviously if $y(x)$ is an eigenfunction corresponding to a given eigenvalue $a_k(q)$ or $b_k(q)$, then so is any multiple of $y(x)$.  This can be seen in the eigenvalue/eigenvector formulation by multiplying the vector of Fourier coefficients by any nonzero constant.  We therefore need to choose a normalization in order to choose a definite eigenfunction.  Sadly, this is done in several different ways in the literature.  When using a particular software package, one has to pay attention to the choices that the authors have made.

\begin{figure}
  \centering
  \subfigure[Convergence when $q=2$]{\label{fig:EigenErrorsq2}\includegraphics[width=60mm]{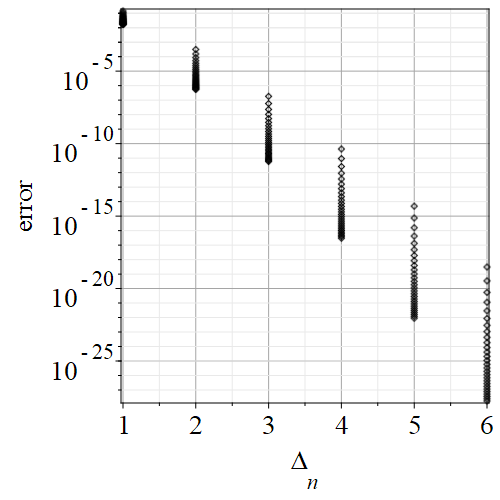}}
\subfigure[Convergence when $q=15+4\,i$]{\label{fig:EigenErrorsq15i4}\includegraphics[width=60mm]{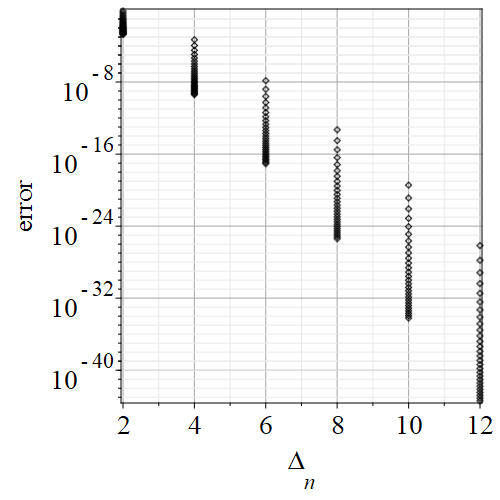}}
  \caption{What dimension matrix is needed to get an accurate value of eigenvalue $a_{2k}$? Here we plot the errors in $a_{2k}(q)$ for $3 \le k \le 30$ by using a matrix of dimension $k + \Delta_n$ where $\Delta_n$ is as indicated on the figure: either $1 \le \Delta_n \le 6$ as on the left, or twice that as on the right. That is, for $a_{6}(2)$ (top symbol in each column of points in the left hand graph) we plot the errors in using matrices of dimension $3+1:6=4$, $5$, $6$, $7$, $8$, and $9$.  This tells us that to get double precision we need dimension $9$ (dimension $8$ almost works). In contrast, for $a_6(15+4\,i)$, (top symbol in each column of points in the right-hand graph) we consider matrices of dimension $3+2:2:12=5$, $7$, $\ldots$, $15$. In this case we need dimension $13$ to get double precision accuracy in this eigenvalue.  Higher-order eigenvalues always need a matrix big enough to contain the desired eigenvalue and all smaller ones; it may be surprising to see that they get more accuracy from any extra dimensions than the smaller ones do. When $k=30$ (lowest points in each column) the accuracy is significantly better; the matrix dimension for the low point on the right is $42$.  All computations done in $60$ Digits. }\label{fig:EigenErrors}
\end{figure}

The eigenvalues of \emph{truncations} of these infinite matrices converge quite quickly to the desired eigenvalues of the Mathieu equation although the rate of convergence depends on the eigenvalue.  See figure~\ref{fig:EigenErrors}.
There are convergent series containing eigenvalues that can be used as numerical checks on the accuracy, but for this graph we simply computed high-precision values of the eigenvalues by a continued fraction method and compared the results to the truncated matrix eigenvalues. As stated, the matrices are real symmetric tridiagonal for real~$q$ (or can be made to be).  As such, there are fast algorithms for their computation, some based on Rayleigh iteration~\cite{parlett1974rayleigh}; in LAPACK there are SSTEV and DSTEV; see~\href{http://www.netlib.org/lapack/explore-html/index.html}{the Netlib repository}.
If~$q$ is complex, then the matrices are \emph{complex} symmetric tridiagonal, which are harder to solve, but for which there are still interesting algorithms~\cite{arbenz2004jacobi}.  There are specialized algorithms to find multiple eigenvalues in the infinite dimension case~\cite{miyazaki2004computation}.

We have not felt the need to resort to fast methods: an ordinary slow ($O(n^3)$) QR eigenvalue algorithm is perfectly fine, because the matrix dimensions are so small in modern terms.  After all, the Fourier series converges spectrally to the Mathieu equation~\cite{shen2009spectral}, and thus relatively few eigenvalues are needed.  However, in an application where there were very many modes needed for many different values of $q$, fast methods would be quite valuable.

The connection to orthogonal polynomials is in the case of complex~$q$ more strained, and may instead be more akin to the skew-symmetric eigenvalue problem: see~\cite{Iserles2019} for some surprises there.

\subsubsection{Asymptotics}
There is some asymptotic work for Mathieu functions in~\cite{whittaker1927course}, and indeed even a little in Mathieu's original work; but the first serious studies of the asymptotics of large eigenvalues was~\cite{ince1927mathieu}.  In this paper Ince first recapitulates some of Mathieu's zero-counting work (it is not clear that Ince knew that Mathieu had done this as well) using, as Mathieu had, Sturm's theory. Next, Ince established what are now the first two terms of DLMF 28.8.1 (we print the first four from there, below): if we denote $h=\sqrt{q}$ and $s=2m+1$, then as $h \to \infty$ with $m=0$, $1$, $2$, $\ldots$,
\[
\left. \begin{matrix}
        a_{m}\left(h^{2}\right)\\
         b_{m+1}\left(h^{2}\right)
       \end{matrix}
\right\}
\sim-2h^{2}+2sh-\frac{1}{8}(s^{2}+1)-\frac{1}{2^{7}h%
}(s^{3}+3s)-\frac{1}{2^{12}h^{2}}(5s^{4}+34s^{2}+9)-\cdots\>.
\]
This was created as a purely real result, but works for at least some complex values of~$q$ as well.  These have been implemented to arbitrary order in some computer algebra systems, for instance Maple.
\subsubsection{Nonperiodicity of the other independent solution}
\begin{figure}
  \centering
  \subfigure[A non-periodic solution]{\label{fig:redbelly}\includegraphics[width=60mm]{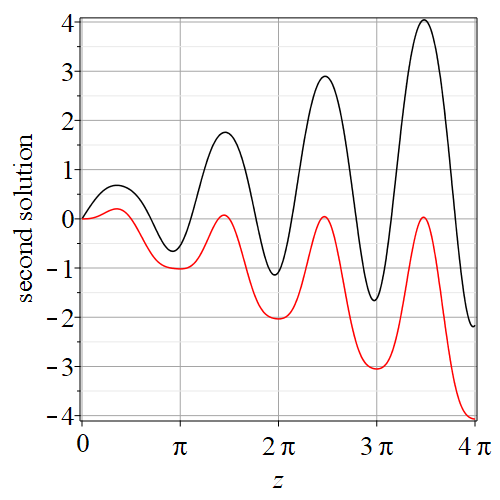}}
  \subfigure[Unstable solutions]{\label{fig:floquet}\includegraphics[width=60mm]{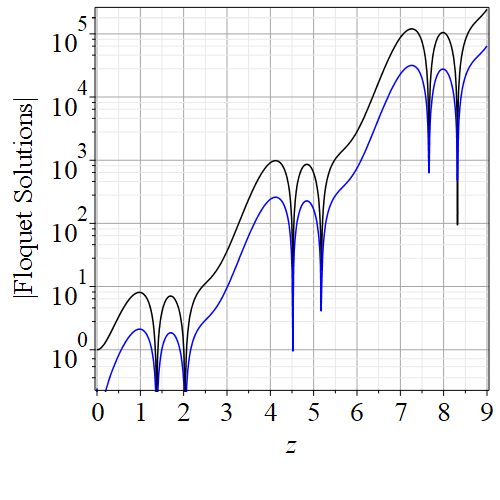}}
  \caption{Some non-periodic solutions of the Mathieu equation. On the left, we have the nonperiodic solution for the Mulholland-Goldstein double point $q \approx 1.4688\,i$ showing secular growth as Ince proved. Black is the real part of the solution, red is the imaginary part. On the right we show absolute values of solutions for $a=4$ and $q=10$, well into an unstable region in $(a,q)$ space by the Floquet theory, showing exponential growth.  Indeed the solutions are of the predicted form $\exp(\mu z)\phi(z)$ where $Re(\mu)>0$ and $\phi(z)$ is periodic; we can see that this particular periodic function has two zeros by the ``spikes'' in the graph.}\label{fig:nonperiodic}
\end{figure}

In~\cite{ince1922proof}, Ince proved that the Mathieu equation (and similarly the Hill equation)  ``can admit but one solution of period $\pi$ or $2\pi$''.  This is generally taken to mean that the other linearly-independent solution cannot be periodic, and indeed this is true for periods $\pi$ and $2\pi$.  See figure~\ref{fig:nonperiodic} where one such secularly-growing solution is graphed.  In~\cite{ince1926second} Ince established for small~$q$ that if, for instance, the periodic solution was $\ce_p(z)$, then the second, necessarily non-periodic, solution would be
\[
y = \se_p(z) + K z \ce_p(z) + \phi(z)
\]
where both $K$ and $\phi(z)$ would be $O(q^p)$ in size and $\phi(z)$ would be periodic.  As we can see in figure~\ref{fig:redbelly} this is true even if~$q$ is not small (in fact we took $q = 1.4688\,i$ approximately, and its eigenvalue $a\approx2.0886$ for this figure; the periodic solution for this value is plotted later, in figure~\ref{fig:GMgeny}). A general theorem to this effect can be found for instance in~\cite{bateman1953higher}.

However, in a throwaway remark in~\cite{blanch1966numerical}, Gertrude Blanch states ``In contrast to this, it is known that there exist other eigenvalues that
give rise to solutions of period $k\pi$, where $k$ is an arbitrary integer greater
than $2$. For these eigenvalues, all solutions are periodic, if one is.'' She gives the reference~\cite{onsager1935solutions} in Chapter 20 of~\cite{abramowitz}.  This is also covered in~\cite{bateman1953higher}.

\subsubsection{A short biography of Ince}
Edward Lindsay Ince (30 November 1891 – 16 March 1941) was, as previously mentioned, a research student of Whittaker and had spent the years 1909--1913 at the University of Edinburgh.  He had a short but colourful life and career, which included what would have then been an exotic episode at the newly-founded Egyptian University of Cairo\footnote{Founded 1908, it is now Cairo University.} from 1926 to 1931.  Returning to England for reasons of health and family, he took various positions, ultimately becoming Head of Technical Mathematics at the University of Edinburgh.  He was elected a Fellow of the Royal Society of Edinburgh in 1923.  He died of leukemia in Edinburgh, aged just 49.  Just before his death he was awarded the Makdougall-Brisbane prize for his work on Lam\'e functions, but he died before he learned of the award.

There are two obituaries of Ince that we know of: one by Whittaker~\cite{whittaker1941edward}, and another by Aitken~\cite{aitken1941dr}.  Both make interesting reading, although there is more mathematics in the one by Whittaker.  The one by Aitken is perhaps more personal, although it does contain the passage ``Ince firmly believed that theoretical solutions of problems, however abstractly elegant, were incomplete
unless the mathematician either tabulated the solving functions himself or rendered them tabulable.
Perusal of his papers will show that in his chosen field of research he achieved both of these objects''.  As an aside, Aitken's biography is itself worth reading; in addition to his mathematical accomplishments he was elected to the Royal Society of Literature for his memoir ``Gallipoli to the Somme: Reflections of a New Zealand Infantryman'', having served in both battles.  In contrast, Ince was not permitted active duty in the First World War owing to ill-health, although he did do a term of National Service.

Returning to mathematics and to Ince, Whittaker tells us that Ince started out reading mathematics under Professor George Chrystal, who we principally know nowadays as the person who wrote the text that inspired Ramanujan; after Chrystal died in 1911, ``Ince continued under his successor'', namely Whittaker himself. Whittaker then carefully describes Ince's achievements with the Mathieu functions, including Ince's proof~\cite{ince1922proof} in 1922 that there could not be two independent solutions with period $\pi$ or $2\pi$ to the same Mathieu equation.  Whittaker echoes Aitken's remark about Ince's sensibility with regard to computation, and amplifies it: ``Ince held that an important part of a pure mathematician's duty is to provide tables for the use of physicists and astronomers, and he was well aware that the possibility of constructing such tables without a colossal expenditure of time and energy depends on the progress of theoretical analysis.''  Whittaker remarks that Ince's 1932 tables of the Mathieu functions their zeros and turning points was ``A splendid piece of work, performed single-handed save for some help by an Egyptian assistant.''  Ince gave the name of his helper, when he acknowledged Mansy Shehata, who was then an Assistant in Pure Mathematics at the Egyptian University in Cairo. Ince also acknowledged grant support in purchasing calculating machines, which seemed to be of significant use.

Ince's work on the Mathieu functions was important in making computation of them practical; Gertrude Blanch, as we shall see next, relied on many of his results.
\subsection{Gertrude Blanch (1897--1996)}
\subsubsection{Numerical computation via continued fractions}
Blanch's first publication on Mathieu functions was~\cite{Blanch:1946:Computation}. This was reprinted in 1950 by the National Bureau of Standards ``with the permission of the editors of the Journal of Mathematics and Physics to meet a continuing demand''.
This refined and improved the continued fraction method that Ince used and in particular allowed the error in the Fourier coefficients to be estimated. She wrote a paper in the Transactions of the American Mathematical Society on the asymptotics of the odd periodic Mathieu functions~\cite{blanch1960asymptotic}, extending and correcting the works of others.
Blanch wrote an influential paper in SIAM Review on continued fractions, namely~\cite{blanch1964numerical}. This paper contained a detailed linearized rounding error analysis for continued fractions, and used both Bessel functions and eigenvalues of Mathieu functions as examples.  She argued that continued fractions gave the preferred method for computing Mathieu function eigenvalues in~\cite{blanch1966numerical}. Then Blanch \& Clemm~\cite{blanch1969double}, still later improved by~\cite{hunter1981eigenvalues}, went on to systematically compute the double eigenvalues\footnote{Blanch 1966 quotes using an ``I.B.M. 1620 and 7094".  We merely note the printed periods in the abbreviation ``I.B.M." used there, as opposed to the simpler trademark IBM used nowadays, and refrain from comparing the speed and memory of the early computers with what is available today (or what is coming).}.

Blanch's use of continued fractions is different from the use of Gauss-type continued fractions for evaluation of other special functions, such as those of hypergeometric type.
Those methods are explained in detail in, for example, Gil \textsl{et al.}~\cite{Gil2007}; who also cite Blanch for her study of the numerical stability of evaluation of continued fractions over the complex plane (as well as give more modern results, of which there are many).  But since the Mathieu equation is not of hypergeometric type, continued fractions cannot be applied directly in this fashion.
Instead, they are used to define a class of nonlinear equations whose roots are the eigenvalues of the Mathieu function; and then the partial quotients give the elements of the corresponding eigenvector.

We believe that Blanch's choice of this particular method of continued fractions was at least partly because of the kind of computing she typically did, which in her own words was ``experimental in nature'' and which by habit were continually checked as the computation went on. [Our computations for this paper were carried out in a similar style.]  In the first part of her career the computations were, like Ince's, carried out by hand and by desk calculating machine.  A significant difference is that while Ince had only one assistant when he was in Cairo, Blanch organized a group of as many as 450 assistants, and later used digital computers. But her computations were in many ways what we might call ``artesenal'' with a great deal of human involvement.

Today people generally prefer software packages that can run without care on the part of the user: in Blanch's words, in a ``robot-like'' manner~\cite{blanch1964numerical}.  Most people want to call a subroutine and be certain that it would never give error messages, would always be fast, and would always return the correct answer.  If one is writing general-purpose software for evaluation of the Mathieu functions, therefore, one might \emph{not} choose continued fractions (as we will see) because their occasional instabilities in the complex plane are somewhat variable; on the real line Blanch had analyzed this behaviour and given a sound rule for routine computation, but not for general complex~$q$.  For our purposes in this paper, however, the continued fraction algorithm is perfect: when carefully monitored, it is fast, flexible, generalizable, and any instability can be controlled simply by using extra precision.

Blanch's variant of the continued fraction algorithm can be described as follows. This treatment looks a little different than the standard connection of three-term recurrence relations, tridiagonal matrices, and orthogonal polynomials~\cite{wilf1962}, but in essence it is similar.
We suppose first that $a$ and~$q$ are given (later we will perform iterations, looking for values of $a$ and sometimes also of~$q$ that make the continued fraction equal to zero).  We will need the auxiliary quantities
\begin{equation}\label{eq:Vm}
  V_m = \frac{a-m^2}{q}
\end{equation}
for $m=0$, $1$, $2$, $\ldots$.  These are not defined if $q=0$, but if $q=0$ then we already know that the eigenvalues are $g^2$ for integers $g$. Equation~\eqref{eq:Vm} is equation (1.05) in~\cite{blanch1966numerical}.
We use this notation in the recurrence relations~\eqref{eq:recurrencebase}--\eqref{eq:Brecurrence}, and further with the auxiliary quantities $c_m$ where $c_0 = 2$ and all other $c_k = 1$, in order to look after the first edge case.  We rewrite the recurrence relations (apart from the base cases) as
\begin{equation}\label{eq:Blanchrecurrence}
  A_{m+2} + c_{m-2}A_{m-2} - V_m A_m = 0\>.
\end{equation}
Putting $G_m = A_m/A_{m-2}$ and dividing equation~\eqref{eq:Blanchrecurrence} by $A_m$ (Blanch was very careful about what happened when any $A_m$ was zero, but here we rely on IEEE arithmetic with signed zeros and infinities to get everything right) we get
\begin{equation}\label{eq:contfracstart}
  G_{m+2} + \frac{c_{m-2}}{G_m} - V_m = 0\>.
\end{equation}
This recurrence can be written either as
\begin{equation}\label{eq:upward}
  G_{m+2} = V_m - \frac{c_{m-2}}{G_m}
\end{equation}
or
\begin{equation}\label{eq:downward}
  G_m = \frac{c_{m-2}}{V_m - G_{m+2}}
\end{equation}
Of course, these recurrences must be started correctly by using the appropriate equations~\eqref{eq:recurrencebase}--\eqref{eq:Brecurrence}.  Running equation~\eqref{eq:downward} until the denominator~$Q$ is so large that $1/|Q^2| \le \varepsilon$ where $\varepsilon$ is our tolerance, starting from some $M > 2$ so that $c_{k}=1$ for all $k \ge M-2$,
\begin{equation}\label{eq:contfracmiddle}
  G_M  = \frac{1}{V_M - \frac{1}{V_{M+2} - \frac{1}{V_{M+4} - \ddots}}}
\end{equation}
and because the $V_m$s grow like $m^2$ this continued fraction converges for all $a$ and all $q\ne 0$.  Call the $G_M$ computed in this way $G_{M,\mathrm{tail}}$.

Now we use equation~\eqref{eq:upward} with increasing $m$ starting from our known edge cases (depending on which class of Mathieu eigenvalue we wish to compute), and \emph{if and only if} $a$ is an eigenvalue then the two values of $G_M$ in the middle will agree.  Call the $G_M$ computed in this way $G_{M,\mathrm{head}}$.  Let
\begin{equation}\label{eq:BlanchT}
T(a,q)= G_{M,\mathrm{head}}(a,q) - G_{M,\mathrm{tail}}(a,q)\>.
\end{equation}
$T(a,q)$ must be zero for $a$ to be an eigenvalue corresponding to~$q$.  The edge cases for $G_{M,\mathrm{head}}$ determine whether this is an $a_{2k}$, $a_{2k+1}$, $b_{2k}$, or $b_{2k+1}$ eigenvalue.  Just \emph{which} integer $k$ depends, as a rule, on whether there is an unambiguous continuous path in~$q$ back to the eigenvalue with that index when $q=0$. Blanch gave a rule, as previously stated, for choosing the $M$ in the middle so as to minimise numerical instability for real~$q$.

There are many methods one could use to find zeros of equation~\eqref{eq:BlanchT}, but since differentiation of $T(a,q)$ with respect to $a$ is simply a matter of differentiating equations~\eqref{eq:upward} and~\eqref{eq:downward} (convergence is uniform in compact neighbourhoods and so differentiation is permissible), we choose Newton's method.  For a given~$q$ and starting with an initial estimate $a^{(0)}$ for the eigenvalue\footnote{In order to find a good initial estimate, people usually use continuation from~$q$ near $0$: one solves for $q=q_n$, and then uses that eigenvalue as an initial estimate for the eigenvalue at $q = q_n + \Delta q$.}, the iteration is
\[
a^{(k+1)} = a^{(k)} - \frac{T(a^{(k)},q)}{T_a(a^{(k)},q)}\>.
\]
Blanch used eigenvalues with slightly different values of~$q$ as initial estimates for the eigenvalues she required, and this worked very well, unless, of course, the eigenvalue was not simple (\emph{i.e.} did not have multiplicity~$1$), in which case even more derivatives of $T$ turn out to be useful.

The eigenvalues of the Mathieu equation are as previously stated almost always simple, but for isolated complex values of~$q$ may have multiplicity~$2$. In particular, if $q=is$ where~$s$ is real and $i^2=-1$, that is if~$q$ is purely imaginary, then as $s$ increases from zero we will necessarily encounter double points: first at approximately $s=1.4688$ and then at approximately $s=6.9289$ (see~\href{http://dlmf.nist.gov/28.7}{section 28.7 of the DLMF}). The double point near $s=1.4688$ was first studied in~\cite{mulholland1929xc} and we will refer to it as the Mulholland-Goldstein double point (we use it as an example, frequently).  There are a countable infinity of double points.  It is proved in~\cite{meixner} that there are no triple points of the Mathieu equation, or simultaneous double points---that is, a value of~$q$ for which two (or more) pairs of eigenvalues merge.

One might be tempted to dismiss double points because they occur ``with probability zero" if one chooses the parameter~$q$ ``at random".  But in applications requiring complex~$q$ the parameter will not usually be chosen at random, and indeed the problem being modelled may call for a continuum of values of the parameter.  In the application that motivated us to study the Mathieu functions, we were interested in \textsl{all} imaginary values of~$q$, and this necessarily included some double points.  In some sense this means that the ``probability zero" events actually occur with ``probability one".  This reversal of expectation is a common occurrence in bifurcation phenomena, for instance~\cite{Golubitsky1985}.

\subsubsection{Double points\label{sec:doubleBlanch}}
Blanch and Clemm used a variation on Newton's method in their systematic search for double points~\cite{blanch1969double}.  Their method was not as efficient as two-dimensional Newton iteration, but it worked and it was the first method to do so.  Instead of just using one derivative with respect to $a$, they used two, and expanding
\[
T(a,q) = T(a^{(k)},q) + T_a(a^{(k)},q)(a-a^{(k)}) + \frac{1}{2}T_{aa}(a^{(k)},q)(a-a^{(k)})^2 + \cdots
\]
they set this to zero and solved the resulting quadratic for the update to $a^{(k)}$, choosing the smaller magnitude square root.  This gives a superlinearly-converging iteration for double roots (even faster for simple roots), and together with interpolation enabled them to compute the smallest $72$ double eigenvalues to what we would now call double precision.  We confirmed their work and recomputed all their roots. We used their tabulated values as initial estimates, and have plotted the results in figure~\ref{fig:BlanchClemmDoubles}.  Blanch and Clemm carried out their computations
on digital computers (we believe models IBM 1620 and IBM 7024).
\subsubsection{A short biography of Blanch}
As previously stated, Gertrude Blanch wrote the chapter on Mathieu functions in the classic handbook~\cite{abramowitz}.  We have cited many of her papers on the Mathieu functions and related matters.  She apparently wrote in 1943 what might be the first modern textbook on numerical analysis, and an updated version in 1982, according to~\cite{grier1997gertrude}.  Unfortunately, we have not seen a copy of either edition.

We draw material for this section from several sources, including a transcript of an interview with her in 1973~\cite{tropp:1973:Interview}, an extensive biography by Grier~\cite{grier1997gertrude}, and the collection of her papers at the Charles Babbage Institute at the University of Minnesota.  Gertrude Blanch's extremely interesting life was well-documented. We have in this short biography concentrated on mathematical aspects of her life and left out important non-mathematical aspects.  The above resources are well worth consulting for a fuller picture.

Gertrude Blanch was born Gittel Kaimowitz in 1898 in Kolno, then part of Russia.  She came to the US in 1907 and attended high school in Brooklyn, graduating in 1914.  She changed her name, after her father died, to an Anglicized version of her mother's family name, Blanc.  She became an American citizen in 1921.  She worked for fourteen years to get enough money to attend university; her employer paid her tuition for New York University, where she graduated \emph{summa cum laude} in 1932.  Apparently following the advice of one of her professors there, Fay Farnum, she then went to Cornell, receiving her PhD in algebraic geometry in 1935 under the guidance of Virgil Snyder and Wallie A.~Hurwitz\footnote{Blanch credits them both, but the Mathematics Genealogy Project only reports Snyder as her advisor.  Snyder also seems to have advised Farnum and ten (by our count) other women for their PhDs, the earliest---Anna van Benschoten---in 1908.}. After a short stint teaching at Hunter College for someone on sabbatical leave, she took an office clerical administration and accounting job.  This administrative job was to prove important for her later work with the Mathematical Tables Project, as detailed in~\cite{grier1997gertrude}.  In order to keep her mathematical interests alive, she took a night course in relativity at Washington Square College given by Arnold Lowan.  When Lowan was asked to create the Mathematical Tables Project under the New Deal Works Progress Administration, he asked Blanch to join.  She became Technical Director, eventually organizing a group of 450 (human) computers.  According to Grier, she deserves much of the credit for the success of the project, and part of that credit is due to her prior business-oriented administrative experience.  She published several papers during this time, including one with Hans Bethe~\cite{blanch1941internal}.

``During her time at the Mathematical Tables Project, she particularly enjoyed working with the Mathieu functions, and these functions would be central to the rest of her career.''~\cite[p.23]{grier1997gertrude}.

When asked how she first got interested in Mathieu functions, she responded:
``Morse\footnote{Philip Morse, at MIT, the first author of~\cite{morse1954methods}.  This monumental work also describes Mathieu functions.} was interested, for example, in Mathieu functions. I got started on Mathieu functions because Morse needed them and there were any number of small things and some special integrals that he came across within his field.''

She remained interested in Mathieu functions for the rest of her career and indeed after her retirement.  Her ultimate academic appointment was as Head of Mathematical Research in the Aerospace Research Laboratory at Wright Paterson Air Force Base in Dayton, Ohio, where she wrote many of her papers.  She became a Fellow of the American Association for Advancement in Science in 1962.  She received the Federal Women's Award from President Lyndon Johnson in 1964. She retired in 1967, and died in 1997, aged 99.
\subsection{Selected other works by other contributors}
There is an excellent concise treatment of Mathieu functions, including the Floquet theory, in Chapter XVI of Volume~3 of~\cite{bateman1953higher}, ``The Bateman Manuscript''.  In particular, it is there that we learn two interesting facts about $\mu = i\pi\nu$, the Floquet characteristic exponent for the Mathieu equation: first, that Poincar\'e found a way to compute it from the two basic solutions $w_I(z)$ and $w_{II}(z)$ using the evenness of the Mathieu equation: at least one of
\[
w_I(z) = \frac{e^{\mu z}\phi(z) + e^{-\mu z}\phi(-z)}{2\phi(0)}
\]
or
\[
w_{II}(z) = \frac{e^{\mu z}\phi(z) - e^{-\mu z}\phi(-z)}{2(\phi'(0)+\mu \phi(0))}
\]
will have nonzero denominator; now differentiate $w_{II}$ and compute $w_I(\pi)$ and $w_{II}'(\pi)$.  Since $\phi(0)=\phi(\pm\pi)$ and
$\phi'(0)=\phi'(\pm\pi)$, we have that
\begin{equation}\label{eq:Poincare}
\cosh \pi \mu = w_{I}(\pi) = w_{II}'(\pi)\>.
\end{equation}
In the DLMF this equation (using cosine and not hyperbolic cosine, because $\nu$ is used instead, where $\mu = i\nu$)
is called the \emph{characteristic equation}, number 28.2.16.  Blanch uses another convention, namely $\mu = i\pi\nu$, defining $\nu$ differently, in~\cite{abramowitz}.
The second interesting fact is that the periodic solutions, that is the Mathieu functions, correspond to the case $\mu = i\,n$ where $n$ is an integer; if $n$ is an even integer the solution is periodic with period $\pi$ and if $n$ is odd the solution is periodic with period $2\pi$.  The case when $\mu = i\,r$ where $r$ is a rational integer generates those mysterious solutions alluded to earlier which are periodic with greater periods. See section 20.3 in~\cite{abramowitz} for more discussion of this fact.

The earlier book~\cite{strutt1932lamesche} by M.~J.~O.~Strutt was critically reviewed in~\cite{goldstein_1933}, which states ``It is a useful book, both for pure mathematicians interested in the theory of special functions, and for applied mathematicians compelled to use the functions in their researches.  It is worthy of consultation by both classes, but it is rather superficial$\ldots$''.

The most extensive and thorough treatment of Mathieu functions is that of Meixner, Sch\"afke, and Wolf~\cite{meixner}, which is based on an earlier book by Meixner and Sch\"afke that we have been unable to get hold of.  The later edition handles significant generalizations of the theory of Mathieu functions, and as previously stated deals completely with the theory of double points.  Nearly every reference after this work cites it; indeed the Bateman Manuscript itself cites (the 1950 book) as ``forthcoming''.  The additional author, G.~Wolf, of the second edition is the author of the DLMF Chapter~28 on Mathieu functions, which updates Blanch's Chapter~20 of~\cite{abramowitz}.

There are many tables of the Mathieu functions. We believe that Blanch's work remains the most extensive, but we can also mention Bickley~\cite{bickley1945tabulation}.  There are tables of integrals and series for Mathieu functions in Prudnikov \textsl{et al.}~\cite{Prudnikov:1990:IS}.  The value of the numerical tables nowadays is of course lessened by the availability of software; the tables of integrals and series may also have been supplanted now by computer algebra systems, at least to some extent.

There are a large number of relatively modern books, papers, and software packages for the computation of Mathieu functions.  Analytical work on the Mathieu function is extensive; to cite a paper almost at random, consider~\cite{naylor1984simplified}.  Some of the works ignore most of the others, and hence rediscover things, but others are very insightful.  We particularly recommend~\cite{chaos2002mathieu} and~\cite{gutierrez2003mathieu} for exceptional visualizations and clarity of exposition.  On the purely computational side, there are three sets of ACM Transactions on Mathematical Software papers discussing implementations, the latest being~\cite{erricolo2003acceleration}.  There are Python (scipy) and third-party Matlab implementations of (real) Mathieu functions.  The Mathematica implementation of complex Mathieu functions may be very good (unfortunately, we have only had limited access to Mathematica so we cannot be sure, but it may even have facilities for computing double eigenvalues) and is described in~\cite{trott2007mathematica}.  The Maple implementation, with which we are most familiar, encodes a substantial amount of analytical work including both~$q$-series and asymptotic series.

\section{Double points\label{sec:double}}
As previously noted, at certain isolated points in the complex~$q$-plane, for instance at the Mulholland-Goldstein point $q^*\approx 1.468768613785142\,i$ (reporting $16$ digits\footnote{This double point was the first found: studied in~\cite{mulholland1929xc} and later computed by~\cite{bouwkamp1948note} to $3$ digits and then to double precision in~\cite{blanch1969double}.}), we have a \emph{double} eigenvalue: $a_0 = a_2 \approx 2.088698902749695$.  Several interesting things happen at double points.  First, the eigenfunctions coalesce: here, $\ce_0(q^*,\alpha) = \ce_2(q^*,\alpha)$, leaving a gap in the completeness of the set of eigenfunctions.  This means that we will have to supplement the set of eigenfunctions in some way in order to expand arbitrary functions in series containing Mathieu functions.

Second, \emph{near} to these double points, the ordering of the eigenvalues becomes ambiguous.  For $\Im(q) < \Im(q^*)$ we have $a_0(q) < a_2(q) < a_4(q) < \cdots$, but at $q^*$ equality occurs.  For $\Im(q) > \Im(q^*)$, both $a_0(q)$ and $a_2(q)$ are complex, and ordering is a matter of convention.  The DLMF adopts the convention in this case that $a_0(q)$ continues as $\Im(q)$ increases by choosing the branch with \emph{negative} imaginary part, while $a_2(q)$ takes the conjugate.  See the visualization in section 28.7 of the DLMF: they have paths for $a_0(q)$ and $a_2(q)$ coming in and merging at the double point, and then emerging at right angles: which way the paths are numbered on emergence is a convention.

The convention given in the DLMF makes sense provided one thinks of~$q$ varying along the imaginary axis, and increasing.  Approaching $q^*$ along some other path in the~$q$-plane may cause puzzlement: the ordering is conventional, no more.

The discussion in~\cite{hunter1981eigenvalues} links the numbering of eigenvalues to the paths used for continuation in $q$, and they describe branch cuts ending at each double point which can be used to disallow paths that would disrupt the conventional numbering.  This analysis becomes more involved the larger the double points get.

Also as~$q$ approaches $q^*$, the coefficients in the Mathieu series expansion for a given function will usually become singular.  For example, consider
\begin{equation}\label{eq:cos2tMathieuexpansion}
  \cos 2\alpha = c_0(q) \ce_0(q,\alpha) + c_2(q) \ce_2(q,\alpha) + \cdots\>.
\end{equation}
By numerical experimentation at high precision\footnote{We are slightly embarrassed to admit to how many figures we took these computations to: we worked at $250$ decimal Digits in Maple, and solved the Mathieu equation with a tolerance of $10^{-120}$.  We then worked with $5$, $8$, $13$, $21$, $34$, and $55$ Digit truncations of $q^*$ and calculated the corresponding $a_0$ and $a_2$ to $120$ Digit accuracy; this enabled us to identify the constants in this section to $50$ Digits or more.  We only report the double precision values. Later, we confirmed these by simply computing the Puiseux expansion of this point.} we find that
\begin{align}\label{eq:singcoeffs}
  c_0 + c_2 &= 1.009185957186356 - 0.1210349964877181\,i \\
  c_0 - c_2 &= \frac{-1.023431886611575 + 0.2551095295356106\,i}{\sqrt{q-q^*}}
\end{align}
when we use the normalization convention that $\ce_{2k}(q,0) = 1$, and moreover that
\begin{align}\label{eq:Puiseuxa2}
  a_0 =& a^* + d\cdot(q-q^*)^{1/2} + O(q-q^*) \\
  a_2 =& a^* - d\cdot(q-q^*)^{1/2} + O(q-q^*)
\end{align}
where $d \approx 1.659487804320256 + 1.659487804320256\,i = 1.659487804320256(1+i)$.
These values were found by using orthogonality, which holds if $q\ne q^*$, and by high-precision computation of
\[
\int_{x=0}^{2\pi} \ce_{2k}^2(q,x)\,dx = O(q-q^*)^{1/2}\>.
\]
Then if $y^* = \ce_{0,2}(q^*,\alpha)$ is the coalesced solution to Mathieu's equation, we can examine the most important contribution to the perturbation by considering the solution to the perturbed equation
\begin{equation}\label{eq:perturbedMathieu}
  y'' + (a^* \pm d\cdot(q-q^*)^{1/2} - 2q\cos(2\alpha))y = 0\>.
\end{equation}
If $y = y^* + u(q^*,\alpha)\cdot d\cdot(q-q^*)^{1/2} + \cdots$, then a short calculation dropping terms of $O(q-q^*)$ and higher gives
\begin{equation}\label{eq:pertsol}
  u'' + (a^* - 2q^*\cos2\alpha)u + y^* = 0\>.
\end{equation}
The function $u$ must be periodic with the same period as $y^*$. It can therefore be computed numerically alongside $y^*$ (by solving a boundary-value problem) or alternatively can be expressed as an integral of Mathieu functions against $y^*$.
This argument is extended and formalized in a short section in~\cite{meixner} starting on p.~$82$.  Here and with just this simple example, we see that $u$ is essentially $\partial y/\partial a$ (found by solving a variational equation).  By combining the equations above, we can see that
\begin{equation}\label{eq:framecos}
  \cos2\alpha = (c_0+c_1)\ce_{0,2}^*(q^*,\alpha) + d(c_0-c_2)\sqrt{q-q^*}\cdot u(q^*,\alpha) + O(q-q^*)^{1/2}\>.
\end{equation}
This arrangement is continuous as $q \to q^*$ because $(c_0-c_2)\sqrt{q-q^*}$ is $O(1)$ in that limit.
This shows explicitly that by adding $\partial \ce_{0,2}(q^*,\alpha)/\partial a$ to the collection of Mathieu functions we are expanding with we ensure completeness of the set and the possibility of the expansion.
We know of no freely-available software package for Mathieu functions that provides for the computation of these extra functions. We discuss methods to compute $u(\alpha)$ in section~\ref{sec:computegeneralized}.

\subsection{Computing double eigenvalues}
We here discuss an effective method for computing the double eigenvalues. When computing $T(a,q)$, we will need to also compute the derivatives $T_a(a,q)$, $T_q(a,q)$,
$T_{aa}(a,q)$ and $T_{aq}(a,q)$.  Then we will be able to carry out a two-dimensional Newton iteration for solving the two equations $T(a,q)=0$ and $T_a(a,q)=0$ simultaneously.  The equations for the iteration looks like this:
\begin{equation}\label{eq:twoDNewton}
  \begin{bmatrix}
    T_a & T_q \\
    T_{aa} & T_{aq}
  \end{bmatrix}\begin{bmatrix}
                 \Delta a \\
                 \Delta q
               \end{bmatrix} = \begin{bmatrix}
                                 -T \\
                                 -T_a
                               \end{bmatrix}
\end{equation}
where all function evaluations and derivative evaluations occur at the current estimates $(a^{(k)}, q^{(k)})$. Then as usual $a^{(k+1)} = a^{(k)}+\Delta a$ and $q^{(k+1)} = q^{(k)} + \Delta q$.  Given sufficiently good initial estimates, this iteration converges quadratically to double points $(a^*,q^*)$. This method was first used by~\cite{hunter1981eigenvalues}, and as previously noted is different from the method of Blanch and Clemm.  Initial estimates are typically obtained by numerical continuation in~$q$.

\begin{remark}
Even if one has $q^*$ to double precision, one cannot naively compute $a^*$ to the same precision: after all, being H\"older continuous but not Lipschitz continuous at the double point it will be sensitive to errors on the order $\sqrt{q-q^*}$. In ordinary numerical parlance, the double eigenvalue is infinitely ill-conditioned, although the H\"older continuity constrains that to some extent.  With only double precision computation, this means one can only expect to know $a^*$ to about single precision, and the double eigenvalue will ordinarily have spuriously split into two simple (but close) eigenvalues.  Their \emph{mean}, as is well-known, will be a double-precision approximation to the true double eigenvalue (we will see that in adding the two Puiseux series expansions the leading error terms cancel).  This numerical split, however, might be used to advantage in computation of independent eigenfunctions: the norms of these erroneous approximate eigenfunctions will be $O(q-q^*)^{1/2}$ and therefore the associated Fourier coefficients in the eigenfunctions will be large,  but perhaps this is tolerable.  The resulting spectral expansion of the function may well be accurate enough for one's purposes.  Perhaps this is the real reason no-one has developed the tools to deal precisely with these extra eigenfunctions in practice; of course in theory this has been understood since at least~\cite{meixner}.
\end{remark}

\subsection{Initial estimates by continuation}
As previously mentioned, to compute $a_g(q)$ or $b_g(q)$ by Newton's method, one typically needs an initial estimate for the eigenvalue. The standard way to do this is to choose a path in the complex $q$ plane from $q=0$ to the desired $q$, and then one knows $a_g(0) = g^2$ and $b_g(0) = g^2$; one increments $q$ by a small amount, and then uses the previous value of $a_g(q)$ or $b_g(q)$ as the initial estimate for a Newton iteration for $a_g(q+\Delta q)$.  This works well except in the neighbourhood of double points, at which it is sensible to switch to two-dimensional Newton iteration, locate the double point carefully, and then step from there in the direction of the desired $q$.

The asymptotic formulae of~\cite{hunter1981eigenvalues} can be used instead to directly estimate the location of large double eigenvalues for subsequent refinement by two-dimensional Newton iteration.  We have not made a systematic study of these formulae, although we have tested using just their simplest approximations from equations 4.14a--4.14d in that paper which give essentially an empirical estimate of the approximate values of $q$ at double eigenvalues; other equations in the paper give methods for numerically computing asymptotically correct values of the double eigenvalues, based on WKBJ approximations and the elliptic integral
\begin{equation}\label{eq:EllipticE}
  I = \int_{0}^{\pi/2} \sqrt{a - 2q \cos2\alpha}\,d\alpha\>.
\end{equation}
This must be equal to $m\pi/2$ if $a = a_m$ or $a=b_m$ is an eigenvalue.
They dryly point out that ``no simple expression for this integral is available'' and indeed Maple's current exact expression for this value contains $7,287$ characters and takes up more than two screens.  We have not investigated if that can be usefully simplified.
Hunter and Guerrieri use numerical evaluation of that integral together with rootfinding in order to compute asymptotic estimates of the location of double points.


\subsection{Computing local series for the eigenvalues by Newton's method}
We may use Newton's method not just to compute a simple eigenvalue, or two-dimensional Newton's method to compute a double eigenvalue, but we can also use (one-dimensional) Newton's method to compute a \emph{series} for an eigenvalue by working in the ring of formal power series.  In the case of a simple eigenvalue at, say, $q=q_s$, we may compute a power series in $(q-q_s)$ for the eigenvalue $a_g(q)$, and simultaneously if we wish for the associated eigenfunction.  If instead we want a series expansion around a \emph{double} eigenvalue at $q=q^*$, then we may compute a \emph{Puiseux} series for the eigenvalues at nearby~$q$, again by Newton's method.  In that case, we will need an initial estimate for $a(q)$ correct to $O(q-q^*)$.  This may seem surprising, if one has only ever used Newton's method purely numerically, but the technique is well-known to the symbolic computation community.  Either of these series can be used for continuation: one computes a series about a given $q$, then uses that series to predict the value of the eigenvalue for a nearby $q+\Delta q$, which can then be corrected by Newton's method at the new point.  This may allow larger $\Delta q$, although the danger of branch switching is always present with too-large a $\Delta q$, and a certain degree of caution is encouraged.

What allows this series computation to work is that Blanch's version of the continued fraction algorithm can be carried out \emph{in series}.  One simply uses series arithmetic when adding, multiplying, or dividing.  This automatically allows the computation of all derivatives needed.  The convergence test only needs to consider the constant term.  More, this allows computation of both local Taylor series for the eigenvalues, that is
\[
a(q) = \sum_{k\ge0 } \alpha_k (q-q_0)^k\>,
\]
by carrying out the Newton iteration in series with $q = q_0 + x$ where $x$ is the series variable. We are solving

\[
T( a(x), q_0+x) = 0
\]
by iterating
\[
a^{(k+1)} = a^{(k)} - \frac{T(a^{(k)},q_0+x)}{T_a(a^{(k)},q_0+x)}
\]
in series; because $x = q-q_0$ we get the desired power series.
In this case, we start with the initial estimate $a^{(0)} = \alpha_0$, and a single Newton iteration gets us $\alpha_0 + \alpha_1 x$ (plus higher order terms that are incorrect and we may ignore), and another iteration gets us $\alpha_0 + \alpha_1 x + \alpha_2 x^2 + \alpha_3 x^3$ (plus higher order terms that are incorrect and we may ignore), and so on.  The initial estimate has error $O(x)$; the first iterate has better error $O(x^2)$; the second has even better error $O(x^4)$, and so on, showing a familiar quadratic convergence; yet somehow nicer than numerical convergence, because more predictable than numerical Newton's method in that after $n$ steps we have error $O(x^{2^n})$, and all lower degree terms are (apart from rounding errors) exactly correct.  See~\cite{geddes1992algorithms} for a proof that this method converges ``in series'' if the second derivative exists, and for a discussion of the linearly convergent iteration using the constant derivative $T_a(\alpha_0,q_0)$ in the denominator instead; that alternative iteration takes more iterations of course but the series arithmetic is cheaper.

A little thought shows that we may also compute \textsl{Puiseux}
series
\[
a(q) = a^* + \sum_{k\ge1 } \alpha_k (q-q^*)^{k/2}\>
\]
for the eigenvalue about double points, again by carrying out Newton iteration in series,
this time with $q = q^* + x^2$ where $x$ is the series variable.

For a survey of Puiseux series, see~\cite{GarciaBarroso2016}.
For a rigorous algorithmic treatment of expansion of solutions of systems of differential equations in Puiseux series about singular points, see~\cite{Cano2020}.  In the treatment of Puiseux series in this paper, which we keep informal so as to maintain readability, we only show how to compute the first few terms of the series, use them as asymptotic series only, and do not demonstrate convergence of the resulting series. One expects, however, by standard results in analysis that the resulting series, if taken to an infinite number of terms, would in fact converge in a disk $|q - q^*| \le \rho$ where $\rho$ was strictly less than the distance to the nearest other double point.

Returning to the problem at hand, we need the initial estimate to be more accurate than we needed for simple Taylor series: we need the first \emph{two} terms correct, namely $a(q) = a^* + \alpha_1 x$ where $\alpha_1$ is found by setting the coefficient of $x^2$ to zero in the following series expansion: $  0 = T(a(x),q^*+x^2) = $
\begin{equation}\label{eq:PuiseuxStarter}
  T(a^*,q^*) + T_a(a^*,q^*)(\alpha_1 x + \cdots) + T_q(a^*,q^*) x^2 + \frac{1}{2} T_{a,a}(a^*,q^*) ( \alpha_1 x)^2 + \cdots\>.
\end{equation}
The constant coefficient $T(a^*,q^*)$ and the linear coefficient $T_a(a^*,q^*)$ are both zero at a double point. The coefficient of $x^2$ is $\alpha_1^2 T_{a,a}(a^*,q^*)/2 + T_q(a^*,q^*)$ and so will be zero if and only if
\begin{equation}\label{eq:PuiseuxFirst}
  \alpha_1 = \pm \left( \frac{-2T_q(a^*,q^*)}{T_{a,a}(a^*,q^*)} \right)^{1/2}\>.
\end{equation}
Since the Mathieu equation has only isolated double points, neither $T_q(a^*,q^*)$ nor $T_{a,a}(a^*,q^*)$ is ever zero\footnote{Certainly $T_{a,a}$ is never zero because there are only double roots, not triple roots.  If however $T_q$ were zero then there would still only be two roots, but in this case $\alpha_1=0$ and $a = a^* + \alpha_2 x^2 + \cdots$ where $\alpha_2$ is one of two nonzero roots of a quadratic equation.  However, we believe that the theorem of~\cite{meixner} guarantees that $T_q$ is never zero so this should never happen, and indeed we never saw it happen.}, so $\alpha_1$ is finite and nonzero.
These distinct choices for $\alpha_1$ lead to distinct series expansions; together these two series describe the eigenvalues that merge as $q \to q^*$.

With the initial estimate $a^{(0)} = a^* + \alpha_1 x$ we may again use Newton iteration, even though this time  $T_a(a(x),q_0+x^2)$ will be $O(x)$ because that derivative is zero when $x=0$.  This means that even if $a^{(k)}$ is correct up to $O(x^m)$, so that the residual $T(a(x),q_0+x^2)$ will be $O(x^m)$, we will lose one power of $x$ from the Newton correction and so $a^{(k+1)}$ will ``only" be correct up to $O(x^{2m-1})$.  Starting with $m=1$ (i.e. just with $a^*$) is therefore not accurate enough; we must have $m=2$ (i.e. start with $a^*+\alpha_1x +O(x^2)$) to get off the ground, and then $2m-1 = 3$ is higher order, and the next step will have $2m-1 = 5$, and then $9$, and so on.
This gives a kind of quadratic convergence---still approximately doubling the number of terms correct with each iteration and after $m$ iterations we will have the series for $a(x)$ correct to $O(x^{2^{m}+1})$---in computation of the Puiseux series.

See Algorithm~\ref{alg:seriescontfrac}, which covers both Taylor series and Puiseux series. This algorithm has been implemented as a Maple procedure and is publically available at \href{https://github.com/rcorless/Puiseux-series-Mathieu-double-points}{Rob Corless's GitHub repository}.
That Newton's method automatically converges in formal power series (including Puiseux series) may be surprising, but it is really the same behaviour as in the numerical world: the initial estimate has to be close enough, i.e. has to have enough correct terms in the series, for convergence to start.  Once it does, convergence is rapid.  The ``asymptotic constant'' which complicates analysis in the numerical world is hidden under the $O(x^m)$ symbol in the formal power series analysis, which makes it simpler.

\begin{algorithm}
\caption{\label{alg:seriescontfrac} Solving $T(a,q)=0$ in series.  This algorithm uses Blanch's algorithm for evaluating $T(a,q)$, except that operations may be carried out in series.  This allows Taylor series solution near regular points, or Puiseux series solution at double points.}
\begin{algorithmic}
  \REQUIRE If Taylor series desired, $q_0$ and a simple eigenvalue $a_0 = a(q_0)$ computed by (say) one-dimensional Newton iteration
  \REQUIRE If Puiseux series desired, a double eigenvalue pair $(a^*,q^*)$ computed by two-dimensional Newton iteration, and $T_{a,a}(a^*,q^*)$ and $T_q(a^*,q^*)$ to compute $\alpha_1 = \pm 2T_q/T_{a,a}$ as in the text.  Choose a sign for $\alpha_1$.
  \REQUIRE Positive integer $N$ for the desired number of terms in the series for $a(x) = a_0 + a_1x + \cdots + a_N x^N$.
  \STATE If Taylor series, put $q \leftarrow q_0 + x$ and $a \leftarrow a_0 $ and $n \leftarrow 1$
  \STATE If Puiseux series, put $q \leftarrow q^* + x^2$ and $a \leftarrow a_0 + \alpha_1 x$ and $n \leftarrow 2$
  \WHILE{$n < N$}
    \STATE $R \leftarrow T(a,q)$ (Trimming leading coefficients $[x^k]$ for $k < n$ b/c rounding errors )
    \STATE If Taylor series, $n \leftarrow \min(2n,N)$
    \STATE If Puiseux series, $n \leftarrow \min(2n-1,N)$
    \STATE $a \leftarrow a - R/T_a(q,a)$ to $O(x^{n})$
  \ENDWHILE
\end{algorithmic}
\end{algorithm}

\begin{remark}
Rounding errors can complicate matters here.  In exact arithmetic, the residual $T(a^{(k)},q(x))$ would be $O(x^m)$ exactly, for some integer $m$.  In practice, the coefficients of the terms $ r_0 + r_1x +\cdots + r_{m-1}x^{m-1}$ are contaminated by rounding errors and while small are typically nonzero.  Especially for the Puiseux series computation, where the derivative starts with a zero constant term and is $O(x)$, this would mean that the change to $a^{(k+1)}$ would have spurious nonzero terms of order $1/x$, $1$, $x$, $\ldots$, $x^{m-1}$.  This can rapidly invalidate the results.  To make the algorithm work, then, one must recognize the rounding errors in the coefficients of the residuals, or simply avoid using terms that one knows ought to be zero.  This is not usually difficult.  In our practice, we used ultra-high precision to check, sometimes working in $100$ or more Digits, that terms that ought to be zero but looked nonzero were really the result of rounding errors and not blunders in programming.  This allowed us to clearly distinguish the effects of rounding errors in our experiments.
\end{remark}

\subsection{Examples of Puiseux series about double points}
For the double eigenvalue
$a^* = 2.088698902749695\ldots$ corresponding to the Mulholland-Goldstein double point $q=q^*=1.46876861378514\ldots\,i$, we have
\begin{equation}\label{eq:Puiseux}
  a = a^* + \alpha_1\sqrt{q-q^*} + \alpha_2(q-q^*) + \alpha_3(q-q^*)^{3/2} + \cdots\>.
\end{equation}
Computation according to the method of the previous section gives that
\begin{align}\label{eq:PuiseuxCoefficients}
  \alpha_1 \approx&\qquad \pm 1.65948780432026\ldots(1 + \,i) \nonumber \\
  \alpha_2 \approx&\qquad  - 0.119150377434444\,i \nonumber \\
  \alpha_3 \approx&\qquad \alpha_1\cdot (- 0.177731786327682\,i) \nonumber\\
  \alpha_4 \approx&\qquad - 0.0383269616582290 \nonumber \\
  \alpha_5 \approx&\qquad \alpha_1\cdot  (0.0107135404169547) \nonumber\\
  \alpha_6 \approx&\qquad - 0.00154061238466389\,i \nonumber\\
  \alpha_7 \approx&\qquad \alpha_1\cdot (0.00273004721440515\,i ) \nonumber \\
  \alpha_8 \approx&\qquad 0.000276547402694740 \nonumber \\
  \alpha_9 \approx&\qquad \alpha_1\cdot (0.000563051707888754) \>.
\end{align}
Puiseux series can be computed about every double point by the method suggested in the last section. We do not believe that such series have been reported in the literature.
We see a certain amount of unexplained regularity in this particular series.  Note that this particular $\alpha_1$ is a multiple of $(1+i)$, which is a consequence of the purely imaginary character of this first double point because $\sqrt{i} = (1+i)/\sqrt{2}$.  In table~\ref{tab:PuiseuxSeries} we tabulate the first few coefficients of some representative series, essentially as an \emph{homage} to all the great tabulators of Mathieu functions.

\subsection{Confirming Blanch \& Clemm} 
\begin{figure}
  \centering
\subfigure[Blanch and Clemm double points]{\label{fig:BlanchClemmPlot}\includegraphics[width=60mm]{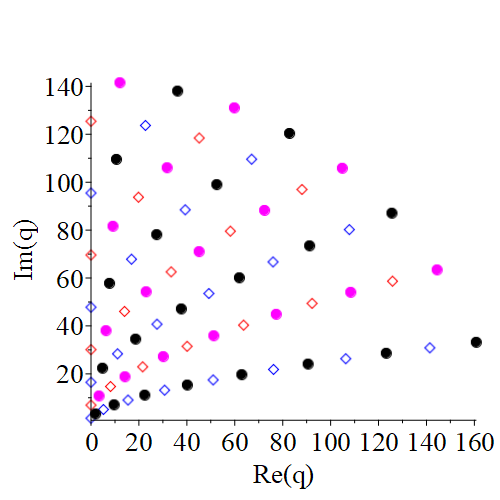}}
\subfigure[double points in all quadrants]{\label{fig:BlanchClemmPlotAllSyms}\includegraphics[width=60mm]{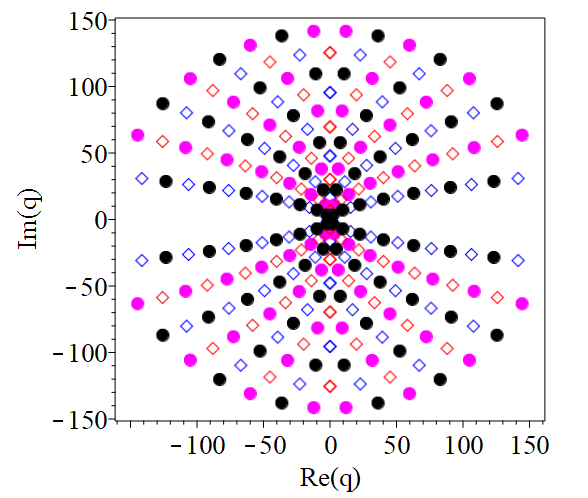}}
  \caption{In~\cite{blanch1969double} some forty double eigenvalues $a_k$ and thirty-two double eigenvalues $b_k$ were tabulated.  We plot their results here in figure~\ref{fig:BlanchClemmPlot}.  In the first quadrant, eigenvalues $a_{2k}$ merging with eigenvalues $a_{2k+2}$ are plotted as blue diamonds.  Eigenvalues $a_{2k+1}$ merging with eigenvalues $a_{2k-1}$ are plotted as black circles. Eigenvalues $b_{2k}$ merging with $b_{2k+2}$ are plotted as red diamonds. Eigenvalues $b_{2k+1}$ merging with $b_{2k-1}$ are plotted as fuchsia circles.
  We see arcs of apparent square-root like curves spreading to positive infinity; we also see oval arcs of similar beads coming from the imaginary axis to the real axis, forming a kind of peacock's tail of double points.
  Extension to all four quadrants in figure~\ref{fig:BlanchClemmPlotAllSyms} follows by conjugate symmetry and by the symmetries $a_{2k}(-q)=a_{2k}(q)$, $b_{2k}(-q)=b_{2k}(q)$, and $a_{2k+1}(-q) = b_{2k+1}(q)$, which last implies that the fuchsia circles and black circles exchange meaning in the left half plane: merging $a_{2k+1}$ and their conjugates are fuchsia circles in the left half plane, black circles in the right half plane; merging $b_{2k+1}$ and their conjugates are black circles in the left half plane, fuchsia in the right half plane. The real axis is special: only at $q=0$ do $a$ and $b$ eigenvalues merge to $k^2$, and they do so while keeping their independent eigenfunctions, which become $\cos k\alpha$ and $\sin k\alpha$ at $q=0$. }\label{fig:BlanchClemmDoubles}
\end{figure}

Consider figure~\ref{fig:BlanchClemmDoubles} where the smallest seventy-two double points\footnote{We took the tabulated values in~\cite{blanch1969double}, used Optical Character Recognition to convert them to computer-readable form, and ran our algorithm (which is the same as that of~\cite{hunter1981eigenvalues} and different from that of~\cite{blanch1969double}) from section~\ref{sec:double} to confirm them.  For an interesting article connecting OCR with~\cite{abramowitz}, see~\cite{Sexton2012}. Here, after correcting several amusing OCR errors including the near-inexplicable occurrence of Russian characters masquerading as numerals, we found (as did~\cite{hunter1981eigenvalues}) that all printed decimals in the tables of~\cite{blanch1969double} were correct. Blanch and Clemm did not plot their double eigenvalues but only tabulated them, but Hunter and Guerrieri did plot some of theirs.  More, they computed farther into the complex plane than did Blanch and Clemm, and found asymptotic formulae. In the figure, we see arcs of apparent square-root like curves spreading to positive infinity; we also see oval arcs of similar beads coming from the imaginary axis to the real axis, forming a kind of peacock's tail of double points.}, are plotted: as we said, there are a lot of them. Morever, as we saw above, for values of~$q$ \emph{close} to double points, the vanishing of the norm of the affected Mathieu functions means that the expansion coefficients must become singular, and thus numerically troublesome: at the very least, there will be cancellation error entailed by the subtraction of large nearly equal quantities.  We know of no discussion of this feature of the Mathieu functions anywhere in the literature.  Of course, this is a familiar phenomenon from other areas of mathematics, such as elementary linear algebra: consider the analogous problem of finding the eigenvectors of the following matrix, and expanding another vector as a linear combination thereof:
\[
\mat{A} = \begin{bmatrix}
            a & 1 \\
            t & a
          \end{bmatrix}
\]
which if $t \ne 0$ has eigenvalues $a \pm \sqrt{t}$, and linearly independent eigenvectors $\mat{v}_0 = [1, \sqrt{t}]^T$ and $\mat{v}_1=[1,-\sqrt{t}]^T$.  Expanding (say) $[1,1]^T = c_0\mat{v}_0 + c_1\mat{v}_1$ requires $c_0+c_1 = 1$ and $c_0\sqrt{t}-c_1\sqrt{t} = 1$ or $c_0 - c_1 = 1/\sqrt{t}$.  This is obviously analogous to the situation above.  It is even more analogous when one considers the \emph{generalized} eigenvector that arises at $t=0$: $\mat{A}\mat{u} = a\mat{u} + [1,0]^T$.  Exactly as in the Mathieu function case above, the numerical difficulties in expanding as a linear combination of eigenvectors show up for small nonzero $t$, but these are alleviated on adding the generalized eigenvector to the mix, and writing instead
\begin{equation}\label{eq:linalgframe}
\begin{bmatrix}
  1 \\
  1
\end{bmatrix} = c_0 \begin{bmatrix}
                      1 \\
                      \sqrt{t}
                    \end{bmatrix}
                   + c_1 \begin{bmatrix}
                      1 \\
                      -\sqrt{t}
                    \end{bmatrix}
                     + c_3\begin{bmatrix}
                            0 \\
                            1
                          \end{bmatrix}
\end{equation}
Now, of course, the set is not linearly independent, and one has to choose the coefficients in a sensible way.

\section{Algorithms for Mathieu functions and Modified Mathieu functions} 
\label{sec:alg}
Once one has the eigenvalue, one needs a way to construct the eigenfunction.  Now, our original motivation for studying this problem required accurate eigenvalues and eigenfunctions as a means to an end: namely the study of blood flow in a vessel of elliptic cross-section.  We consulted the literature and the available software and were not confident in the selection: there was a wide variety available, of varying quality and applicability.  Most frequently, the Mathieu functions were computable for real~$q$ only, and we needed them for complex values of~$q$.  For example, the Mathieu functions in SciPy fall into this category.  We found no software that promised to deal with double eigenvalues (except possibly Mathematica, but we do not have a license for Mathematica).

We decided to thoroughly investigate the available software and algorithms and see what they could do.  To that end we decided to work from the ground up: we would investigate \emph{all known} methods to compute Mathieu functions.
Our goal was to \emph{understand what was going on}, not necessarily invent our own method.  However, as a comparison to existing software, we did construct an arbitrary-precision method that is not too inefficient, which we could assess for \emph{accuracy} internally.  This gave us a good standard with which to measure other methods' accuracy.

We started with the simplest method.
\subsection{An impractical algorithm\label{sec:simpleTaylor}}
The most straightforward method for computing an entire function is to use its Taylor series at a convenient point, say the origin.  We are guaranteed, because the series for an entire function always converges, that by taking enough terms and using enough precision in our arithmetic we can get an accurate answer. For instance, the basic Mathieu function $w_{I}(a,q,z)$, called \texttt{MathieuC} in Maple, has the Taylor series beginning
\begin{equation}
    w_{I}(a,q,z) = 1 + \frac{a-2q}{2!}z^2 + \frac{(a-2q)^2-8q}{4!}z^4 + \cdots
\end{equation}
and since the function is entire, this series converges for all $z$.  However, the series is impractical, (as is well-known to numerical analysts) which we demonstrate explicitly now.

Taking (say) the modest values $a=29/20$ and $q=3/5$, we still use several cpu minutes (on a 2017 vintage Windows tablet computer) to compute all the terms in the above series, using exact arithmetic, up to $O(z^{800})$.  This expense might be considered as pre-computation cost, and ignored, and so we do.
This truncated series can then be used to compute, for instance
\begin{equation}
w_{I}(a,q,4.0i)=-0.13260687218535639758416285435423\textcolor{red}{86957557534433}\>,
\end{equation}
using 46 Digits of precision in the computation; the final nonzero term of the truncated series is about $-1.4\cdot10^{-47}$ and because all subsequent terms are smaller\footnote{The series is not an alternating series because the sign pattern is more complicated.  This is not a rigorous error estimate, therefore.}, we would hope that the computed value is accurate.
But the final 12 Digits, coloured red, are incorrect, because some intermediate terms in the series are about $10^{12}$ in size and rounding errors in those terms are revealed by the cancellation of the large terms. This is a well-known effect, of course, discussed in many textbooks and educational papers; see e.g.~\cite{sevyeri2018runge}, which discusses it as an effect of the ill-conditioning of the truncated Taylor series.

The situation gets very much worse as $|z|$ increases.  Already by $z=4.36i$ the best accuracy we can achieve with the $800$ term series is double precision, because although in exact arithmetic the truncation error is about $10^{-17}$, the condition number is about $10^{20}$ and so we have to carry $20$ extra digits; by $z=4.45i$ we can only achieve single precision, and that by using $22$ extra digits; and by $4.51i$ only half precision by using $24$ extra digits. By $z=5.0i$ it is already true that $800$ terms are \textsl{not enough} in the series, no matter what precision we use because the truncation error is too big.  Moreover, even if we had enough terms, because the condition number is about $10^{40}$, the computation would need about $40$ extra decimal digits of precision anyway.

The final conclusion is that this method is unaffordable. It can be rescued by using analytic continuation instead of a single series (which of course is the underlying basis for most numerical methods to integrate ODE), and we look at that method in section~\ref{sec:numericalintegration}.  Most researchers instead choose to jump straight to a spectral method for the periodic Mathieu functions, expanding in Fourier series; and then by a trick these can be used also for the modified Mathieu functions.
We will discuss this method in section~\ref{sec:spectralmethod}.
\subsection{Analytic continuation (marching)\label{sec:numericalintegration}} 
One straightforward thing to try, valid for both the Mathieu functions and for the modified Mathieu functions, is numerical solution of the Mathieu equation.  We could in principle use any method that allows integration along a complex path, such as any Runge-Kutta method or multistep method.  Because these solutions are to be later used as functions, requiring evaluation at any point in their domain, we will need good interpolants and modern codes automatically supply these.

However, because the equation is linear, and because its variable coefficient has a known Taylor series algorithm, it is a straightforward exercise to develop a specialized numerical solution by marching Taylor series, a method known as \textsl{analytic continuation} if infinite Taylor series are used and as a \textsl{Taylor series method} if a finite order is used at each step.  Taylor series methods are often introduced in numerical analysis textbooks and then dismissed practically in the same breath as being too costly or insufficiently general, but these methods do not actually suffer from those faults when properly implemented: see for example~\cite{nedialkov2005solving} which shows how to use them to solve DAE but also gives many references to papers using them to solve initial-value problems (IVP) for ODE.  See also~\cite{Gil2007} for a discussion of their use for computation of special functions.  Taylor series methods are especially attractive when the differential equations are very simple, but they work very well on most smooth problems.
For such problems, Taylor series methods are of cost \emph{polynomial} in the number of bits of accuracy required, on finite intervals, as the number of bits of accuracy required goes to infinity~\cite{ilie2008adaptivity}; in contrast, fixed-order methods such as any given Runge--Kutta method are of cost \emph{exponential} in the number of bits of accuracy required. So Taylor series methods are actually a quite natural method to try when high accuracy is wanted, which might surprise the casual user of mathematical software for solving ordinary differential equations.

One practical advantage of such methods is that they come with free interpolants, and another advantage is they come with an inexpensive estimate for the \emph{residual}.  A third advantage is that the residual can be accurately measured afterwards (as opposed to estimated beforehand) to validate the solution.  These methods can be implemented as variable stepsize and as variable order.  In the case of stiff problems they can be implemented as implicit methods~\cite{barton1980taylor}.  Indeed, these methods are interesting for other problems, not just the Mathieu equation.

So, that's what we did.
In fact, we have implemented (in Maple) what is called a \emph{Hermite-Obreschkoff} method, which is implicit and uses Taylor series at both ends of the marching step, and is more numerically stable than explicit Taylor series methods.  Implicit methods are more effective for stiff problems, as is well-known~\cite{soderlind2015stiffness}.

The Mathieu equation is not usually \emph{stiff}, per se, but rather is frequently \emph{oscillatory} (especially along the imaginary $z$-axis, i.e.~for modified Mathieu functions).  This causes a pure Taylor series method to suffer some instability. In contrast, the Hermite-Obreschkoff method, which is implicit (something like a generalization of the implicit midpoint method), performs more satisfactorily.

We implemented this, not because we thought it would be the \emph{best} method to evaluate Mathieu functions, but because we thought the implementation would be useful in our quest for understanding of Mathieu functions, and moreover provide an \emph{independent} check on other methods that we explore.  The fact that such methods are possibly interesting for other problems is merely a bonus.

\subsubsection{Specifics of the Hermite-Obreschkoff method we use}
This description is quite short because the basic technology is widely understood and our implementation is not the main focus of this paper.  We provide details only for reproducibility.
The key piece is the recurrence relations for the Taylor coefficients of the expansion of $y(t_n+\Delta t) = \sum_{k\ge0} w_k \Delta t^k$.  Simultaneously, we need the Taylor coefficients for $\cos2t$, which necessitates the Taylor coefficients for $\sin2t$.  This gives
\begin{align}\label{eq:TaylorRecurrence}
  C_{k+1} =& -\frac{2}{k+1}S_k \nonumber \\
  S_{k+1} =& \frac{2}{k+1} C_k \nonumber \\
  w_{k+2} =& -\left(a w_k - \frac{2 q}{(k+1)(k+2)} \sum_{j=0}^{k} C_j w_{k-j}\right)\>.
\end{align}
$C_0 = \cos(2 t_n)$ and $S_0 = \sin(2t_n)$, while $w_0 = y(t_n)$ and $w_1 = y'(t_n)$.  This recurrence needs to be scaled if we are integrating on a complex path.

We interpolate over the interval $t_n \le t \le t_{n+1}$ by using what we call a ``blend'': this is nothing more than Hermite interpolation using the Taylor coefficients at each end of the step~\cite{corless2020blends}.  A string of such blends put together as a piecewise polynomial is called a ``string of blends'' because it is tied together at ``knots''.  Such interpolants are quite remarkably stable numerically, even at very high order, and can be evaluated in cost linear in the degree of the interpolant.  The high order means that they are potentially competitive with spectral methods in efficiency for smooth problems, when high accuracy is desired~\cite{ilie2008adaptivity}.  They are simple to integrate and differentiate, and reasonably simple to multiply together, and this gives inexpensive ways to evaluate integrals containing these numerical solutions, e.g. $\int_{0}^{\pi} f(z)\ce_3(q,z)\,dz$. There are also simple methods based on companion matrices to find zeros of blends; alternatively, there is an iterative scheme of higher order than Newton's method ($1+\sqrt{3} \approx 2.732$ instead of $2$) for the same cost per iteration~\cite{corless2020inverse}.  One can imagine implementing an analogue of Chebfun~\cite{Battles2004} using blends, and this might be interesting because the approximation properties are different (mostly not as good, but in some niche circumstances when many derivatives are easily computed at isolated points in the complex plane, possibly competitive). An analogue of Chebfun using sinc functions~\cite{Richardson2011} has some interesting properties; perhaps an analogue of Chebfun using blends would be worth investigating.  A final advantage for this method is that it works directly on the Mathieu equation and the modified Mathieu equation both, so that both Mathieu functions and modified Mathieu functions can be computed with the same code merely by changing the initial conditions.

The Hermite interpolant (``blend'') essentially doubles the order of the method for the same number of Taylor coefficients taken at each step, and moreover greatly improves stability: when the degrees of the Taylor coefficients are the same at each end (as we use) this is quite stable~\cite{nedialkov2005solving}. Indeed on the Dahlquist test problem, a balanced implicit Taylor series method is $A$-stable (this was known as early as 1991~\cite{kirlinger1991implicit}). Because the problem is linear, the method winds up being only linearly implicit.

To take a step, we predict the stepsize by the PI step control suggested in~\cite{gustafsson1988api}, which uses data from the previous two steps (N.B.: at the beginning we use a predictor based on Taylor series alone).  Then we generate two separate Taylor series at $t_n+\Delta t$, one (say $y_c(t)$) with initial conditions $y(t_n+\Delta t) = 1$ and $y'(t_n+\Delta t) = 0$ and another (say $y_s(t)$) with $y(t_n+\Delta t) = 0$ and $y'(t_n+\Delta t) = 1$; we then use collocation at the points $\tau_1 = t_n + \Delta t/4$ and $\tau_2 = t_n + 3\Delta t/4$ (these are Chebyshev-Lobatto points): we blend the Taylor series at $t_n$ (which we knew already) with a linear combination $z = \alpha y_c(t) + \beta y_s(t)$ of the new ones at $t_n + \Delta t$ with the linear combination chosen to make the residual $r(t) = z'' - (a-2q\cos(t))z$ equal to zero at $\tau_1$ and $\tau_2$.  This gives us two linear equations in the two unknowns $\alpha$ and $\beta$.  Solution of the $2$-by-$2$ system is hand-coded in. The second derivative $z''(t)$ of the blend $z(t)$ is computed by program differentiation (actually hand-coded, so not technically ``automatic'' differentiation: call it ``semi-automatic differentiation'').

We then \emph{measure} the residual of the resulting blend at $s = t_n + \Delta t/2$.  This is asymptotically (as $\Delta t \to 0$) the location of the maximal residual of this blend.  If this residual is small enough, smaller than the specified tolerance, we accept the step and use this estimate in the PI control to predict the size of the next step.

This therefore gives us an independent, reliable, and reasonably efficient special-purpose numerical solver for the Mathieu equation.  In our experience it performs well, especially because we can measure the residual $r(t)$ separately at many points, to provide reassurance that we have obtained the exact solution to a differential equation very near to the Mathieu equation.  Trivially, we have solved
\[
y'' + (a - 2q\cos2t)y = r(t)
\]
denoting the residual by $r(t)$, and we have chosen stepsizes to ensure that $|r(t)|$ is small.
By the Alexeev-Gr\"obner nonlinear variation of constants formula this means that the global error will also be small: there exists a function $G$ so that (with the same initial conditions for the reference solution $y(t)$ and the computed solution $z(t)$ solving the above)
\[
y(t) - z(t) = \int_{\tau=0}^t G(t,\tau) r(\tau)\,d\tau\>.
\]
Indeed because the Mathieu equation is linear we may write $G(t,\tau)$ explicitly as a Green's function:
\begin{equation}\label{eq:GreensFunction}
  G(t,\tau) = w_I(\tau)w_{II}(t) - w_{II}(\tau)w_I(t)\>.
\end{equation}
We will use this again later.
The Wronskian of the Mathieu equation is $1$.
We did not plot $G(t,\tau)$ but instead merely estimated its effects by integration at different tolerances.  Outside of regions of exponential growth, \emph{i.e.} where the real part of the characteristic exponent is zero, the Mathieu equation is well-conditioned.  Even when $\Re(\mu) > 0$ this method allows us to keep the global error \emph{relatively} small over short enough integration paths.

\begin{figure}
  \centering
\subfigure[$\Re(\mu)$ in the real $(a,q)$ plane]{\label{fig:ReFloquet3D2}\includegraphics[width=60mm]{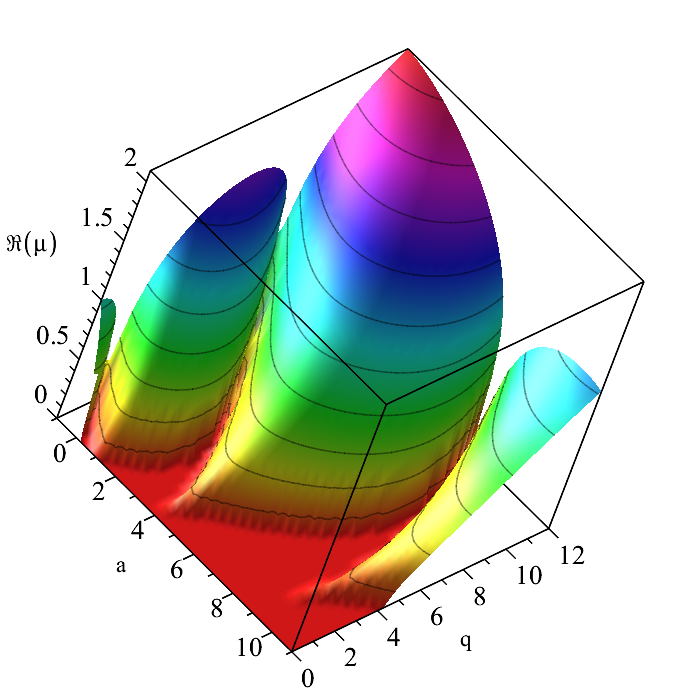}}
\subfigure[Contours where $\Re(\mu)=0$]{\label{fig:ReFloquet}\includegraphics[width=60mm]{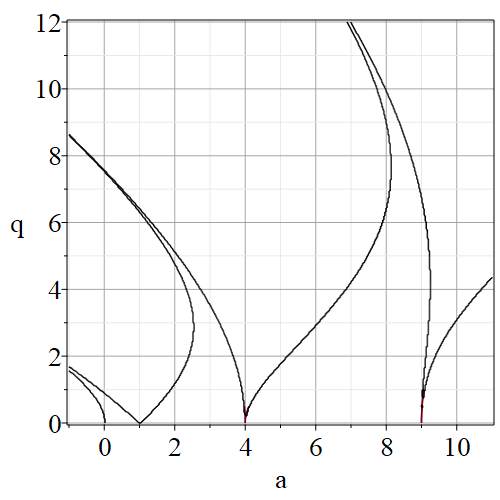}}
\newline
\subfigure[$\Re(\mu)$ in the $(a,s)$ plane where $q=is$]{\label{fig:ImFloquet3D}\includegraphics[width=60mm]{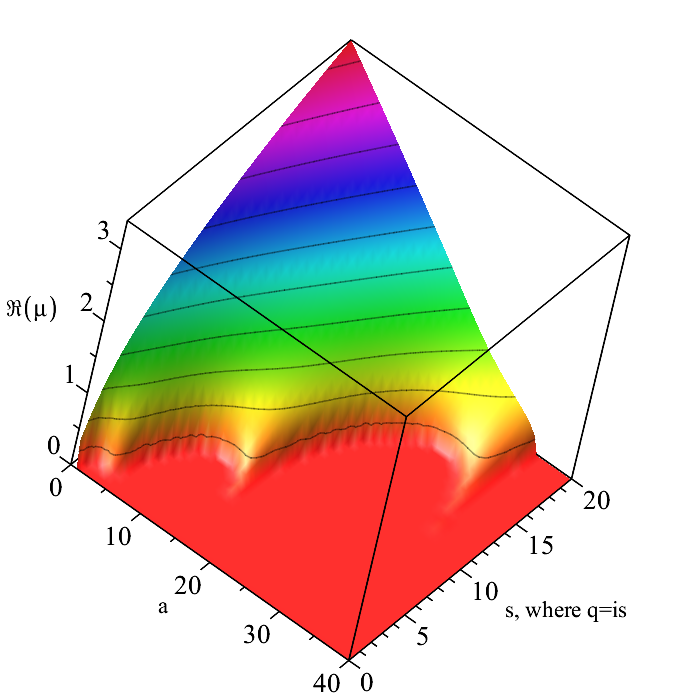}}
\subfigure[Contours where $\Re(\mu)=0$ showing double points.]{\label{fig:FloquetWithDoubles}\includegraphics[width=60mm]{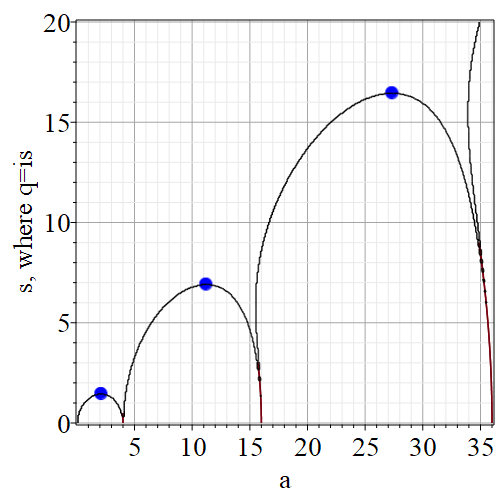}}
  \caption{Stability regions and characteristic exponent $\mu$ of solutions of the Mathieu equations. Here we track $\Re(\mu)$ where the Floquet solutions are $\exp(\mu z)\phi(z)$ and $\exp(-\mu z)\phi(-z)$ with $\phi(z)$ periodic with period $\pi$.  Regions near the $a$ axis are stable as~$q$ is increased in a purely real fashion, except only neutrally so at $q=0$ and $a=g^2$ for integers $g$; regions near the $a$ axis are stable as $s$ is increased where $q=is$, except as before only neutrally so at $a=g^2$. Near those difficult-to-contour regions we supplemented the graph with Taylor series expansions of the characteristic curves, in red. Double eigenvalues are indicated with blue dots.}\label{fig:FloquetHO4}
\end{figure}

In short, we wrote a special-purpose numerical ODE solver to evaluate the Mathieu functions.  By varying the initial conditions, we may compute either of the basic solutions, $w_I(t)$ or $w_{II}(t)$; once one has an eigenvalue given~$q$ this already allows computation of any of the Mathieu functions.  We may compute the other independent (nonperiodic) solution as we did for figure~\ref{fig:redbelly}.  By integrating on a complex path, we may compute any of the modified Mathieu functions.  The modified Mathieu functions, which are also needed already for the solution of the vibrating elliptic membrane, are of course not periodic.
We can compute Floquet solutions as we did for figure~\ref{fig:floquet}, and estimate the Floquet exponent numerically~\cite{fairgrieve1991ok} and compare it with Poincar\'e's formula for $\mu$ (equation~\eqref{eq:Poincare}), see figure~\ref{fig:FloquetHO4}.  It is also possible to use this method with the Green's function in equation~\eqref{eq:GreensFunction} to compute the generalized eigenfunctions needed for double eigenvalues (see the discussion in section~\ref{sec:computegeneralized}).

In spite of these capabilities, we are \emph{not} recommending our implementation of this method as a general-purpose Mathieu solver, both because we have not made it bulletproof and because we believe that for most purposes it would not be as fast as a spectral solver based on Fourier series---after all, with a Fourier method \emph{all} the Fourier coefficients are essentially immediately available once the eigenvectors have been computed---but for our purposes, which included an independent check on existing software, it was adequate.  By varying the precision at which we computed and by taking tolerances to be extremely small we were able to buy accurate solutions by paying for computing cycles.  That said, it is a variable order method that can tolerate quite high order\footnote{We have run it routinely with order $20$, and sometimes as high as $80$.  That means that the residual error on a subinterval of width $h$ is $O(h^{80})$, asymptotically as $h \to 0$.  But in practice the code can take quite large steps at this order, so the notion of an asymptotic order is not as helpful as one might think.  What makes this work is that the \emph{error coefficients} are also very small, and of course the solutions to the problem are analytic.} without numerical difficulty, and so is quite fast.  We have not made a detailed speed comparison with a spectral method.
However, for the Mathieu functions on which we concentrate this paper---which are periodic---one expects that the spectral method of expanding in Fourier series to be more efficient than any possible marching method. For the \emph{modified} Mathieu functions, the issue is not so clear-cut.  Again, though, we wanted an independent and reliable method as a comparator for other methods, and in this the Hermite-Obreschkoff code we implemented is perfectly satisfactory.

\subsubsection{Complexity and other issues}
One interesting complication is that when using the Hermite-Obreschkoff or Taylor method (or, indeed, when using \textsl{any} method), accurate solution becomes very expensive for large $|z|$.  This is because solutions to the modified Mathieu equation become \textsl{highly oscillatory} along the real axis (so the solutions to the Mathieu equation become highly oscillatory along the imaginary axis). Indeed the asymptotics of the solution to the modified Mathieu equation contain a term $\exp( i\,2\sqrt{q}\cosh(z) )$ (see Formula 28.25.1 in the DLMF; we ignore the denominator, which grows only more slowly).  This is confirmed by inspection of the formulae in Chapter 28.23 of the DLMF (\hyperlink{https://dlmf.nist.gov/28.23}{https://dlmf.nist.gov/28.23}) where we see the arguments $2\sqrt{q}\cosh(z)$ and $2\sqrt{q}\sinh(z)$ appearing inside the Bessel functions used in the expansions for the modified Mathieu functions.

Resolving any solution for graphical purposes requires computing a fixed number of points in each cycle, say $5$ or $10$.   One therefore sees that the number of computed points required to resolve the solution grows exponentially with the real part of $z$, because the number of cycles grows exponentially with the width of the interval.  We therefore see that the cost to directly compute---in this sense---the solution accurately must grow exponentially with the argument. This in part was why the simple Taylor series at the origin failed already by $z=5.0i$, in section~\ref{sec:simpleTaylor}.

This does not contradict the theorem in~\cite{ilie2008adaptivity} because that only holds on a fixed finite interval, in the limit as the required accuracy goes to infinity.

Off the real or imaginary axes, if $z=x+iy$ has nonzero $x$ and $y$, then the solution not only oscillates but grows \textsl{doubly exponentially} as $x$ grows, containing the terms $\exp(\pm 2\sqrt{q}\sinh(x)\sin(y))$. This fact does not seem to have been explicitly remarked on elsewhere, perhaps because the information is readily visible in the formula, and the user is expected to understand this without being told.  That fact suggests that most applications of the Mathieu functions must have only small $|z|$ or else either purely real $z$ or purely imaginary $z$; that is, either the real Mathieu functions or the real modified Mathieu functions will be the relevant functions.  Asymptotics cannot recover the phase information, of course---so if you want an accurate solution for large $z$, you will have to pay a significant price for it (and know $z$ to exponential accuracy, in a certain sense).

This conclusion may be surprising, so take the simpler function $S(x) = \cos( \cosh x )$ (which occurs when setting $q=1/4$ in formula 28.25.1 of the DLMF and taking only one part of the complex exponential and ignoring all slower-varying terms).  If $x$ is about $1000$ or so, how many bits of $x$ are necessary to know before you can extract one bit of information, namely the sign, of $S(x)$?  Since $\cos\theta = 0$ when $\theta = (2k+1)\pi/2$ for some integer $k$ it follows that we must know $x$ so well that we can detect that it is between $x_k$ with $\cosh x_k = (2k+1)\pi/2$ and $x_{k+1}$ with $\cosh x_{k+1} = (2k+3)\pi/2$.
Now $x_{k+1}-x_k = \mathrm{inv}\cosh( (2k+3)\pi/2 ) - \mathrm{inv}\cosh( (2k+1)\pi/2 ) \sim 1/k - 1/k^2 + \ldots$ as $k \to \infty$.  But how large is $k$?  If $x_k$ is about $1000$, then $(2k+1)\pi/2$ is about $\cosh 1000$, or $k$ is about $2^{1440}$.  So $1/k$ is about $2^{-1440}$. To detect the difference between $x_{k+1}$ and $x_k$, then, we need to carry twice that, or $2880$ bits. That corresponds to about $180$ hexadecimal digits, or about $240$ decimal digits.  For $x=10^4$ instead, we need to know about $2400$ decimal digits of $x$; all this just to get the sign of $S(x)$ correct. Yet we know the asymptotics of this function perfectly well---it's just a cosine with exponentially increasing frequency.  We remark that the cost of arithmetic with big floats grows \emph{at least} linearly with the length of the float; FFT multiplication is a logarithmic factor more costly (and naive multiplication grows quadratically in cost).

This discussion suggests that any numerical method that we use will only be helpful and efficient for $|z|$ bounded by a modest constant, say $2\pi$.  For larger values of $\Re{z}$ one must use the asymptotic formula instead, and give up on phase information.

\begin{remark}
The highly oscillatory nature of the solutions to the modified Mathieu equation also induce an instability for large $|z|$ in the Taylor series method, akin to the instability induced for stiff problems~\cite{corless2019backward,soderlind2015stiffness}.
If the stepsize is not aggressively reduced as $|z|$ increases, then eventually this instability takes over and the numerical solution becomes meaningless and rapidly overflows.
One would hope that instead using an implicit Taylor series method would alleviate this problem, and to a certain extent it does, but even with an implicit method the stepsize is forced to be so small for accuracy that failure is assured in practice when $|z|$ is at all large.
\end{remark}

\begin{remark}
The Bessel function expansions in \hyperlink{https://dlmf.nist.gov/28.23}{https://dlmf.nist.gov/28.23}, and similarly those of~\cite{erricolo} which we will discuss presently, seem to offer a way around the difficulty, at least when $a$ is an eigenvalue of the problem and the solution is periodic.  Accurate computation of the series coefficients, together with accurate computation of the appropriate Bessel functions, seem to offer an inexpensive way to accurately evaluate a modified Mathieu function.  Inspection of the derivative, however, shows that the large argument $z=x+iy$ also needs to be known to exponential accuracy in order to find an accurate value of the Mathieu function in question, especially near a zero.  The difficulty seems to be intrinsic.  Since hardly anyone complains about this in the literature, we conclude that most people only want the values of Mathieu functions and modified Mathieu functions for small or at most moderately large values of $z$, or magnitude information and not phase information when $z$ is large.
\end{remark}

Mathieu himself thought there might be issues for large $z$ and introduced the change of variable $\nu = \cos(z)$, which leads to the following \emph{algebraic} differential equation (ADE) (equation 28.2.3 in the DLMF).  Recall that an ADE is generally different from a DAE.
\begin{equation}
    \left( {\nu}^{2}-1 \right) {\frac {d^{2}}{d{\nu}^{2}}}y \left(
\nu \right)    +  \nu {\frac d{d\nu}}y \left( \nu \right)
+   \left(  \left( 4\,{\nu}^{2}-2 \right) q-a \right) y \left( \nu
 \right)
= 0\>.
\end{equation}
This algebraic differential equation (and indeed also the similar one in equation 28.2.2 in the DLMF that arises on $\zeta = \sin^2 z$) has some interesting computational properties: for one, they are what is called $D$-finite or \emph{holonomic}, meaning the recurrence relation for the Taylor series terms is of fixed order, and which means accurate computation can be done asymptotically quickly~\cite{van2001fast}.  See also~\cite{mezzarobba2010numgfun,mezzarobba2012note,benoit2017rigorous}.

Indeed near the imaginary axis this DE is also numerically easier---one may take more nearly equal integration steps in the new variable, instead of the ever-decreasing ones necessary in the original variable $z$---but somehow this is just ``sweeping the problem under the rug'' because the value of $\nu$ itself becomes very large; if the cost of the stepsize selection is low (and the control is effective) one should get very nearly the same numerical performance in the original variable, except that the recurrence relation for the Taylor series terms in the original variable depends on all previous terms, not just a fixed number as for D-finite functions and so each step is also more expensive.  However, the presence of the singularities $\nu=\pm 1$ complicates matters---indeed the initial condition in $\alpha$ at $\alpha=0$ corresponds to a singular point $\nu=1$ and so something special must be done to get the integration started, and moreover the branching structure of the nonlinear transformation also makes its presence felt because the inverse transformation is multivalued.  Assessment of the resulting solution by the method of residuals also needs an extra step.  So although all these obstructions can be dealt with, it is not actually clear which of these two approaches is numerically best. Experiments seem necessary.  Since our purposes were served by the Hermite-Obreschkoff method in the original variable, we did not pursue this.

\subsection{Spectral methods\label{sec:spectralmethod}}
However, the method of choice, at least for most implementers, for computation of Mathieu functions is the use of what is effectively a spectral method.  This is \emph{prima facie} valid only for the case when $a$ is an eigenvalue and the solution is periodic.  One computes the eigenvalues by the matrix method as in section~\ref{sec:Ince}, and then the resulting eigenvector gives the coefficients in the Fourier series expansion of the Mathieu function (if using the continued fraction instead, the recurrence relations can be used, although one has to take numerical care).  The Fourier coefficients decay extremely rapidly, as is usual for Fourier expansion of smooth functions.  To give an example to show just \emph{how} rapidly, here is Equation 28.4.24 from the DLMF (here the superscripts refer to which eigenfunction and the subscripts refer to which coefficient in its Fourier expansion):
\begin{equation}
\frac{A^{2n}_{2m}(q)}{A^{2n}_{0}(q)}=\frac{(-1)^{m}}{(m!)^{2}}\left(\frac{q}{4%
}\right)^{m}\frac{\pi\left(1+O\left(m^{-1}\right)\right)}{w_{\mbox{\tiny II}}(%
\frac{1}{2}\pi;a_{2n}\left(q\right),q)}\>.
\end{equation}
This holds as $m \to \infty$, for fixed $n$.  This states that the $m$th Fourier coefficient ultimately decays like $(q/4)^m/(m!)^2$, much faster than exponentially.  The other three types of Fourier coefficients decay similarly rapidly.  This very rapid decay means that good approximations can be made with only a few terms of the Fourier series.

For the corresponding \emph{modified} Mathieu functions these Fourier series also converge, but now only slowly.  As an alternative, one can use the same Fourier coefficients in a Bessel function expansion, for instance equation 28.23.2 of the DLMF:
\[
\mathrm{me}_{\nu}\left(0,h^{2}\right){\mathrm{M}^{(j)}_{\nu}}\left(z,h\right)=%
\sum_{n=-\infty}^{\infty}(-1)^{n}c_{2n}^{\nu}(h^{2}){\cal C}_{\nu+2n}^{(j)}(2h%
\cosh z)\>,
\]
where the modified Mathieu function on the left can be approximated by the series on the right; the notation therein is different from the notation in our paper, but the $\cal C$s are Bessel functions and the $c_{2n}$s are the Fourier coefficients for the ordinary even period-$\pi$ Mathieu functions.  These are not the only series one might use. In~\cite{erricolo} (and in~\cite{abramowitz} and the DLMF) we find the following expansions, which the authors claim are rapidly convergent:
\begin{align}\label{eq:ModifiedMathieuBesselExpansions}
  \Ce_{2n}(q,x)   &= \frac{(-1)^n}{A_0}\sqrt{\frac{\pi}{2}} \sum_{k \ge 0} (-1)^k A_{2k} J_k(s)J_k(t) \nonumber\\
  \Ce_{2n+1}(q,x) &= \frac{(-1)^n}{A_1}\sqrt{\frac{\pi}{2}} \sum_{k \ge 0} (-1)^k A_{2k+1} \left(J_{k+1}(s)J_k(t) + J_k(s)J_{k+1}(t)  \right)\nonumber\\
  \Se_{2n}(q,x)   &= \frac{(-1)^n}{B_0}\sqrt{\frac{\pi}{2}} \sum_{k \ge 1} (-1)^k B_{2k} \left(J_{k+1}(s)J_{k-1}(t) - J_{k-1}(s)J_{k+1}(t)  \right)\nonumber\\
  \Se_{2n+1}(q,x) &= \frac{(-1)^n}{B_1}\sqrt{\frac{\pi}{2}} \sum_{k \ge 0} (-1)^k B_{2k+1} \left(J_{k+1}(s)J_{k}(t) - J_{k}(s)J_{k+1}(t)  \right)         \>.
\end{align}
Here $s = \sqrt{q}\exp(x)$ and $t = \sqrt{q}\exp(-x)$.
We find these series to be preposterous: the $A_k$ and the $B_k$ are the Fourier coefficients of the corresponding Mathieu functions, namely the \emph{same} coefficients as in the Fourier series for the corresponding Mathieu function. Now they are to be used in a completely different, almost alien-looking, series?  But these preposterous formulae are both correct and useful, and go back at least to~\cite{dougall1915solution}. Erd\'elyi thought the `coincidence' of these series coefficients to be significant~\cite{erdelyi1942certain}.
Here we have translated to the notation of this paper---although the normalization used in the above does not agree with the normalization here, even though the authors of~\cite{erricolo} claim to use the same normalization that we do; instead the formulas are the same as those in~\cite{bateman1953higher}. We tried this, and it worked well.
Indeed, such a Fourier-Bessel method is similar to those advocated by almost everyone, from \cite{alhargan1996complete} to~\cite{ziener}.
Against this Fourier-Bessel method it is not clear that a straightforward numerical solution such as the one we have implemented would be competitive, but for instance the authors of~\cite{Schneider:1999:modified} seem to think that something like it might be. Several authors indeed claim that numerical evaluation of Bessel functions for large arguments and high order is difficult or inefficient, but we do not believe this statement: in our experience the standard recurrence method performs well.  There do seem to be numerical instabilities in some Bessel series that can cause difficulty, although these can be mitigated by careful choices among the expansions~\cite{van2007accurate}. We have not done a detailed comparison of methods.

In general one has to be a bit careful with rounding error if the recurrence relations are used to compute the Fourier coefficients: the relations can be unstable, as noted by many authors.  A practical solution is to use forward recurrence for the first few and backward recurrence for the rest.  This seems to have been first advocated by Blanch~\cite{Blanch:1946:Computation}.  Similar considerations apply if numerical eigenvectors are used (after all, the matrix is simply an arrangement of the recurrence relations).  A rule of thumb is that if one is working to $d$ digits with the forward recurrence, then the Fourier coefficients will decay down to about $10^{-d/2}$ times the magnitude of the largest coefficient and then rounding error will start to impact the results thereafter.  In our experiments we simply used ultra-high precision and didn't worry about rounding errors at all.

\subsection{Computing a generalized eigenfunction\label{sec:computegeneralized}}
At least four ways suggest themselves to compute the generalized eigenfunction $u = \partial y/\partial a$ at a double point $q^*$ with double eigenvalue $a^*$, which is a solution of equation~\eqref{eq:pertsol} that satisfies periodic boundary conditions. We duplicate that equation here for convenience (recall $'$ indicates $d/d\alpha$):
\begin{equation}\label{eq:pertsoldu}
  u'' + (a^* - 2q^*\cos2\alpha)u + y^* = 0\>.
\end{equation}
The first and simplest way is undoubtedly what people actually use: one pretends that the eigenvalue is not actually a double one---typically because of rounding error it would have split anyway into $a^* + d\sqrt{q-q^*}+\cdots$ and $a^* - d\sqrt{q-q^*} + \cdots$ where~$q$ is a floating-point approximation to $q^*$ anyway---and then use the computed eigenfunctions from the matrix method, each with norm $O((q-q^*)^{1/2})$ and simply live with the errors. That does not sound like professional practice, but if it is done knowingly then we suspect that it will usually give perfectly reasonable answers.  If done \emph{unknowingly} then we disapprove, but the criminals will likely get away with it.

The second way is to compute a generalized eigenvector of the infinite tridiagonal matrix $\mat{A}$ for the problem. This is scarcely harder than the crude approach above: first, one averages the two computed approximate eigenvectors for the double eigenvalue split pair, and averages the eigenvalues, to get a more accurate double eigenvalue $a^*$ and eigenvector~$\mat{v}^*$.  For convenience in the exposition below, suppose that the eigenvectors of $\mat{A}$ are numbered $\mat{v}_1$ (corresponding to eigenvalue $a_1$), $\mat{v}_2$ (corresponding to eigenvalue $a_2$), and so on.
Suppose further that we have reordered the eigenvalue and eigenvector pairs so that the pair of eigenvalues that arose on splitting the double eigenvalue numerically are put in positions $1$ and $2$, and similarly put their associated eigenvectors in columns $1$ and $2$ of the matrix of eigenvectors, and number them $\mat{v}_1$ and $\mat{v}_2$. Put $a^* = (a_1 + a_2)/2$ and $\mat{v}^* = (\mat{v}_1 + \mat{v}_2)/2$.  Averaging is well-known to produce good estimates of double eigenvalues and eigenvectors.
If, of course, by some miracle the eigenvalue routine doesn't split the double eigenvalue and instead produces a single, accurate, double eigenvalue and only one corresponding eigenvector, then we use that.

Next, one solves
\[
\left(a^* \mat{I} - \mat{A}\right)\mat{u} = - \mat{v}^*\>.
\]
This system is singular, but $\mat{v}^*$ is in its range and this is not difficult; one could use the SVD, for instance\footnote{It is straightforward to set up the recurrence relations for this generalized eigenvector: they are merely forced versions of the recurrences in equations~\ref{eq:recurrencebase}--\ref{eq:tridseodd}; but as mentioned these recurrences are known to be unstable sometimes, while in contrast the SVD has the virtue of answering the question in a manner that relieves all doubts.}.  Since the matrix is of low dimension, the expense of the SVD is no obstacle.  Then one uses the entries of $\mat{u}$ to construct the generalized eigenfunction $u(z)$ in the same manner one constructs the eigenfunction $v_1(z)$ from the eigenvector~$\mat{v}^*$.

To expand a given function $f(z)$ as a sum of these eigenfunctions, notice that the ``norm'' of $v_1$ is zero, but the bilinear form of $v_1(z)$ with $u(z)$ is nonzero.  The generalized eigenfunction and $v_1(z)$ are each orthogonal to all other eigenfunctions, however. Put
\begin{equation}\label{eq:genexpansion}
  f(z) = \alpha v_1(z) + \beta u(z) + \sum_{j=3}^{N} \gamma_k v_k(z)\>.
\end{equation}
Then the $\gamma_k$ are easily found by orthogonality as usual:
\[
\gamma_k = \frac{\int_{z=0}^{p} f(z) v_k(z)\,dz}{\int_{z=0}^{p} v_k^2(z)\,dz} \qquad k=3, 4, \ldots, N\>.
\]
We remind you that this bilinear form does \emph{not} use the conjugate.  It is not an inner product. The ``norm'' of a nonzero function can be negative, complex, or indeed zero.  Such ``norms'' are called \emph{indefinite} norms, in for instance~\cite{arbenz2004jacobi}.

To find $\alpha$ and $\beta$ we use the fact that while the norm of $v_1(z)$ is zero, the norm of the generalized eigenfunction $u(z)$ is not zero, and also
\[
\int_{0}^{p} v_1(z)u(z)\,dz \neq 0\>.
\]
Thus the two equations
\begin{align}\label{eq:solvegencoeff}
\int_{0}^{\pi} f(z) v_1(z)\,dz &= \alpha\cdot 0 + \beta \int_{0}^{\pi} u(z)v_1(z)\,dz \nonumber\\
\int_{0}^{\pi} f(z) u(z)\,dz &= \alpha\int_{0}^{\pi} v_1(z)u(z)\,dz + \beta\int_{0}^{\pi} u^2(z)\,dz
\end{align}
give us a triangular two-by-two system (indeed with constant diagonal) to solve for the unknown coefficients.

\begin{figure}
  \centering
  \subfigure[coalesced eigenfunctions]{\label{fig:GMgeny}\includegraphics[width=60mm]{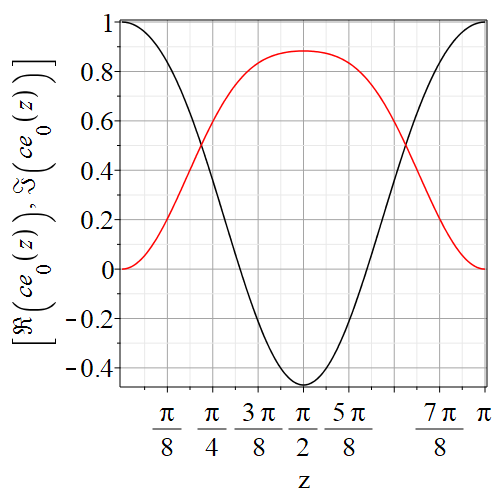}}
  \subfigure[generalized eigenfunction]{\label{fig:GMgenu}\includegraphics[width=60mm]{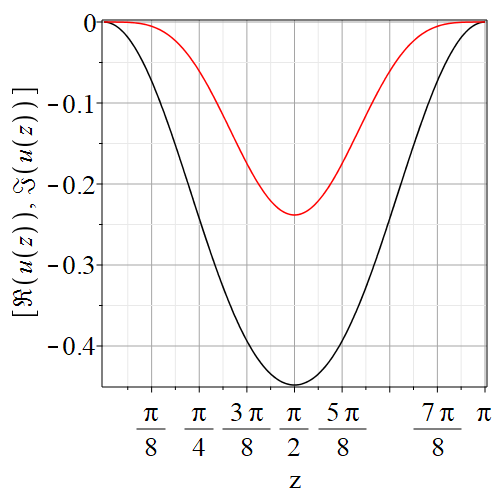}}
  \newline
  \subfigure[coalesced eigenfunctions]{\label{fig:GMgenys}\includegraphics[width=60mm]{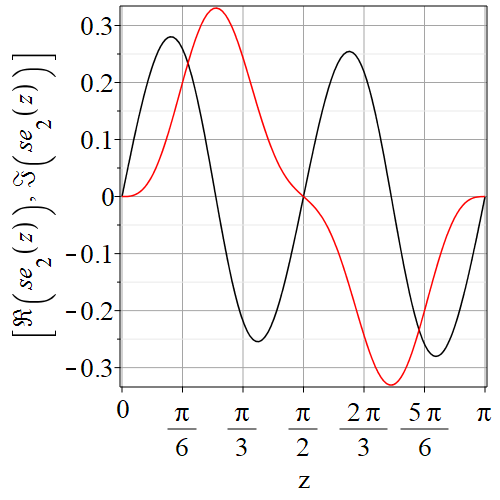}}
  \subfigure[generalized eigenfunction]{\label{fig:GMgenus}\includegraphics[width=60mm]{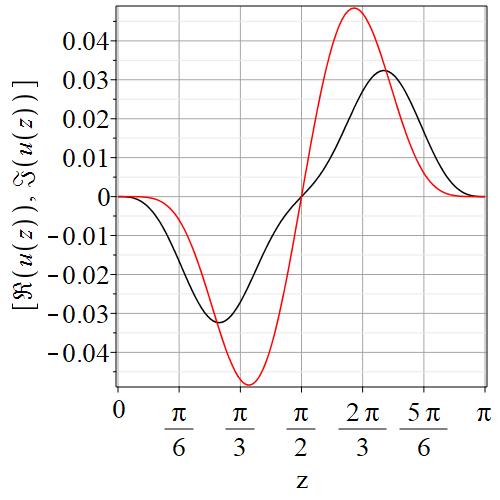}}
  \caption{Top row left: Real and imaginary parts of the coalesced eigenfunctions $v_1(z)=\ce_0(q,z) = \ce_2(q,z)$ corresponding to the Mulholland-Goldstein double point $q \approx 1.4688\,i$ (real part in black, imaginary part in red). On the right, we have the corresponding generalized eigenfunction obtained by solving $y'' + (a-2q\cos2z)y + v_1=0$.
  Bottom row left: Real and imaginary parts of the coalesced eigenfunctions $v_1(z)=\se_2(q,z) = \se_4(q,z)$ corresponding to the next-largest pure imaginary double point $q = 6.92895\ldots\,i$  with eigenvalue approximately $11.1905$. On the right, we have the corresponding generalized eigenfunction obtained by solving $y'' + (a-2q\cos2z)y + v_1=0$.}\label{fig:GMgen}
\end{figure}

For example, consider $q = 1.468768613785142\,i$, the Mulholland-Goldstein double point again, and its associated eigenvalue $a = 2.08869890274970$ (computed this time by averaging the computed eigenvalues of the $N$ by $N$ matrix, where we took $N=25$: its split eigenvalues were $2.088698902749696\textcolor{red}{{}^{72}_{27}} \pm 8.31667446021810974\cdot 10^{-8}\,i$), where we have displayed the distinct real digits in red.  We averaged the corresponding eigenvectors to get $\mat{v}^* = [c_1, c_2, \ldots, c_{N}]$, and put
\[
v_1(z) = \frac{c_1}{\sqrt{2}} + \sum_{k=2}^{20} c_k\cos(2(k-1)z)\>.
\]
The $\sqrt{2}$ is needed because the symmetrizing trick for the matrix of equation~\eqref{eq:tridceeven} gives an extra $\sqrt{2}$ in the $0$th coefficient.  We then enforced $v_1(0)=1$ by scaling. We index from $1$ in the above equation because that is usual for matrices.  The real and imaginary parts of this are plotted in figure~\ref{fig:GMgeny}; $v_1(z)$ is, of course, an approximation for $\ce_0(q,z) = \ce_2(q,z)$, the coalesced eigenfunction.  By examining its residual $v_1'' + (a-2q\cos2z)v_1$ we see that it is accurate to $10^{-14}$ (plot not shown). We verified that $v_1(z)$ has numerically zero norm, as well:
\[
\int_0^\pi v_1^2(z)\,dz \approx -9.19\cdot 10^{-16} - 3.62\cdot 10^{-15}\,i\>.
\]

We then use the SVD to solve the singular system $(a^*\mat{I}-\mat{A})\mat{u} = -\mat{v}^*$, and form
\begin{equation}\label{eq:exampleu}
u(z) = \frac{u_1}{\sqrt{2}} + \sum_{k=2}^{n} u_k\cos(2(k-1)z)\>.
\end{equation}
We then removed a multiple of $v_1(z)$ so that $u(0) = 0$.
The real and imaginary parts of this generalized eigenfunction are plotted in figure~\ref{fig:GMgenu}.

\begin{figure}
  \centering
  \subfigure[Decay of coefficients]{\label{fig:GMexamplecoefs}\includegraphics[width=60mm]{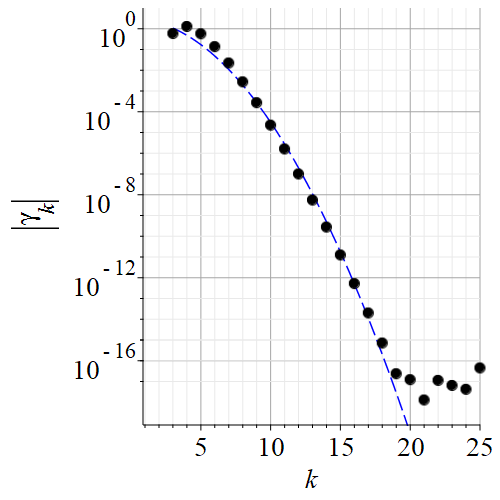}}
  \subfigure[Approximation error]{\label{fig:GMexampleerror}\includegraphics[width=60mm]{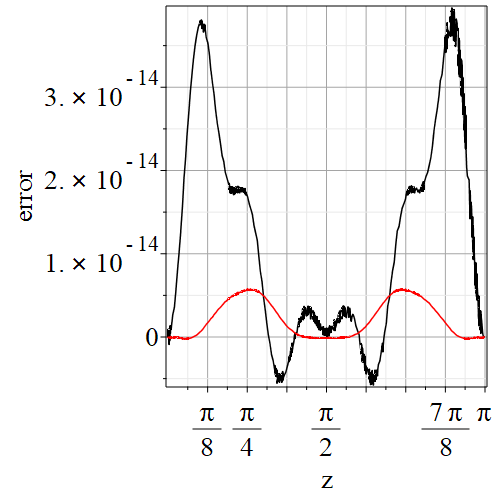}}
  \caption{(Left) size of coefficients in the expansion of $f(z) = \exp(\cos2z)\cos6z$ in terms of Mathieu functions at the Mulholland-Goldstein double point $q = 1.468768\ldots\,i$; the coefficients seem to decay like  (blue dashed line) $\exp(1.11-0.115k^2)$. (Right) the difference $f(z) - \alpha v_1(z) - \beta u(z) - \sum_{k=3}^{20}\gamma_k v_k(z)$. Here $\alpha \approx 0.057266- 0.015745\,i$ and $\beta \approx 0.19855- 0.042167\,i$. (black for real part, red for imaginary part). }\label{fig:GM}
\end{figure}

We now use these eigenfunctions to expand a smooth function of period $\pi$.  We took as an example
\begin{equation}\label{eq:examplefunc}
  f(z) = e^{\cos 2z}\cos6z \approx \alpha v_1(z) + \beta u(z) + \sum_{k=3}^{20} \gamma_k v_k(z) \>.
\end{equation}
As described above, we computed the coefficients of eigenfunctions $v_k(z)$ for $k=3$, $4$, $\ldots$, $20$ by orthogonality.  The final five eigenvalues and eigenvectors were not needed.
Then because
\[
\int_{0}^{\pi} f(z) v_1(z)\,dz = \alpha\cdot 0 + \beta \int_{0}^{\pi} u(z)v_1(z)\,dz
\]
we may identify $\beta \approx 0.3152 + 0.1086\,i$ .
Now because
\[
\int_{0}^{\pi} f(z) u(z)\,dz = \alpha\int_{0}^{\pi} v_1(z)u(z)\,dz + \beta\int_{0}^{\pi} u^2(z)\,dz
\]
where both of the integrals on the right are nonzero and we now know $\beta$, this gives
$\alpha \approx 0.1536 - 0.08560\,i$.
As you can see from the graph of the magnitudes of the computed $\gamma_k$ for $3 \le k \le 25$ in figure~\ref{fig:GMexamplecoefs}, these are appreciable.  Notice also the exponential decay of the coefficients; this is typical for a spectral expansion of a smooth function.
We plot the error $f(z) - \alpha v_1(z) - \beta u(z) - \sum_{k\ge 3} \gamma_k v_k(z)$ in figure~\ref{fig:GMexampleerror} where we see that it is less than $5\cdot10^{-14}$.

The \emph{third} way that we were thinking that one could compute these eigenfunctions seems a little more complicated: we solve the initial-value problem sweeping forward for $v_1(z)$ using the Hermite-Obreschkoff method described earlier, which chooses the mesh; we record the local Taylor series for the two local functions satisfying $y(z_j) = 1$, $y'(z_j) = 0$ and $y(z_j)=0$, $y'(z_j)=1$.  We then solve the boundary value problem for $u(z)$ on that interval by imposing periodic boundary conditions and using collocation at two points in each interval.  If there were $M$ subintervals, this gives an almost block diagonal matrix\footnote{Each collocation point will give an equation involving four unknowns, the $\alpha$s and $\beta$s of the endpoints of the interval containing the collocation point.} of $2M$ equations in the $2M$ unknowns, namely the coefficients $\alpha_k$ and $\beta_k$ of the linear combination of the two solutions at each interior node, together with $\alpha_0 = \alpha_M$ and $\beta_0 = \beta_M$.  This sounds involved, but in fact it is straightforward.
We suspect that few people will actually implement this method, however; we haven't, yet, either.

A fourth way, similar but perhaps even simpler, which we actually did do, is to use the Green's function from equation~\eqref{eq:GreensFunction}; we already have methods for integrating strings of blends and for multiplying strings of blends, so all this requires is the ability to transfer $w_I$ and $w_{II}$ onto the same string of blends. We found it simplest just to compute them at the same time.
The general solution of equation~\eqref{eq:pertsoldu} is (suppressing the dependence of $w_I(z;a,q)$ on $a$ and~$q$ and similarly $w_{II}$ for brevity)
\[
u(z) = \alpha w_I(z) + \beta w_{II}(z) + w_{II}(z)\int_{0}^{z} w_I(\zeta)y^*(\zeta)\,d\zeta  - w_{I}(z)\int_{0}^{z} w_{II}(\zeta)y^*(\zeta)\,d\zeta \>.
\]
If we are solving for a generalized eigenfunction for $\ce_{2g}(q^*,z)$ then $y^* = \ce_{2g}(q^*,z) = w_I(z; q^*, a^*)$ and this is already periodic so $\alpha = 0$ and therefore $u(0)=0$; notice that $w_{II}(0)=0$; moreover the integral to $\pi$ for $w_I(\zeta)y^*(\zeta)$ is also zero, so that at $z=\pi$
\[
0 = u(\pi) = \beta w_{II}(\pi) - w_{I}(\pi) \int_{0}^{\pi} w_{II}(\zeta)y^*(\zeta)\,d\zeta \>.
\]
Since $w_{I}(\pi)=0$ because it, being the eigenfunction in question in this example, is periodic, we see that $\beta=0$ as well.  Thus all we need to identify the eigenfunction is the integral against the Green's function.
We tried this, and it confirmed the results in the top row of figure~\ref{fig:GMgen}. The bottom row in that figure was computed this way.

\section{Concluding remarks\label{sec:concl}}
We have presented a historical survey of the computation of the Mathieu functions, which are defined as the period $\pi$ and period $2\pi$ solutions of the Mathieu differential equation~\eqref{eq:mathieueq}. Our original motivation for this undertaking was to solve a problem of pulsatile blood flow in a vessel of elliptic cross-section. To actually do that we used the spectral method, but with the brute force of multiple precision to power through the double-eigenvalue difficulty.  That was inelegant, so we investigated the alternatives.

To carry out this investigation, we implemented our own procedures (in Maple) for the computation of Mathieu functions (and generalized eigenfunctions) in order to explore some of the difficulties involved.  Our code is at present, like we imagine Blanch's to have been, ``artesanal'' and was intended only for use with careful human supervision.  For experimentation, of course, this is a feature, not a bug.  The task of constructing fully general, bulletproof code for the computation of Mathieu functions is one that calls for dedicated effort and analysis.  We hope that this present paper encourages a team to undertake the task.  We know of no such code in existence currently.

In 1957 G.~Temple called Mathieu functions ``indispensable but intractable instruments of mathematical physics''. The word ``intractable'' has a technical meaning nowadays, and certainly computing Mathieu function is not intractable in that sense.  But they are somewhat involved, and the major remaining question is how best to compute the sometimes-needed generalized eigenfunctions.
Of course, we must not forget about the advances in direct numerical solution of the underlying PDE: the methods of~\cite{gander2019class} may be preferable in many applications over expansion in Mathieu functions.

Mathieu's work on the nodal lines of an elliptic drum occurred a little after Ernst Chladni (1756--1827) demonstrated the existence of nodal lines on freely vibrating plates, apparently following work of Hooke; for a lovely discussion of these see~\cite{gander2012chladni}.

Perhaps the most interesting results of our readings of the literature to us were, first, the discovery that Mathieu had anticipated Lindstedt's anti-secularity perturbation method by over a decade\footnote{In~\cite{vanderPol:1943:stability} we find that in the Astronomy literature some people called the Mathieu equation the ``Lindstedt equation'', so the ideas of Mathieu and of Lindstedt may be more connected than we know.}; and that Blanch's algorithm for computing eigenvalues could be implemented (easily!) in series, thereby allowing one to compute series for eigenvalues which would allow greater surety in continuation methods (Newton iteration starting with estimates from nearby~$q$ is usually used) or to compute Puiseux series about double points.  We believe that the computation of these Puiseux series is new to this paper.  The notion of generalized eigenfunctions for the Mathieu equation is not new to this paper (it is in~\cite{meixner}, as previously stated) but we believe that the details of their computation are presented for the first time here.

\medskip\par\noindent
\section*{Acknowledgments}
We were grateful for advice from John May, which sped up our Maple implementation of Mathieu's anti-secular perturbation method.  We also thank Erik Postma and J\"urgen Gerhard for many comments on the
computation of Mathieu functions in Maple, and Marcus Webb at Manchester for discussions on completeness. Martin Gander provided comments on a draft which allowed us to improve the paper.  He also reminded us of \textsl{Chladni figures}.  We are also grateful for the time taken by the referees of this paper, during the pandemic; refereeing is pretty thankless at the best of times, but in 2020/21 it was even worse.  Their comments helped us to improve the paper substantially. Tom Cuchta helped to `beautify' table~\ref{tab:PuiseuxSeries}, and Owen Maresh introduced us for the purpose of said beautification; thank you both!
The Department of Applied Mathematics provided useful support for this project, as did the Rotman Institute of Philosophy at Western.
RMC thanks the Isaac Newton Institute for Mathematical
Sciences and the staff of both the University Library and the Betty and Gordon Moore Library at Cambridge for support and hospitality during the programme
Complex Analysis: Tools, techniques, and applications,
when some of the work on this project was undertaken.

\bibliographystyle{siamplain}

\begin{thebibliography}{10}

\bibitem{abramowitz}
{\sc M.~Abramowitz and I.~A. Stegun}, {\em Handbook of mathematical functions:
  with formulas, graphs, and mathematical tables}, vol.~55, Courier
  Corporation, 1964.

\bibitem{aitken1941dr}
{\sc A.~Aitken}, {\em Dr. {EL} {I}nce}, Nature, 148 (1941), pp.~309--310.

\bibitem{alhargan1996complete}
{\sc F.~A. Alhargan}, {\em A complete method for the computations of {Mathieu}
  characteristic numbers of integer orders}, SIAM review, 38 (1996),
  pp.~239--255.

\bibitem{arbenz2004jacobi}
{\sc P.~Arbenz and M.~E. Hochstenbach}, {\em A {Jacobi--Davidson} method for
  solving complex symmetric eigenvalue problems}, SIAM J. on Sci. Comp., 25
  (2004), pp.~1655--1673.

\bibitem{GarciaBarroso2016}
{\sc E.~R.~G. Barroso, P.~D.~G. P{\'{e}}rez, and P.~Popescu-Pampu}, {\em
  Variations on inversion theorems for {N}ewton{\textendash}{P}uiseux series},
  Mathematische Annalen, 368 (2016), pp.~1359--1397,
  \url{https://doi.org/10.1007/s00208-016-1503-1}.

\bibitem{barton1980taylor}
{\sc D.~Barton}, {\em On {Taylor} series and stiff equations}, ACM Transactions
  on Mathematical Software (TOMS), 6 (1980), pp.~280--294.

\bibitem{Battles2004}
{\sc Z.~Battles and L.~N. Trefethen}, {\em An extension of {MATLAB} to
  continuous functions and operators}, {SIAM} Journal on Scientific Computing,
  25 (2004), pp.~1743--1770, \url{https://doi.org/10.1137/s1064827503430126}.

\bibitem{benoit2017rigorous}
{\sc A.~Benoit, M.~Jolde{\c{s}}, and M.~Mezzarobba}, {\em Rigorous uniform
  approximation of {D}-finite functions using {Chebyshev} expansions}, Math.
  Comp., 86 (2017), pp.~1303--1341.

\bibitem{bickley1945tabulation}
{\sc W.~Bickley}, {\em The tabulation of {M}athieu functions}, Mathematical
  Tables and Other Aids to Computation, 1 (1945), pp.~409--419.

\bibitem{Blanch:1946:Computation}
{\sc G.~Blanch}, {\em On the computation of {M}athieu functions}, Journal of
  Mathematics and Physics, 25 (1946), pp.~1--20,
  \url{https://doi.org/10.1002/sapm19462511}.

\bibitem{blanch1960asymptotic}
{\sc G.~Blanch}, {\em The asymptotic expansions for the odd periodic {M}athieu
  functions}, Transactions of the American Mathematical Society, 97 (1960),
  pp.~357--366.

\bibitem{blanch1964numerical}
{\sc G.~Blanch}, {\em Numerical evaluation of continued fractions}, Siam
  Review, 6 (1964), pp.~383--421.

\bibitem{blanch1966numerical}
{\sc G.~Blanch}, {\em Numerical aspects of {Mathieu} eigenvalues}, Rendiconti
  del Circolo Matematico di Palermo, 15 (1966), pp.~51--97.

\bibitem{blanch1969double}
{\sc G.~Blanch and D.~Clemm}, {\em The double points of {Mathieu}’s
  differential equation}, Mathematics of Computation, 23 (1969), pp.~97--108.

\bibitem{blanch1941internal}
{\sc G.~Blanch, A.~Lowan, R.~Marshak, and H.~Bethe}, {\em The internal
  temperature-density distribution of the sun.}, The Astrophysical Journal, 94
  (1941), p.~37.

\bibitem{bolmont:2015:Mathieu}
{\sc E.~Bolmont, P.~Nabonnand, and L.~Rollet}, {\em Les ambitions {P}arisiennes
  contrari\'ees d'\'{E}mile {M}athieu (1835--1890)}, 2015,
  \url{https://images.math.cnrs.fr/Les-ambitions-parisiennes-contrariees-d-Emile-Mathieu-1835-1890.html}.

\bibitem{Bourget1866}
{\sc J.~Bourget}, {\em M\'emoire sur le mouvement vibratoire des membranes
  circulaires}, Annales scientifiques de l'\'Ecole Normale Sup\'erieure, 1e
  s{\'e}rie, 3 (1866), pp.~55--95, \url{https://doi.org/10.24033/asens.19}.

\bibitem{bouwkamp1948note}
{\sc C.~J. Bouwkamp}, {\em A note on {Mathieu} functions}, Proc. Kon.
  Nederland. Akad. Wetensch. v51,  (1948), pp.~891--893.

\bibitem{Cano2020}
{\sc J.~Cano, S.~Falkensteiner, and J.~R. Sendra}, {\em Algebraic, rational and
  puiseux series solutions of systems of autonomous algebraic {ODEs} of
  dimension one}, Mathematics in Computer Science,  (2020),
  \url{https://doi.org/10.1007/s11786-020-00478-w}.

\bibitem{chaos2002mathieu}
{\sc L.~Chaos-Cador and E.~Ley-Koo}, {\em Mathieu functions revisited: matrix
  evaluation and generating functions}, Revista mexicana de f{\'\i}sica, 48
  (2002), pp.~67--75.

\bibitem{corless2020inverse}
{\sc R.~M. Corless}, {\em Inverse cubic iteration}, arXiv preprint
  arXiv:2007.06571,  (2020).

\bibitem{corless2020pure}
{\sc R.~M. Corless}, {\em Pure tone modes for a 5:3 elliptic drum}, 2020,
  \url{https://arxiv.org/abs/2008.06936}.

\bibitem{corless2019backward}
{\sc R.~M. Corless and N.~Fillion}, {\em Backward error analysis for
  perturbation methods}, in Algorithms and complexity in mathematics,
  epistemology, and science, Springer, 2019, pp.~35--79,
  \url{https://doi.org/10.1007/978-1-4939-9051-1_3}.

\bibitem{corless2020blends}
{\sc R.~M. Corless and E.~Postma}, {\em Blends in {Maple}}, arXiv preprint
  arXiv:2007.05041,  (2020).

\bibitem{sevyeri2018runge}
{\sc R.~M. Corless and L.~Rafiee~Sevyeri}, {\em The {Runge} example for
  interpolation and {Wilkinson}’s examples for rootfinding}, SIAM Review, 62
  (2020), pp.~231--243.

\bibitem{dougall1915solution}
{\sc J.~Dougall}, {\em The solution of {M}athieu's differential equation},
  Proceedings of the Edinburgh Mathematical Society, 34 (1915), pp.~176--196.

\bibitem{duan1991approximate}
{\sc B.~Duan and M.~Zamir}, {\em Approximate solution for pulsatile flow in
  tubes of slightly noncircular cross-sections}, Utilitas Mathematica, 40
  (1991), pp.~13--26.

\bibitem{duhem1892emile}
{\sc P.~Duhem}, {\em {\'E}mile {M}athieu, his life and works}, Bulletin of the
  American Mathematical Society, 1 (1892), pp.~156--168.

\bibitem{erdelyi1942certain}
{\sc A.~Erd{\'e}lyi}, {\em On certain expansions of the solutions of
  {Mathieu}'s differential equation}, Mathematical Proceedings of the Cambridge
  Philosophical Society, 38 (1942), pp.~28--33.

\bibitem{bateman1953higher}
{\sc A.~e.~a. Erd{\'e}lyi}, {\em Higher Transcendental Functions}, McGraw-Hill,
  1953.

\bibitem{erricolo2003acceleration}
{\sc D.~Erricolo}, {\em Acceleration of the convergence of series containing
  {M}athieu functions using {S}hanks transformation}, IEEE Antennas and
  Wireless Propagation Letters, 2 (2003), pp.~58--61.

\bibitem{erricolo}
{\sc D.~Erricolo and G.~Carluccio}, {\em Algorithm 934: Fortran 90 subroutines
  to compute {M}athieu functions for complex values of the parameter}, ACM
  Transactions on Mathematical Software (TOMS), 40 (2013), p.~8.

\bibitem{fairgrieve1991ok}
{\sc T.~F. Fairgrieve and A.~D. Jepson}, {\em {OK Floquet} multipliers}, SIAM
  journal on numerical analysis, 28 (1991), pp.~1446--1462.

\bibitem{frenkel2001algebraic}
{\sc D.~Frenkel and R.~Portugal}, {\em Algebraic methods to compute {Mathieu}
  functions}, Journal of Physics A: Mathematical and General, 34 (2001),
  p.~3541.

\bibitem{gander2012chladni}
{\sc M.~J. Gander and F.~Kwok}, {\em Chladni figures and the {T}acoma bridge:
  motivating {PDE} eigenvalue problems via vibrating plates}, SIAM Review, 54
  (2012), pp.~573--596.

\bibitem{gander2019class}
{\sc M.~J. Gander and H.~Zhang}, {\em A class of iterative solvers for the
  {H}elmholtz equation: factorizations, sweeping preconditioners, source
  transfer, single layer potentials, polarized traces, and optimized schwarz
  methods}, SIAM Review, 61 (2019), pp.~3--76.

\bibitem{geddes1992algorithms}
{\sc K.~O. Geddes, S.~R. Czapor, and G.~Labahn}, {\em Algorithms for computer
  algebra}, Springer, 1992.

\bibitem{Gil2007}
{\sc A.~Gil, J.~Segura, and N.~M. Temme}, {\em Numerical Methods for Special
  Functions}, Society for Industrial and Applied Mathematics, Jan. 2007,
  \url{https://doi.org/10.1137/1.9780898717822}.

\bibitem{goldstein_1933}
{\sc S.~Goldstein}, {\em Lam\'esche, {Mathieusche} und verwandte {Funktionen}
  in {Physik} und {Technik}. by {M.J.O. Strutt}.}, The Mathematical Gazette, 17
  (1933), p.~59–60, \url{https://doi.org/10.2307/3607965}.

\bibitem{Golubitsky1985}
{\sc M.~Golubitsky and D.~G. Schaeffer}, {\em Singularities and Groups in
  Bifurcation Theory}, Springer New York, 1985,
  \url{https://doi.org/10.1007/978-1-4612-5034-0}.

\bibitem{grier1997gertrude}
{\sc D.~A. Grier}, {\em Gertrude {B}lanch of the mathematical tables project},
  IEEE Annals of the History of Computing, 19 (1997), pp.~18--27.

\bibitem{gustafsson1988api}
{\sc K.~Gustafsson, M.~Lundh, and G.~S{\"o}derlind}, {\em A {PI} stepsize
  control for the numerical solution of ordinary differential equations}, BIT
  Numer Math, 28 (1988), pp.~270--287.

\bibitem{gutierrez2003mathieu}
{\sc J.~C. Guti{\'e}rrez-Vega, R.~Rodr{\i}guez-Dagnino, M.~Meneses-Nava, and
  S.~Ch{\'a}vez-Cerda}, {\em Mathieu functions, a visual approach}, American
  Journal of Physics, 71 (2003), pp.~233--242.

\bibitem{heine1878handbuch}
{\sc E.~Heine}, {\em Handbuch der Kugelfunktionen I, II}, Berlin: Reimer, 1878.

\bibitem{hunter1981eigenvalues}
{\sc C.~Hunter and B.~Guerrieri}, {\em The eigenvalues of {Mathieu's} equation
  and their branch points}, Studies in Applied Mathematics, 64 (1981),
  pp.~113--141.

\bibitem{ilie2008adaptivity}
{\sc S.~Ilie, G.~S{\"o}derlind, and R.~M. Corless}, {\em Adaptivity and
  computational complexity in the numerical solution of odes}, Journal of
  Complexity, 24 (2008), pp.~341--361.

\bibitem{ince1922proof}
{\sc E.~Ince}, {\em A proof of the impossibility of the coexistence of two
  {M}athieu functions}, Proc. Camb. Phil. Soc, 21 (1922), pp.~117--120.

\bibitem{ince1926second}
{\sc E.~Ince}, {\em The second solution of the {Mathieu} equation},
  Mathematical Proceedings of the Cambridge Philosophical Society, 23 (1926),
  pp.~47--49.

\bibitem{ince1927mathieu}
{\sc E.~Ince}, {\em The {Mathieu} equation with numerically large parameters},
  Journal of the London Mathematical Society, 1 (1927), pp.~46--50.

\bibitem{ince1933xxii}
{\sc E.~Ince}, {\em Tables of the elliptic-cylinder functions}, Proceedings of
  the Royal Society of Edinburgh, 52 (1933), pp.~355--423.

\bibitem{Iserles2019}
{\sc A.~Iserles and M.~Webb}, {\em Orthogonal systems with a skew-symmetric
  differentiation matrix}, Foundations of Computational Mathematics, 19 (2019),
  pp.~1191--1221, \url{https://doi.org/10.1007/s10208-019-09435-x}.

\bibitem{kirlinger1991implicit}
{\sc G.~Kirlinger and G.~F. Corliss}, {\em On implicit {T}aylor series methods
  for stiff {ODEs}}, tech. report, Argonne National Lab., IL (United States),
  1991.

\bibitem{Knuth1992}
{\sc D.~E. Knuth}, {\em Two notes on notation}, The American Mathematical
  Monthly, 99 (1992), pp.~403--422,
  \url{https://doi.org/10.1080/00029890.1992.11995869}.

\bibitem{korner1989fourier}
{\sc T.~W. K{\"o}rner}, {\em Fourier analysis}, Cambridge University Press,
  1989.

\bibitem{levi1988stability}
{\sc M.~Levi}, {\em Stability of the inverted pendulum—a topological
  explanation}, SIAM review, 30 (1988), pp.~639--644.

\bibitem{Lindstedt1882}
{\sc A.~Lindstedt}, {\em Bemerkungen zur {I}ntegration einer gewissen
  {D}ifferentialgleichung}, Astronomische Nachrichten, 103 (1882),
  pp.~257--268, \url{https://doi.org/10.1002/asna.18821031702}.

\bibitem{lubkin1943stability}
{\sc S.~Lubkin and J.~Stoker}, {\em Stability of columns and strings under
  periodically varying forces}, Quarterly of Applied Mathematics, 1 (1943),
  pp.~215--236.

\bibitem{mathieu1868memoire}
{\sc {\'E}.~Mathieu}, {\em M{\'e}moire sur le mouvement vibratoire d'une
  membrane de forme elliptique.}, Journal de math{\'e}matiques pures et
  appliqu{\'e}es, 13 (1868), pp.~137--203.

\bibitem{moir2021memoir}
{\sc {\'E}.~Mathieu}, {\em Memoir on the vibratory movement of an elliptical
  membrane}, 2021, \url{https://arxiv.org/abs/2103.02730}.
\newblock translated by Robert H.~C.~Moir.

\bibitem{meixner}
{\sc J.~Meixner, F.~W. Sch{\"a}fke, and G.~Wolf}, {\em Mathieu functions},
  Springer, 1980.

\bibitem{mezzarobba2010numgfun}
{\sc M.~Mezzarobba}, {\em {NumGfun}: a package for numerical and analytic
  computation with {D}-finite functions}, in Proceedings of the 2010
  International Symposium on Symbolic and Algebraic Computation, 2010,
  pp.~139--145.

\bibitem{mezzarobba2012note}
{\sc M.~Mezzarobba}, {\em A note on the space complexity of fast {D}-finite
  function evaluation}, in Int. Workshop on Computer Algebra in Scientific
  Computing, Springer, 2012, pp.~212--223.

\bibitem{miyazaki2004computation}
{\sc Y.~Miyazaki, N.~Asai, Y.~Kikuchi, D.~Cai, and Y.~Ikebe}, {\em Computation
  of multiple eigenvalues of infinite tridiagonal matrices}, Mathematics of
  computation, 73 (2004), pp.~719--730.

\bibitem{morse1954methods}
{\sc P.~M. Morse and H.~Feshbach}, {\em Methods of theoretical physics},
  American Journal of Physics, 22 (1954), pp.~410--413.

\bibitem{mulholland1929xc}
{\sc H.~Mulholland and S.~Goldstein}, {\em The characteristic numbers of the
  {M}athieu equation with purely imaginary parameter}, The London, Edinburgh,
  and Dublin Philosophical Magazine and Journal of Science, 8 (1929),
  pp.~834--840.

\bibitem{naylor1984simplified}
{\sc D.~Naylor}, {\em On simplified asymptotic formulas for a class of
  {M}athieu functions}, SIAM journal on mathematical analysis, 15 (1984),
  pp.~1205--1213.

\bibitem{nedialkov2005solving}
{\sc N.~S. Nedialkov and J.~D. Pryce}, {\em Solving differential-algebraic
  equations by {Taylor} series (i): Computing {Taylor} coefficients}, BIT
  Numerical Mathematics, 45 (2005), pp.~561--591.

\bibitem{onsager1935solutions}
{\sc L.~Onsager}, {\em Solutions of the {Mathieu} Equation of Period 4 Pi and
  Certain Related Functions}, PhD thesis, Yale University, 1935.

\bibitem{parlett1974rayleigh}
{\sc B.~N. Parlett}, {\em The {R}ayleigh quotient iteration and some
  generalizations for nonnormal matrices}, Mathematics of Computation, 28
  (1974), pp.~679--693.

\bibitem{Prudnikov:1990:IS}
{\sc A.~P. Prudnikov, {\relax{Yu}}.~A. Brychkov, and O.~I. Marichev}, {\em
  Integrals and Series: More Special Functions, {V}ol. 3}, Gordon and Breach
  Science Publishers, New York, 1990.
\newblock Translated from the Russian by G. G. Gould.

\bibitem{pryce1993numerical}
{\sc J.~D. Pryce}, {\em Numerical solution of {Sturm-Liouville} problems},
  Oxford University Press, 1993.

\bibitem{Richardson2011}
{\sc M.~Richardson and L.~N. Trefethen}, {\em A sinc function analogue of
  {C}hebfun}, {SIAM} Journal on Scientific Computing, 33 (2011),
  pp.~2519--2535, \url{https://doi.org/10.1137/110825947}.

\bibitem{rubin1964anecdote}
{\sc H.~Rubin}, {\em Anecdote on power series expansions of {M}athieu
  functions}, Journal of Mathematics and Physics, 43 (1964), pp.~339--341.

\bibitem{salvy1994gfun}
{\sc B.~Salvy and P.~Zimmermann}, {\em Gfun: a {Maple} package for the
  manipulation of generating and holonomic functions in one variable}, ACM
  Transactions on Mathematical Software (TOMS), 20 (1994), pp.~163--177.

\bibitem{Schneider:1999:modified}
{\sc M.~{Schneider} and J.~{Marquardt}}, {\em Fast computation of modified
  {M}athieu functions applied to elliptical waveguide problems}, IEEE
  Transactions on Microwave Theory and Techniques, 47 (1999), pp.~513--516.

\bibitem{Sexton2012}
{\sc A.~P. Sexton}, {\em Abramowitz and {S}tegun {\textendash} a resource for
  mathematical document analysis}, in Lecture Notes in Computer Science,
  Springer Berlin Heidelberg, 2012, pp.~159--168,
  \url{https://doi.org/10.1007/978-3-642-31374-5_11}.

\bibitem{shen2009spectral}
{\sc J.~Shen and L.-L. Wang}, {\em On spectral approximations in elliptical
  geometries using {Mathieu} functions}, Mathematics of Computation, 78 (2009),
  pp.~815--844.

\bibitem{soderlind2015stiffness}
{\sc G.~S{\"o}derlind, L.~Jay, and M.~Calvo}, {\em Stiffness 1952--2012: Sixty
  years in search of a definition}, BIT Numerical Mathematics, 55 (2015),
  pp.~531--558.

\bibitem{strutt1932lamesche}
{\sc M.~J. Strutt}, {\em Lam{\'e}sche, {M}athieusche und verwandte Funktionen
  in Physik und Technik}, vol.~3, Springer-Verlag, 1932.

\bibitem{temple1956edmund}
{\sc G.~F.~J. Temple}, {\em Edmund {Taylor} {Whittaker}, 1873-1956}, 1956.

\bibitem{teschl2012ordinary}
{\sc G.~Teschl}, {\em Ordinary differential equations and dynamical systems},
  vol.~140, American Mathematical Soc., 2012.

\bibitem{Townsend2015}
{\sc A.~Townsend and S.~Olver}, {\em The automatic solution of partial
  differential equations using a global spectral method}, Journal of
  Computational Physics, 299 (2015), pp.~106--123,
  \url{https://doi.org/10.1016/j.jcp.2015.06.031}.

\bibitem{tropp:1973:Interview}
{\sc H.~Tropp}, {\em Interview with {Gertrude Blanch}}, 1973,
  \url{https://sova.si.edu//details/NMAH.AC.0196#ref86}.

\bibitem{trott2007mathematica}
{\sc M.~Trott}, {\em The {M}athematica guidebook for symbolics}, Springer
  Science, 2007.

\bibitem{van2007accurate}
{\sc A.~Van~Buren and J.~Boisvert}, {\em Accurate calculation of the modified
  {Mathieu} functions of integer order}, Quarterly of applied mathematics, 65
  (2007), pp.~1--23.

\bibitem{van2001fast}
{\sc J.~van~der Hoeven}, {\em Fast evaluation of holonomic functions near and
  in regular singularities}, Journal of Symbolic Computation, 31 (2001),
  pp.~717--744.

\bibitem{vanderPol:1943:stability}
{\sc B.~van~der Pol and M.~J.~O. Strutt}, {\em On the stability of the
  solutions of {Mathieu's} equation}, The London, Edinburgh, and Dublin
  Philosophical Magazine and Journal of Science, 5 (1928), pp.~18--38,
  \url{https://doi.org/10.1080/14786440108564441}.

\bibitem{whittaker1912functions}
{\sc E.~T. Whittaker}, {\em On the functions associated with the elliptic
  cylinder in harmonic analysis}, in Proc. Fifth International Congress of
  Mathematicians, vol.~1, 1912, pp.~366--371.

\bibitem{whittaker1941edward}
{\sc E.~T. Whittaker}, {\em Edward {L}indsay {I}nce: 1891—1941}, Journal of
  the London Mathematical Society, 1 (1941), pp.~139--144.

\bibitem{whittaker1927course}
{\sc E.~T. Whittaker and G.~N. Watson}, {\em A course of modern analysis}, CUP,
  1927.

\bibitem{wilf1962}
{\sc H.~S. Wilf}, {\em Mathematics for the Physical Sciences}, Dover, 1962.

\bibitem{zamir}
{\sc M.~Zamir}, {\em Hemo-dynamics}, Springer, 2016.

\bibitem{ziener}
{\sc C.~Ziener, M.~R{\"u}ckl, T.~Kampf, W.~Bauer, and H.~Schlemmer}, {\em
  Mathieu functions for purely imaginary parameters}, Journal of Computational
  and Applied Mathematics, 236 (2012), pp.~4513--4524.

\end{thebibliography}

\bigskip
\bigskip\goodbreak
\appendix
\section{Pulsatile flow in tube of elliptic cross-section\label{app:pulsatiledetails}}
The notation in this appendix differs slightly from that of the main paper, and instead agrees with that of our other study (currently in progress).

The equation governing the oscillatory component of the flow takes the general form~\cite{zamir}

\begin{equation}
\label{eq:basic}
    \frac{\partial u}{\partial t}
    + \frac{1}{\rho} \frac{\partial p}{\partial z} = \frac{\mu}{\rho} \left( \frac{\partial^{2} u}{\partial x^{2}} + \frac{\partial^{2} u}{\partial y^{2}} \right)\>,
\end{equation}
where $x,y$ are rectangular coordinates within the cross section of the tube, $z$ is along the axis of the tube and $p$ is pressure. It is at this point that the geometry of the cross sectional boundary of the tube dictates the choice of coordinate system in which the governing equation is to be solved.

{\textbf{In the case of a tube of circular cross section}}, the governing equation \eqref{eq:basic} takes the form
\begin{equation}
\frac{\partial u_{c}}{\partial t} + \frac{1}{\rho} \frac{\partial p}{\partial z} = \frac{\mu}{\rho} \left( \frac{\partial^{2} u_{c}}{\partial r^{2}} + \frac{1}{r} \frac{\partial u_{c}}{\partial r} \right)\>,
\end{equation}
where subscript `$c$' is being used to associate the results with a tube of circular cross section.

For an oscillatory pressure gradient of the form
\begin{equation}
    \frac{\partial p}{\partial z} = k_{0}e^{i\omega t}
\end{equation}


and by separation of variables
\begin{equation}
    u_{c} \left( r, t \right) = U_{c} \left( r \right) e^{i\omega t}
\end{equation}
the equation becomes
\begin{equation} \label{eq:bessel}
    \frac{d^{2} U_{c}}{d r^{2}} + \frac{1}{r} \frac{d U_{c}}{d r} - \frac{i \Lambda_{c}}{a^{2}} U_{c} = \frac{k_{0}}{\mu}\>,
\end{equation}
where
\begin{equation} \label{eq:nonDimensionCircle}
    \Lambda_{c} = \frac{\rho \omega a^{2}}{\mu}
\end{equation}
is a nondimensional frequency parameter and $a$ is the radius of the tube.

Equation~\eqref{eq:bessel} is a form of a Bessel equation with general solution
\begin{equation}
    U_{c} (r) = \frac{i k_{0} a^{2}}{\mu \Lambda_{c}} + AJ_{0}\left( \zeta \right) + B Y_{0} \left( \zeta \right)
\end{equation}
where $A$ and $B$ are arbitrary constants and $J_{0}$ and $Y_{0}$ are Bessel functions of order zero and of the first and second kind, respectively, satisfying the standard Bessel equations
\begin{align}
    \frac{d^{2} J_{0}}{d \zeta^{2}} + \frac{1}{\zeta} \frac{d J_{0}}{d \zeta} + J_{0} &= 0 \\
     \frac{d^{2} Y_{0}}{d \zeta^{2}} + \frac{1}{\zeta} \frac{d Y_{0}}{d \zeta} + Y_{0} &= 0 \>.
\end{align}
The new variable $\zeta$ is related to the radial coordinates by
\begin{equation}
    \zeta \left( r \right) = \Omega \frac{r}{a}\>,
\end{equation}
where $\Omega$ is a frequency parameter related to the nondimensional frequency parameter
\begin{equation}
    \Omega = \left( \frac{i - 1}{\sqrt{2}} \right) \sqrt{\Lambda_{c}}\>.
\end{equation}

%

{\textbf{In the case of a tube of elliptic cross section}} the boundary conditions suggest a transformation to elliptic coordinates
\begin{equation}
    x = d \cosh \xi \cos \eta, \ y = d \sinh \xi \sin \eta
\end{equation}
where $2d$ is the focal distance and $\xi$, $\eta$ are the elliptic coordinates, and
equation~\eqref{eq:basic} becomes
\begin{equation} \label{eq:afterLaplacian}
    \frac{\partial u_{e}}{\partial t} + \frac{1}{\rho} \frac{\partial p}{\partial z} = \frac{\mu}{\rho} \frac{2}{d^{2} \left( \cosh 2 \xi - \cos 2 \eta \right)} \left( \frac{\partial^{2} u_{e}}{\partial \xi^{2}} + \frac{\partial^{2} u_{e}}{\partial \eta^{2}} \right)\>,
\end{equation}
where the subscript `$e$' is now used to associate the results with a tube of elliptic cross section.

For an oscillatory pressure gradient of the form
\begin{equation}
    \frac{\partial p}{\partial z} = k_{0} e^{i\omega t}
\end{equation}
we use separation of variables
\begin{equation}
    u_{e}\left( \xi, \eta, t \right) = w \left( \xi, \eta \right)e^{i\omega t}
\end{equation}
so that equation~\eqref{eq:afterLaplacian} can be formulated as an inhomogeneous Helmoltz equation
\begin{equation} \label{eq:inhomo}
    \frac{2}{d^{2} \left( \cosh 2 \xi - \cos 2 \eta \right)} \left( \frac{\partial^{2} u_{e}}{\partial \xi^{2}} + \frac{\partial^{2} u_{e}}{\partial \eta^{2}} \right) - \frac{i \rho \omega}{\mu} w = \frac{k_{0}}{\mu}\>.
\end{equation}
Using the translation
\begin{equation}
    w \left( \xi, \eta \right) = v \left( \xi, \eta \right) - \frac{k_{0}}{i \rho \omega}\>,
\end{equation}
the inhomogeneous term of equation~\eqref{eq:inhomo} becomes
\begin{equation} \label{eq:homo}
    \left( \frac{\partial^{2} v}{\partial \xi^{2}} + \frac{\partial^{2} v}{\partial \eta^{2}} \right) - \frac{i}{2} \Lambda_{e} \left( \cosh 2\xi - \cos 2 \eta \right) v = 0
\end{equation}
where
\begin{equation}
    \Lambda_{e} = \frac{\rho \omega d^{2}}{\mu}
\end{equation}
is the elliptic equivalent of the nondimensional frequency parameter.

Applying separation of variables using
\begin{equation}
    v \left( \xi, \eta \right) = f \left( \xi \right) g \left( \eta \right)\>,
\end{equation}
equation~\eqref{eq:homo} yields two separate Mathieu equations
\begin{align}
    \frac{d^{2} g}{d \eta^{2}} + \left( s + 2 q \cos 2 \eta \right) g &= 0 \label{eq:ordinary} \\
    \frac{d^{2} f}{d \xi^{2}} - \left( s + 2 q \cosh 2 \xi \right) f &= 0 \label{eq:modified}
\end{align}
where $s$ is a separating constant and
\begin{equation}
    q = \frac{i \Lambda_{e}}{4}\>.
\end{equation}

The foregoing analysis demonstrates clearly that the boundary and boundary conditions play a key role in the transition from Bessel equations to Mathieu equations in the case of pulsatile flow in a tube. However, while this explains the mathematical aspects and origin of the transition from Bessel to Mathieu equations in this particular application, it does not provide a hint as to the corresponding origin of this transition in the context of the physics of the flow. One of the aims of our study was to understand the link between the mathematical and the physical aspects and origin of this transformation.

\begin{figure}
\centering
\includegraphics[width=10cm]{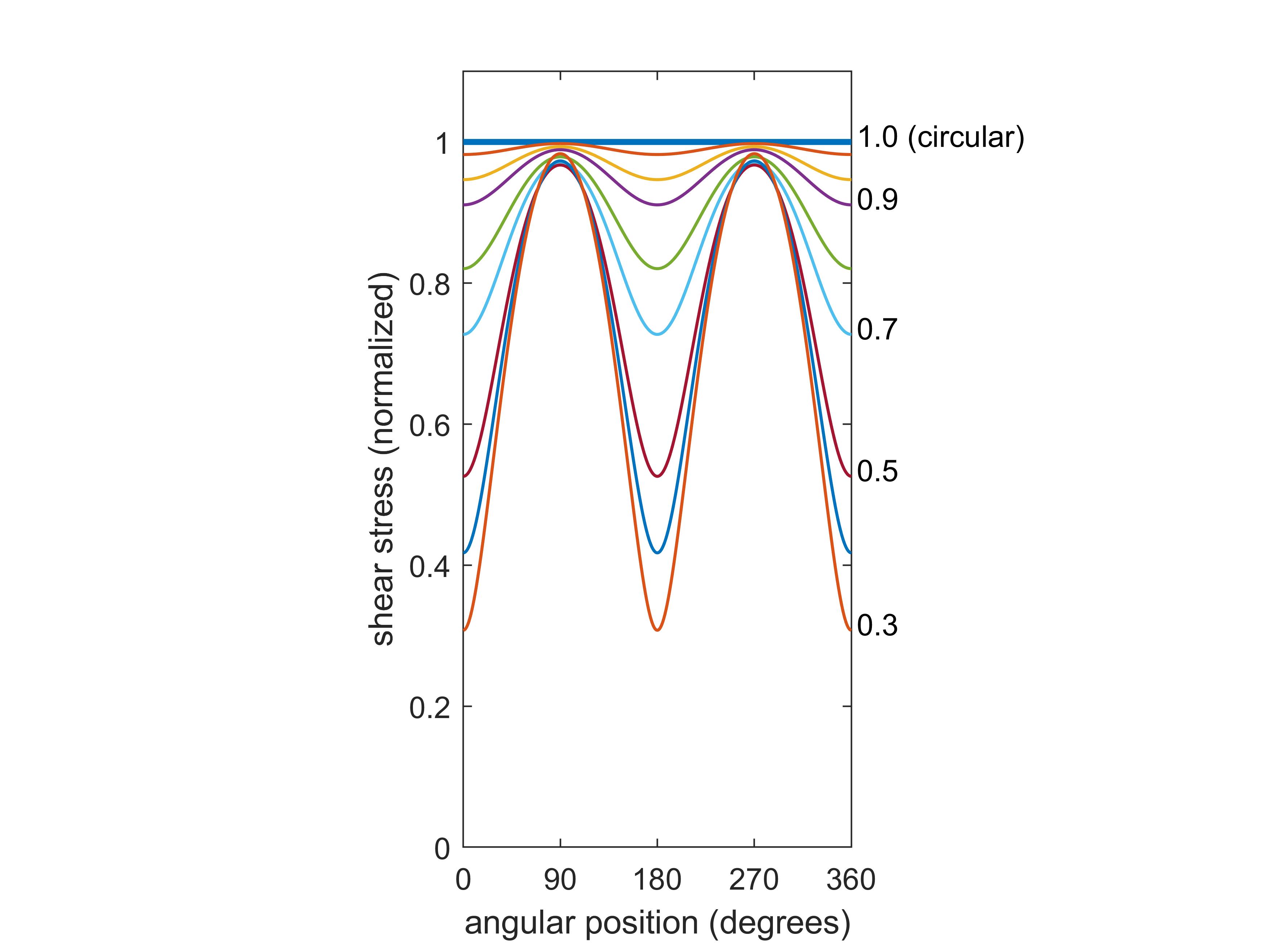}
\caption{Shear stress at the boundary of a tube in pulsatile flow. The curves represent a range of tubes with different cross sections ranging from circular to increasingly elliptic. The numbers on the right identify the tubes in terms of the ratio of their minor to major axes of their cross section. The y-axis on the left of the figure represents that ratio of the shear stress normalized in terms of the shear stress in a tube of circular cross section.}
\label{fig:ShearStress}
\end{figure}

While, at first, properties of the flow in a slightly elliptic tube appear as a small perturbation of the flow in a circular tube (and can indeed be analyzed in this manner~\cite{duan1991approximate}), a closer look at the shear stress on the boundary of the tube shows a more significant change that occurs as the cross section of the tube changes from circular to elliptic as shown in Figure~\ref{fig:ShearStress}. The figure shows that no matter how small the change from circular to elliptic cross section is, the two vertices\footnote{A \emph{vertex} of an ellipse is an endpoint of the major axis; a \emph{co-vertex} is an endpoint of the minor axis.} of the ellipse lead to periodicity in the distribution of shear stress on the boundary. This periodicity is the origin of the transition from Bessel to Mathieu equations in the description of the flow.

\section{Orthogonality of Periodic Mathieu Functions\label{app:orthogonality}}
What follows is a derivation of the orthogonality relation~\eqref{eq:orthogonality}, according to the usual Sturm--Liouville theory. This exercise is straightforward material from a typical undergraduate curriculum, but is included here because the indefiniteness of the ``norm'' supplied by the bilinear form~\ref{eq:orthogonality} may be surprising to the reader.  We reproduce~\eqref{eq:orthogonality} here for convenience:
\begin{align}
    \left\langle y_k, y_\ell \right\rangle := \int_0^{2\pi} y_k(x) y_\ell(x)\,dx = \mathrm{const}\cdot[\mathcal{C}\ {\bf\mathrm{and}}\  k=\ell]\>.
    \label{eq:orthogonalitycopy}
\end{align}
We start as usual with the supposition that we have two separate eigenfunctions, $y_m(x)$ and $y_n(x)$, of Mathieu's equation, corresponding to different eigenvalues $\lambda_m$ and $\lambda_n$:
\begin{align}\label{eq:mathieubilin}
  y_m''(x) - q\cos(2x)y_m(x) &= \lambda_m y_m(x) \nonumber\\
  y_n''(x) - q\cos(2x)y_n(x) &= \lambda_n y_n(x)\>.
\end{align}
Here either of the eigenvalues $\lambda_m$ and $\lambda_n$ can be an $a_k(q)$ or $b_k(q)$, so long as they are different from each other.  Multiply the first equation by $y_n(x)$ and the second by $y_m(x)$ and subtract to get
\begin{equation}\label{eq:lostlinear}
  y_n(x)y_m''(x) - y_m(x)y_n''(x) = (\lambda_m - \lambda_n)y_m(x)y_n(x)\>.
\end{equation}
Now integrate over the period, and use integration by parts and periodicity:
\begin{align}\label{eq:intparts}
(\lambda_m-\lambda_n)\int_{0}^{2\pi} y_m(x)y_n(x)\,dx &= \int_0^{2\pi}y_n(x)y_m''(x) - y_m(x)y_n''(x)\,dx \nonumber\\
   &= \left.\left(y_n(x)y_m'(x)-y_m(x)y_n'(x)\right)\right\Vert_0^{2\pi}\nonumber\\
   &= 0\>.
\end{align}
Since the eigenvalues are distinct, this ensures that the bilinear form is zero.

If we instead tried to use the genuine inner product $\langle u,v\rangle := \int_{0}^{2\pi } u(x)\overline{v}(x)\,dx$ and norm $\|u\|_2 = \langle u,u \rangle^{1/2}$, instead of the bilinear form and indefinite ``norm,'' we fail.  The Mathieu functions, when $q$ is not real, are not orthogonal with respect to this inner product,
as can be verified by a straightforward computation with some non-real $q$, say $q=1.0i$. Computing the eigenvalues $a_2(1.0i)$ and $a_4(1.0i)$, which indeed have distinctly different numerical values, and numerically computing the integral
\[
\int_{0}^{2\pi} \overline{\ce}_4(x,1.0i)\ce_2(x,1.0i)\,dx\>,
\]
we get a complex number of magnitude about $0.5138$.  That inner product is not zero.  In contrast, we find that the bilinear form (without the complex conjugate) is zero to numerical accuracy, as it should be.

Where does the proof fail, when $q$ is not real?  
In trying the method of the proof, we would first have to conjugate one of the equations to start with; say the second one.
\begin{align}\label{eq:mathieuorth}
  y_m''(x) - q\cos(2x)y_m(x) &= \lambda_m y_m(x) \nonumber\\
  \overline{y}_n''(x) - \overline{q}\cos(2x)\overline{y_n(x)} &= \overline{\lambda}_n \overline{y}_n(x)\>.
\end{align}
Since we will integrate from $x=0$ to $x=2\pi$ again, $x$ can be taken as real.  Now multiply the first equation by $\overline{y}_n(x)$ and the second by $y_m(x)$ and subtract: but now, the linear term does not cancel---unless, of course, $q$ is real:
\begin{equation}\label{eq:keptlinear}
  \overline{y}_n(x)y_m''(x) - y_m(x)\overline{y}_n''(x) + \cos2x \left(\overline{q}-q\right)y_m(x)\overline{y}_n(x)  = (\lambda_m - \overline{\lambda}_n)y_m(x)\overline{y}_n(x)\>.
\end{equation}
Without that cancellation, the integrals of the next step tell us nothing useful.  And indeed as we saw above by direct computation for a specific non-real $q$, the naively desired result---namely, orthogonality under a true inner product instead of a bilinear form---isn't true, and so no possible proof could exist.

\section{Confocal Ellipses verses Fixed Aspect Ratio Ellipses and Bessel functions as a limit of Mathieu functions\label{app:Bessel2Mathieu}}

\begin{figure}
  \centering
  \subfigure[Confocal ellipses and a circle]{\label{fig:confocal}\includegraphics[width=60mm]{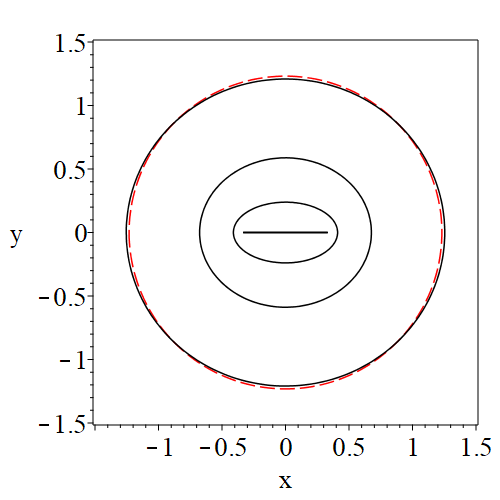}}
  \subfigure[Fixed aspect ellipses and a circle]{\label{fig:fixedar}\includegraphics[width=60mm]{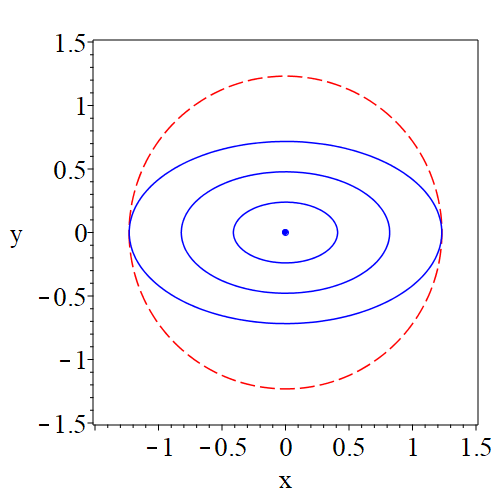}}
  \caption{A comparison of confocal ellipses (left) with fixed aspect ratio ellipses (right). The confocal ellipses have foci at $\pm 1/3$.  Their parametric equations are $x = c\cosh(\beta)\cos(\alpha)$, $y=c\sinh(\beta)\sin(\alpha)$.  The four ellipses shown have $\beta = [0, 2/3, 4/3, 2]$. The circle (red dashed line) was chosen to have radius $r = c\exp(2)/2$, which as the mean of the semi-major and semi-minor axes of the largest ellipse, corresponds well.  We see that the largest confocal ellipse shown here is appreciably circular.  Larger $\beta$ would generate even more circular-looking ellipses.  In contrast, the fixed aspect-ratio ellipses all have the same aspect ratio as the smallest nonsingular confocal ellipse shown on the left, namely $a/b = \tanh(2/3) \approx 0.58$. }\label{fig:confocalvsfixedar}
\end{figure}

To help people visualize the difference between a family of \emph{confocal} ellipses, all of which have the same set of foci, and a family of \emph{fixed aspect ratio} ellipses, which is likely what most people imagine when they think of a family of ellipses, we made figure~\ref{fig:confocalvsfixedar}.  The point is to demonstrate how quickly confocal ellipses become circular-looking.

Algebraically, the aspect ratio of a confocal ellipse is $a/b = \tanh(\beta)$ where the confocal parameterization is $x = c\cosh(\beta)\cos(\alpha)$, $y=c\sinh(\beta)\sin(\alpha)$.  Asymptotically, $a/b \approx 1 - 2\exp(-2\beta)$ so we see that a confocal family becomes circular exponentially quickly.

\begin{figure}
  \centering
\subfigure[$\ce_0(q,i\eta)$ (black solid line) for $q=1/10$ and unscaled $\Re(I_M(\varepsilon\sin(i\eta))$ (red dashed line) ]{\label{fig:MathieuAs}\includegraphics[width=60mm]{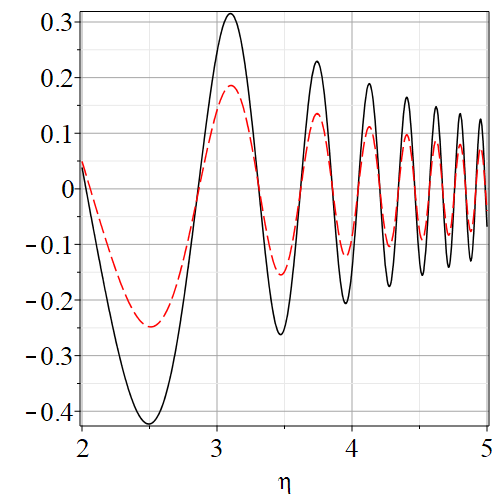}}
\subfigure[Difference $\ce_0 - (0.58698)\Re(I_M)$]{\label{fig:BesselAs}\includegraphics[width=60mm]{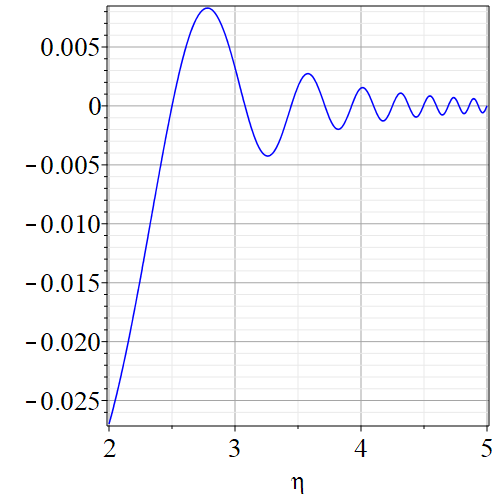}}
  \caption{In the limit as $q \to 0$ and $\eta \to \infty$ keeping $\xi = 2\sqrt{q}\sinh\eta$ constant, a Mathieu function $\ce_0(q,z)$ rapidly approaches a Bessel function $\Re I_M(2\sqrt{q}\sin(z))$. In this figure, $q=1/10$ and $M\approx 0.4527i$.  Here $z=i\eta$ is purely imaginary. We have ignored the $O(\varepsilon^2)$ term of the outer solution. Even so, the match seems good. }\label{fig:Mathieu2Bessel}
\end{figure}

This helps in understanding the limiting case in which the solutions of the modified Mathieu equation approach Bessel functions.  We modify the treatment slightly in~\cite{whittaker1927course} in what follows.  We start from the Mathieu equation but replace $q = \varepsilon^2/4$ (they use $-\varepsilon^2/4$) to get
\begin{equation}\label{eq:modifiedMathieuWW}
  \frac{d^2y}{dz^2} + \left(a - \tfrac12 \varepsilon^2 \cos(2z) \right)y = 0\>,
\end{equation}
so $\varepsilon = \sqrt{4q} = 2 h$.  We change variables with $\xi = \varepsilon\sin(z)$, so that with $M^2 = a - \varepsilon^2/2$ we get, using $\cos2z = 1 - 2\sin^2 z = 1 - 2\xi^2/\varepsilon^2$,
\begin{equation}\label{eq:AlgebraicDEBesselLike}
  \xi^2 \frac{d^2y}{d\xi^2} + \xi\frac{dy}{d\xi} - (\xi^2+M^2)y(\xi) = -\varepsilon^2 \frac{d^2y}{d\xi^2}\>.
\end{equation}
If $\varepsilon$ is small, we recognize this as a small (admittedly singular) perturbation of a form of Bessel's equation, so the outer solution, away from $\xi = 0$, will be
\[
y(\xi) = C_1 I_M(\xi) + C_2 K_M(\xi) + O(\varepsilon^2)\>,
\]
where $I$ and $K$ denote Bessel functions. The next term, which is $O(\varepsilon^2)$, in the perturbation expansion is somewhat complicated, though small in size when we computed it, but in fact we don't  need it.  The result is plotted in figure~\ref{fig:Mathieu2Bessel} and compared with the relevant Mathieu function.
Now, as $\varepsilon \to 0$, for $\xi$ to remain $O(1)$ we must have $\sin(z) \to \infty$; that is, this matching will only be valid for large imaginary $z$.  This is the double limit mentioned earlier.  Owing to the exponential growth of $\sin$, however, it won't have to be \emph{that} large.  We already saw in figure~\ref{fig:confocal} that with $c=1/3$ the confocal ellipse is pretty circular already for $z = 2$; and indeed already by $q=1/10$ and for $\ce_0(q,i\eta)$ (that is, we use $a_0 = -0.00495\ldots$) we see a marked resemblance to $\Re(I_M(\xi))$ in figure~\ref{fig:Mathieu2Bessel}.  Indeed if we choose a point ($\eta = 2.5$) and choose a normalization factor $C = 0.58698$ so that the asymptotic approximation agrees at that point then the error $\ce_0(q,i\eta) - C\Re(I_M(\varepsilon\sin(i\eta)))$ gets very small very quickly as shown in figure~\ref{fig:BesselAs}.  Notice that here $\varepsilon\approx 0.632$, which isn't even all that small.  The various complex numbers in this example have shuffled the usual behaviour amongst the Bessel functions, and we have simply taken a real part for comparison; but the real part of the solution to a real linear differential equation, in this case the modified Mathieu equation, is also a solution.

\section{Comparing Mathieu's perturbation series to Maple's\label{app:MathieuPerturbForm}}
In order to compare the~$q$-series produced by Maple's series command with the
hand computation of Mathieu, we have to ensure that the same normalization is enforced. We also had to check our work carefully to ensure that the Maple results were correct.  Since Mathieu's normalization was to make the coefficient of $\cos g\alpha$ equal to unity, and Maple's normalization is to make $\int_0^p \ce_g^2(\alpha)\,d\alpha = \pi$ except when $g=0$ when it is $2\pi$, we compute Maple's series and then divide by the coefficient of $\cos g\alpha$. Similarly for the $\se_g(\alpha)$ functions.  By independently computing residuals, we assure ourselves of the correctness of those computations.  The Maple code for this appendix can be found at \href{https://github.com/rcorless/MathieuPerturbationMethod}{Rob Corless's GitHub Repository}.

For $\ce_g(q,t)$ with symbolic $g$ we get, on dividing by the coefficient of
$\cos g t$, the following terms in the series. The coefficient of~$q$ is, provided $g > 1$,
\begin{equation}
{\frac {\cos \left( g-2 \right)\alpha }{4\,(g-1)}}-{\frac {
\cos \left( g+2 \right)\alpha  }{4\,(g+1)}}\>.
\end{equation}
This agrees perfectly with what Mathieu had.
The coefficient of $q^2$ is, provided $g > 2$,
\begin{equation}
{\frac {\cos \left( (g-4)\,\alpha \right) }{ 32\,\left( g-2\right)
   \left( g-1 \right) }}+{\frac {\cos \left( (g+4)\,
\alpha \right) }{ 32\,\left( g+2\right)  \left( g+1 \right) }}
\end{equation}
Again this agrees perfectly with what Mathieu had.
The coefficient of $q^3$ is, provided $g>3$,
\begin{align}
{\frac {\cos \left( g-6\right)\alpha  }{  384\,\left((g-1)
 \right)  \left( g-2 \right)  \left( g-3 \right) }}&+{\frac { \left( {g
}^{2}-4\,g+7 \right) \cos \left( g-2\right)\alpha  }{
 128\,\left( g+1 \right)  \left( g-2 \right)  \left( g-1 \right) ^{3
}}}\nonumber\\
-{\frac { \left( {g}^{2}+4\,g+7 \right) \cos \left( g+2
\right)\alpha  }{ 128\,\left( g+2 \right)  \left( g-1 \right)
 \left( g+1 \right) ^{3}}}&-{\frac {\cos \left( g+6
 \right)\alpha }{ 384\,\left( g+3 \right)  \left( g+2 \right)  \left( g+1
 \right) }}
\end{align}
Again this is in agreement with Mathieu, although he wrote $128$ as $2^7$ and $384$ as $2^7\cdot3$. It is at this term that the difference between Mathieu's form of the series and a modern series expansion of $\ce_g(q,\alpha)$ is most noticeable, because his work eliminated the $\cos g\alpha$ term at this order (and altered the other coefficients).
The coefficient of $q^4$ is, provided $g>4$,
\begin{align}
{\frac {\cos \left( g-8\right)\alpha  }{ 6144\,\left( g-1
 \right)  \left( g-2 \right)  \left( g-3 \right)  \left( g-4 \right) }
}&+{\frac { \left( {g}^{2}-5\,g+10 \right) \cos \left( g-4\right)
\alpha  }{ 768\,\left( g-2 \right)  \left( g-3 \right)
 \left( g+1 \right)  \left( g-1 \right) ^{3}}}\nonumber\\
 +{\frac { \left( {g}^{2}
+5\,g+10 \right) \cos \left( g+4\right)\alpha  }{ 768
\,\left( g-1 \right)  \left( g+3 \right)  \left( g+2 \right)  \left( g+1
 \right) ^{3}}}
 &+{\frac {\cos \left( g+8\right)\alpha  }{
 6144\,\left( g+4 \right)  \left( g+3 \right)  \left( g+2 \right)
 \left( g+1 \right) }}\>.
\end{align}
Mathieu wrote $6144$ as $2^{11}\cdot 3$ and $768$ as $2^8\cdot 3$; he also wrote ${g}^{3}+7\,{g}^{2}+20\,g+20$ as the numerator for the $\cos(g+4)\alpha$ term (and similar for the $\cos(g-4)\alpha$ and had an extra factor $g+2$ in the denominator.  Since ${g}^{3}+7\,{g}^{2}+20\,g+20 = \left( g+2 \right)  \left( {g}^{2}+5\,g+10 \right) $, we see that his result was again correct though not in simplest form.
The coefficient of $q^5$ is (and instead of editing it for elegance, this time we leave it as automatically generated by Maple), provided $g>5$,
\begin{align}
&{\frac {\cos \left( \alpha\,g-10\,\alpha \right) }{ \left( 122880\,g-
614400 \right)  \left( g-1 \right)  \left( g-2 \right)  \left( g-3
 \right)  \left( g-4 \right) }}\nonumber\\
&+{\frac { \left( {g}^{2}-6\,g+13
 \right) \cos \left( \alpha\,g-6\,\alpha \right) }{ \left( 8192\,g-
16384 \right)  \left( g-3 \right)  \left( g-4 \right)  \left( g+1
 \right)  \left( g-1 \right) ^{3}}}\nonumber\\
&+{\frac { \left( {g}^{6}-5\,{g}^{5}
+8\,{g}^{4}-8\,{g}^{3}+47\,{g}^{2}+13\,g+232 \right) \cos \left(
\alpha\,g-2\,\alpha \right) }{ \left( 3072\,g-6144 \right)  \left( g-3
 \right)  \left( g+2 \right)  \left( g+1 \right) ^{3} \left( g-1
 \right) ^{5}}}\nonumber\\
&-{\frac { \left( {g}^{6}+5\,{g}^{5}+8\,{g}^{4}+8\,{g}^{
3}+47\,{g}^{2}-13\,g+232 \right) \cos \left( \alpha\,g+2\,\alpha
 \right) }{ \left( 3072\,g-6144 \right)  \left( g+3 \right)  \left( g+
2 \right)  \left( g-1 \right) ^{3} \left( g+1 \right) ^{5}}}\nonumber\\
&-{\frac {
 \left( {g}^{2}+6\,g+13 \right) \cos \left( \alpha\,g+6\,\alpha
 \right) }{ \left( 8192\,g-8192 \right)  \left( g+4 \right)  \left( g+
3 \right)  \left( g+2 \right)  \left( g+1 \right) ^{3}}}\nonumber\\
&-{\frac {\cos
 \left( \alpha\,g+10\,\alpha \right) }{ \left( 122880\,g+614400
 \right)  \left( g+4 \right)  \left( g+3 \right)  \left( g+2 \right)
 \left( g+1 \right) }}
\end{align}
Mathieu wrote $122880$ as $2^{11}\cdot 3\cdot 4\cdot 5$, and similarly other large numbers in factored form.  He had ${g}^{4}+11\,{g}^{3}+49\,{g}^{2}+101\,g+78= \left( g+3 \right)  \left(
g+2 \right)  \left( {g}^{2}+6\,g+13 \right)
$ in the numerator of the $\cos(g+6)\alpha$ term and an extra $(g+3)(g+2)$ in the denominator.  Similarly for the $\cos(g-6)\alpha$ term.  For the $\cos(g-4)\alpha$ term he had
${g}^{7}+7\,{g}^{6}+18\,{g}^{5}+24\,{g}^{4}+63\,{g}^{3}+81\,{g}^{2}+206
\,g+464
$ which is $g+2$ times the numerator printed above; of course he had an extra factor $g+2$ in the denominator to cancel it.

Finally, he has an extra factor $(g^2-4)$ in the denominator of the $q^6$ ($h^{12}$) term of the \emph{eigenvalue} and an apparently incorrect numerator, $9\,{g}^{5}+22\,{g}^{4}-203\,{g}^{2}-116$.  But if you replace the $5$th power with a $6$th (and, really, reading the PDF of this manuscript, it's hard to tell whether it should be a $6$ anyway), this factors into the correct form:
$9\,{g}^{6}+22\,{g}^{4}-203\,{g}^{2}-116 = (g^2-4)(9\,{g}^{4}+58\,{g}^{2}+29)$.  Actually, on the line above Mathieu's final form for $R$, the power is more clearly a $6$: this isn't even a typo, just something hard to read given the printing and transcribing process.

The end of the story is that Mathieu was able to give the correct generic perturbation series to $\ce_g(h^2,\alpha)/F$ up to and including terms of order $q^5$ ($h^{10}$), on the understanding that the factor $F$ was chosen to make all coefficients of $\cos g\alpha$ equal to zero apart from the first one.

\par\noindent
\textbf{Special series}.
The generic series is good only for ``large enough" $g$.  For specific small $g$, and indeed for any fixed $g$ if one wants to compute enough terms, special computations have to be done.  We here give the results that Mathieu attempted, and comment on some minor errors in his paper.

He began with $g=2$.  We get (again removing the coefficients of $\cos2\alpha$ as he did)
\begin{align}
    \frac{\ce_2(q,\alpha)}{F} &= \cos \left( 2\,\alpha \right) + \left( {\frac{1}{4}}-{\frac {\cos
 \left( 4\,\alpha \right) }{12}} \right) q+{\frac {\cos \left( 6\,
\alpha \right) }{384}}{q}^{2}\nonumber\\
&+ \left( -{\frac{5}{192}}-{\frac {43\,
\cos \left( 4\,\alpha \right) }{13824}}-{\frac {\cos \left( 8\,\alpha
 \right) }{23040}} \right) {q}^{3}\nonumber\\
 &+ \left( {\frac {293\,\cos \left( 6
\,\alpha \right) }{2211840}}+{\frac {\cos \left( 10\,\alpha \right) }{
2211840}} \right) {q}^{4}\nonumber\\
&+ \left( {\frac{1363}{221184}}+{\frac {21041
\,\cos \left( 4\,\alpha \right) }{79626240}}-{\frac {167\,\cos \left(
8\,\alpha \right) }{66355200}}-{\frac {\cos \left( 12\,\alpha \right)
}{309657600}} \right) {q}^{5}\nonumber\\
&+ \left( -{\frac {139453\,\cos \left( 6\,
\alpha \right) }{12740198400}}+{\frac {629\,\cos \left( 10\,\alpha
 \right) }{22295347200}}+{\frac {\cos \left( 14\,\alpha \right) }{
59454259200}} \right) {q}^{6}+O \left( {q}^{7} \right)
\end{align}
where the factor $F$ (not computed by Mathieu, but needed by us to compare a modern series to his results) is
\begin{equation}
    F = 1-{\frac {19\,{q}^{2}}{288}}+{\frac {51191\,{q}^{4}}{2654208}}-{\frac
{88995077\,{q}^{6}}{12740198400}}\>.
\end{equation}
Notice first that Ince was correct: there is a real value of~$q$ (near $q=2.37$ by use of \texttt{fsolve}) for which that factor $F$ is zero,  and therefore this normalization is not universally possible\footnote{Changing the order at which we do the computation can make the zero move (when the order is $O(q^{20})$, the value of $q$ that makes $F$ vanish is about $1.8$), or even vanish ($F$ only has complex roots if we work to $O(q^{13})$); but that does not invalidate Ince's point.}.
Here we also see apparent arithmetic errors: Mathieu has $287$ and not $293$ as we have in the $O(q^4)$ term. He has $21059$ where we have $21041$ in the $O(q^5)$ term, and $41$ instead of $167$. The other two terms at that order are correct. He did not report the $O(q^6)$ term.

For the \emph{eigenvalue} at $g=2$ Mathieu reports the correct expansion,
\begin{equation}
    a = 4+{\frac{5}{12}}{q}^{2}-{\frac{763}{13824}}{q}^{4}+{\frac{1002401}{
79626240}}{q}^{6}+O \left( {q}^{7} \right)\>,
\end{equation}
except he has $1002419$ instead of $1002401$.

For $g=4$, Mathieu reports everything correctly up until the constant ($\cos 0 \alpha$) term of the $O(q^4)$ term: he gets $-1/92160$ where we get $1/34560$, a different sign. In the $O(q^5)$ term we get
\begin{equation}
    -{\frac {53\,\cos \left( 2\,\alpha \right) }{124416000}}-{\frac {4037
\,\cos \left( 6\,\alpha \right) }{2419200000}}-{\frac {53\,\cos
 \left( 10\,\alpha \right) }{1032192000}}-{\frac {\cos \left( 14\,
\alpha \right) }{1857945600}}
\end{equation}
while Mathieu gets
\begin{equation}
-\tfrac{1}{1857945600}\cos14\alpha
-\tfrac{53}{1032192000}\cos10\alpha
-\tfrac{4037}{2419200000}\cos6\alpha
-\tfrac{439}{62208000}\cos2\alpha
\end{equation}
which has evident discrepancies at the $\cos2\alpha$ and $\cos14\alpha$ terms but is otherwise correct.

For the eigenvalue, we get
\begin{equation}
a = 16+{\frac{1}{30}}{q}^{2}+{\frac{433}{864000}}{q}^{4}-{\frac{5701}{
2721600000}}{q}^{6}+O \left( {q}^{7} \right)
\end{equation}
which is nearly the same as Mathieu's,
\begin{equation}
R=16+\tfrac{1}{30}h^4+\tfrac{433}{864000}h^8
-\tfrac{189983}{21772800000}h^{12}+\cdots
\end{equation}
except he erroneously reports ${{189983}/{21772800000}}$ as the coefficient of $q^6$.  This is an irreducible fraction and not equal to the correct coefficient.

For $g=1$
\begin{align}
    \frac{\ce_1(q,\alpha)}{F} &=  \cos \left( \alpha \right) -{\frac {\cos \left( 3\,\alpha \right) }{8
}}q+ \left( -{\frac {\cos \left( 3\,\alpha \right) }{64}}+{\frac {\cos
 \left( 5\,\alpha \right) }{192}} \right) {q}^{2} \nonumber\\
 &+ \left( {\frac {\cos
 \left( 5\,\alpha \right) }{1152}}-{\frac {\cos \left( 3\,\alpha
 \right) }{1536}}-{\frac {\cos \left( 7\,\alpha \right) }{9216}}
 \right) {q}^{3}\nonumber\\
 &+ \left( -{\frac {\cos \left( 7\,\alpha \right) }{
49152}}
+{\frac {11\,\cos \left( 3\,\alpha \right) }{36864}}+{\frac {
\cos \left( 5\,\alpha \right) }{24576}}+{\frac {\cos \left( 9\,\alpha
 \right) }{737280}} \right) {q}^{4}\nonumber\\
 &+ \left( {\frac {\cos \left( 9\,
\alpha \right) }{3686400}}-{\frac {7\,\cos \left( 5\,\alpha \right) }{
393216}}+{\frac {49\,\cos \left( 3\,\alpha \right) }{589824}}-{\frac {
\cos \left( 7\,\alpha \right) }{983040}}-{\frac {\cos \left( 11\,
\alpha \right) }{88473600}} \right) {q}^{5}\nonumber\\
&+ \left( -{\frac {\cos
 \left( 11\,\alpha \right) }{424673280}}+{\frac {17\,\cos \left( 7\,
\alpha \right) }{39321600}}+{\frac {55\,\cos \left( 3\,\alpha \right)
}{9437184}}-{\frac {719\,\cos \left( 5\,\alpha \right) }{141557760}}\right. \nonumber\\
&\qquad\qquad\qquad\qquad+\left.{
\frac {\cos \left( 9\,\alpha \right) }{70778880}}+{\frac {\cos \left(
13\,\alpha \right) }{14863564800}} \right) {q}^{6}+O \left( {q}^{7}
 \right)\>.
\end{align}
The correction factor is
\begin{equation}
    F = 1-{\frac{1}{128}}{q}^{2}-{\frac{1}{512}}{q}^{3}-{\frac{37}{294912}}{q
}^{4}+{\frac{121}{1769472}}{q}^{5}+{\frac{8105}{339738624}}{q}^{6}
+\cdots
\end{equation}
The eigenvalue is
\begin{equation}
    a = 1+q-{\frac{1}{8}}{q}^{2}-{\frac{1}{64}}{q}^{3}-{\frac{1}{1536}}{q}^{4
}+{\frac{11}{36864}}{q}^{5}+{\frac{49}{589824}}{q}^{6}+O \left( {q}^{7
} \right)
\>.
\end{equation}

Mathieu gets
\begin{align}
\frac{\ce_1(h^2,\alpha)}{F} &= \cos\alpha-\tfrac{h^2}{8}\cos3\alpha
+h^4\left(-\tfrac{1}{192}\cos5\alpha-\tfrac{1}{64}\cos3\alpha\right)\nonumber\\
&-h^6\left(\tfrac{1}{9216}\cos7\alpha
-\tfrac{1}{1152}\cos5\alpha+\tfrac{1}{1536}\cos3\alpha\right)\nonumber\\
&+h^8\left(\tfrac{1}{737280}\cos9\alpha
-\tfrac{1}{49152}\cos7\alpha\right.\nonumber\\
&\qquad\quad+\left.\tfrac{1}{24576}\cos5\alpha
+\tfrac{11}{36864}\cos3\alpha\right)+\cdots
\end{align}
and for the eigenvalue gets
\begin{equation}
R=1+h^2-\tfrac{1}{8}h^4-\tfrac{1}{64}h^6-\tfrac{1}{1536}h^8+\tfrac{11}{36864}h^{10}+\cdots\>.
\end{equation}
All terms are in complete agreement with our results.

For $g=3$ we have
\begin{align}
    \frac{\ce_3(q,\alpha)}{F} &= \cos \left( 3\,\alpha \right) + \left( {\frac {\cos \left( \alpha
 \right) }{8}}-{\frac {\cos \left( 5\,\alpha \right) }{16}} \right) q+
 \left( {\frac {\cos \left( \alpha \right) }{64}}+{\frac {\cos \left(
7\,\alpha \right) }{640}} \right) {q}^{2} \nonumber\\
&+ \left( {\frac {\cos \left(
\alpha \right) }{1024}}-{\frac {7\,\cos \left( 5\,\alpha \right) }{
20480}}-{\frac {\cos \left( 9\,\alpha \right) }{46080}} \right) {q}^{3
}\nonumber\\
&+ \left( -{\frac {\cos \left( 5\,\alpha \right) }{16384}}-{\frac {
\cos \left( \alpha \right) }{4096}}+{\frac {17\,\cos \left( 7\,\alpha
 \right) }{1474560}}+{\frac {\cos \left( 11\,\alpha \right) }{5160960}
} \right) {q}^{4}+O \left( {q}^{5} \right)
\end{align}
where
\begin{equation}
    F = 1-{\frac {5\,{q}^{2}}{512}}-{\frac {{q}^{3}}{512}}-{\frac {1621\,{q}^{
4}}{13107200}}+{\frac {9\,{q}^{5}}{131072}}
+ O(q^6)\>.
\end{equation}
The eigenvalue is
\begin{equation}
    a = 9+{\frac{1}{16}}{q}^{2}+{\frac{1}{64}}{q}^{3}+{\frac{13}{20480}}{q}^{
4}-{\frac{5}{16384}}{q}^{5}-{\frac{1961}{23592960}}{q}^{6}+O \left( {q
}^{7} \right)\>.
\end{equation}
Mathieu gets
\begin{align*}
  P_2 = & \cos3\alpha
+h^2\left(-\tfrac{1}{16}\cos5\alpha+\tfrac{1}{12}\cos\alpha\right) \\
   & +h^4\left(\tfrac{1}{640}\cos7\alpha+\tfrac{1}{64}\cos\alpha\right) \\
   & +h^6\left(\tfrac{-1}{46080}\cos9\alpha
-\tfrac{7}{20480}\cos5\alpha
+\tfrac{1}{768}\cos\alpha\right) \\
   & +h^8\left(\tfrac{1}{2^{14}\cdot 3^2\cdot 5\cdot7}\cos11\alpha
-\tfrac{17}{2^{15}\cdot3^2\cdot5}\cos7\alpha\right. \\
   & \qquad\quad-\left.\tfrac{1}{2^{14}}\cos5\alpha
-\tfrac{1}{2^{13}}\cos\alpha\right)+\cdots \>;
\end{align*}
$$
R=9+\tfrac{1}{16}h^4+\tfrac{1}{64}h^6+\tfrac{59}{61440}h^8
-\tfrac{3}{16384}h^{10}+\cdots.
$$
Mathieu's eigenfunction seems already wrong at $q^4$ ($h^8$) in several coefficients: $1/12$ instead of $1/8$, $1/768$ instead of $1/1024$.  Even his eigenvalue has an incorrect $O(q^4)$ coefficient.

Similarly, Mathieu's series for $\se_1(q,\alpha)/F$ is correct in every term, whilst his series for $\se_3(q,\alpha)/F$ is wrong already at $O(q^4)$, as is his series for the eigenvalue.

Even so, Mathieu's algebra goes on for many pages.  That there are so few errors might be gratifying to his shade.
Perhaps what would be even more gratifying is to be recognized at last as one of the pioneers of anti-secularity in perturbation methods.

\vfill
\pagebreak
\section{A Table of Double Eigenvalues and Puiseux Series\label{app:DoubleEigTable}}
\begin{flushright}
``The object of a table is to present in a concise and orderly manner information that could not be presented so clearly in any other way.''\\
 --- United States Government Printing Manual, 1959, Chapter 14, section 14.1.
\end{flushright}
{
\begin{table}[H]
  \centering
  \caption{Leading Puiseux series coefficients from $a = a^* + \sum_{k\ge1} \alpha_k (q-q^*)^{k/2}$, the expansion about the double point $q^*$, given for all the double points computed by Blanch and Clemm.  Both the values of the selected double point parameter $q^*$ and its corresponding eigenvalue $a^*$ are given.  The first column, $m$, indicates the ``type'' of continued fraction according to the scheme of Blanch.  The coefficients presented were computed in $32$ decimal Digits in Maple and verified at that precision, then rounded to the precision shown below. This table is presented
  as an \emph{homage} to all the great table-makers of the past.  A machine-readable version of this table (containing all $32$ decimals for each entry, and series coefficients up to $\alpha_6$) is available at \href{https://github.com/rcorless/Puiseux-series-Mathieu-double-points}{Rob Corless's GitHub page}, together with the Maple Workbook used to generate them.
  }\label{tab:PuiseuxSeries}
{
\ttfamily
\tiny
\setlength{\tabcolsep}{8pt} 
\renewcommand{\arraystretch}{0.9} 
\begin{tabular}{ cccccc }
$m$ & $q^*$ & $a^*$ & $\pm\alpha_1$ & $\alpha_2$ & $\pm\alpha_3$ \\
\strut
0 & 0.000e+00+1.469e+00i & 2.089e+00-0.000e-01i & 1.659e+00+1.659e+00i & 0.000e-01-1.192e-01i & 2.928e-01-2.928e-01i \\
1 & 1.931e+00+3.238e+00i & 6.176e+00+1.232e+00i & 1.924e+00+2.009e+00i & 3.831e-01-1.995e-01i & 1.788e-01-1.824e-01i \\
0 & 5.174e+00+5.104e+00i & 1.280e+01+2.763e+00i & 2.118e+00+2.234e+00i & 5.765e-01-1.849e-01i & 1.423e-01-1.632e-01i \\
1 & 9.688e+00+7.045e+00i & 2.193e+01+4.490e+00i & 2.280e+00+2.410e+00i & 6.915e-01-1.625e-01i & 1.281e-01-1.554e-01i \\
0 & 0.000e-01+1.647e+01i & 2.732e+01+0.000e+00i & 2.902e+00+2.902e+00i & 0.000e-01-4.027e-01i & 5.069e-02-5.069e-02i \\
0 & 1.546e+01+9.043e+00i & 3.354e+01+6.363e+00i & 2.423e+00+2.557e+00i & 7.687e-01-1.427e-01i & 1.213e-01-1.501e-01i \\
1 & 4.852e+00+2.233e+01i & 3.841e+01+2.533e+00i & 3.099e+00+3.146e+00i & 1.773e-01-4.071e-01i & 3.153e-02-5.554e-02i \\
1 & 2.247e+01+1.109e+01i & 4.764e+01+8.351e+00i & 2.551e+00+2.687e+00i & 8.249e-01-1.266e-01i & 1.174e-01-1.457e-01i \\
0 & 1.108e+01+2.834e+01i & 5.203e+01+5.552e+00i & 3.262e+00+3.346e+00i & 3.092e-01-3.940e-01i & 2.361e-02-6.270e-02i \\
0 & 3.074e+01+1.317e+01i & 6.421e+01+1.043e+01i & 2.668e+00+2.805e+00i & 8.680e-01-1.134e-01i & 1.147e-01-1.419e-01i \\
1 & 1.866e+01+3.447e+01i & 6.816e+01+8.962e+00i & 3.405e+00+3.517e+00i & 4.101e-01-3.746e-01i & 2.123e-02-6.892e-02i \\
1 & 4.024e+01+1.530e+01i & 8.326e+01+1.260e+01i & 2.777e+00+2.912e+00i & 9.026e-01-1.027e-01i & 1.128e-01-1.385e-01i \\
0 & 0.000e+00+4.781e+01i & 8.066e+01-0.000e-01i & 3.765e+00+3.765e+00i & 0.000e-01-4.496e-01i & 2.233e-02-2.233e-02i \\
0 & 2.755e+01+4.072e+01i & 8.679e+01+1.270e+01i & 3.533e+00+3.668e+00i & 4.895e-01-3.538e-01i & 2.151e-02-7.368e-02i \\
0 & 5.098e+01+1.745e+01i & 1.048e+02+1.484e+01i & 2.878e+00+3.012e+00i & 9.311e-01-9.371e-02i & 1.112e-01-1.354e-01i \\
1 & 7.754e+00+5.776e+01i & 9.877e+01+3.830e+00i & 3.925e+00+3.961e+00i & 1.141e-01-4.507e-01i & 1.311e-02-2.774e-02i \\
1 & 3.773e+01+4.707e+01i & 1.079e+02+1.672e+01i & 3.651e+00+3.803e+00i & 5.534e-01-3.333e-01i & 2.303e-02-7.711e-02i \\
1 & 6.296e+01+1.962e+01i & 1.288e+02+1.714e+01i & 2.974e+00+3.105e+00i & 9.552e-01-8.618e-02i & 1.098e-01-1.326e-01i \\
0 & 1.691e+01+6.785e+01i & 1.194e+02+8.203e+00i & 4.066e+00+4.133e+00i & 2.087e-01-4.439e-01i & 7.973e-03-3.352e-02i \\
0 & 4.921e+01+5.353e+01i & 1.316e+02+2.099e+01i & 3.761e+00+3.926e+00i & 6.061e-01-3.141e-01i & 2.507e-02-7.950e-02i \\
0 & 7.618e+01+2.183e+01i & 1.553e+02+1.949e+01i & 3.064e+00+3.194e+00i & 9.759e-01-7.976e-02i & 1.086e-01-1.300e-01i \\
1 & 2.744e+01+7.809e+01i & 1.425e+02+1.304e+01i & 4.194e+00+4.288e+00i & 2.882e-01-4.329e-01i & 5.451e-03-3.887e-02i \\
1 & 6.196e+01+6.008e+01i & 1.577e+02+2.548e+01i & 3.864e+00+4.040e+00i & 6.503e-01-2.964e-01i & 2.726e-02-8.111e-02i \\
1 & 9.064e+01+2.405e+01i & 1.842e+02+2.190e+01i & 3.149e+00+3.277e+00i & 9.941e-01-7.423e-02i & 1.075e-01-1.276e-01i \\
0 & 0.000e-01+9.548e+01i & 1.621e+02+0.000e+00i & 4.465e+00+4.465e+00i & 0.000e-01-4.687e-01i & 1.290e-02-1.290e-02i \\
0 & 3.932e+01+8.847e+01i & 1.682e+02+1.829e+01i & 4.311e+00+4.429e+00i & 3.555e-01-4.197e-01i & 4.580e-03-4.353e-02i \\
0 & 7.598e+01+6.671e+01i & 1.863e+02+3.017e+01i & 3.962e+00+4.146e+00i & 6.879e-01-2.801e-01i & 2.942e-02-8.214e-02i \\
0 & 1.063e+02+2.630e+01i & 2.156e+02+2.436e+01i & 3.231e+00+3.357e+00i & 1.010e+00-6.942e-02i & 1.064e-01-1.254e-01i \\
1 & 1.065e+01+1.095e+02i & 1.872e+02+5.128e+00i & 4.603e+00+4.633e+00i & 8.400e-02-4.691e-01i & 7.185e-03-1.703e-02i \\
1 & 5.253e+01+9.897e+01i & 1.964e+02+2.391e+01i & 4.420e+00+4.558e+00i & 4.132e-01-4.056e-01i & 4.740e-03-4.747e-02i \\
1 & 9.127e+01+7.342e+01i & 2.174e+02+3.504e+01i & 4.054e+00+4.246e+00i & 7.203e-01-2.654e-01i & 3.147e-02-8.273e-02i \\
1 & 1.233e+02+2.856e+01i & 2.495e+02+2.686e+01i & 3.309e+00+3.433e+00i & 1.024e+00-6.519e-02i & 1.054e-01-1.234e-01i \\
0 & 2.272e+01+1.237e+02i & 2.149e+02+1.082e+01i & 4.729e+00+4.786e+00i & 1.573e-01-4.651e-01i & 3.469e-03-2.127e-02i \\
0 & 6.706e+01+1.096e+02i & 2.270e+02+2.986e+01i & 4.522e+00+4.678e+00i & 4.632e-01-3.912e-01i & 5.532e-03-5.075e-02i \\
0 & 1.078e+02+8.020e+01i & 2.510e+02+4.008e+01i & 4.143e+00+4.340e+00i & 7.486e-01-2.519e-01i & 3.336e-02-8.302e-02i \\
0 & 1.414e+02+3.085e+01i & 2.859e+02+2.939e+01i & 3.384e+00+3.506e+00i & 1.037e+00-6.146e-02i & 1.045e-01-1.215e-01i \\
1 & 3.618e+01+1.380e+02i & 2.451e+02+1.703e+01i & 4.845e+00+4.927e+00i & 2.216e-01-4.582e-01i & 1.195e-03-2.532e-02i \\
1 & 8.289e+01+1.203e+02i & 2.602e+02+3.612e+01i & 4.619e+00+4.790e+00i & 5.069e-01-3.771e-01i & 6.700e-03-5.344e-02i \\
1 & 1.256e+02+8.705e+01i & 2.871e+02+4.528e+01i & 4.228e+00+4.429e+00i & 7.736e-01-2.397e-01i & 3.509e-02-8.306e-02i \\
1 & 1.608e+02+3.314e+01i & 3.247e+02+3.197e+01i & 3.457e+00+3.577e+00i & 1.049e+00-5.813e-02i & 1.036e-01-1.197e-01i \\
2 & 0.000e-01+6.929e+00i & 1.119e+01+0.000e+00i & 2.359e+00+2.359e+00i & 0.000e-01-3.396e-01i & 9.681e-02-9.681e-02i \\
3 & 3.397e+00+1.075e+01i & 1.877e+01+1.884e+00i & 2.585e+00+2.644e+00i & 2.442e-01-3.528e-01i & 6.169e-02-9.150e-02i \\
2 & 8.152e+00+1.469e+01i & 2.889e+01+4.195e+00i & 2.762e+00+2.860e+00i & 4.050e-01-3.341e-01i & 4.995e-02-9.596e-02i \\
3 & 1.422e+01+1.875e+01i & 4.152e+01+6.826e+00i & 2.914e+00+3.038e+00i & 5.175e-01-3.086e-01i & 4.711e-02-1.000e-01i \\
2 & 0.000e+00+3.010e+01i & 5.048e+01+0.000e+00i & 3.361e+00+3.361e+00i & 0.000e-01-4.324e-01i & 3.199e-02-3.199e-02i \\
2 & 2.158e+01+2.291e+01i & 5.665e+01+9.716e+00i & 3.050e+00+3.191e+00i & 6.003e-01-2.835e-01i & 4.761e-02-1.025e-01i \\
3 & 6.304e+00+3.800e+01i & 6.507e+01+3.182e+00i & 3.536e+00+3.577e+00i & 1.389e-01-4.344e-01i & 1.931e-02-3.785e-02i \\
3 & 3.021e+01+2.716e+01i & 7.427e+01+1.282e+01i & 3.173e+00+3.326e+00i & 6.639e-01-2.608e-01i & 4.929e-02-1.037e-01i \\
2 & 1.400e+01+4.605e+01i & 8.220e+01+6.883e+00i & 3.688e+00+3.762e+00i & 2.494e-01-4.252e-01i & 1.308e-02-4.453e-02i \\
2 & 4.011e+01+3.148e+01i & 9.438e+01+1.611e+01i & 3.287e+00+3.449e+00i & 7.145e-01-2.407e-01i & 5.127e-02-1.040e-01i \\
3 & 2.306e+01+5.424e+01i & 1.018e+02+1.102e+01i & 3.822e+00+3.925e+00i & 3.387e-01-4.107e-01i & 1.057e-02-5.056e-02i \\
3 & 5.127e+01+3.587e+01i & 1.170e+02+1.957e+01i & 3.393e+00+3.562e+00i & 7.558e-01-2.231e-01i & 5.321e-02-1.038e-01i \\
2 & 0.000e-01+6.960e+01i & 1.179e+02+0.000e+00i & 4.129e+00+4.129e+00i & 0.000e-01-4.608e-01i & 1.661e-02-1.661e-02i \\
2 & 3.345e+01+6.256e+01i & 1.240e+02+1.553e+01i & 3.945e+00+4.071e+00i & 4.121e-01-3.941e-01i & 1.016e-02-5.556e-02i \\
2 & 6.369e+01+4.032e+01i & 1.421e+02+2.316e+01i & 3.494e+00+3.666e+00i & 7.903e-01-2.077e-01i & 5.498e-02-1.032e-01i \\
3 & 9.204e+00+8.159e+01i & 1.395e+02+4.479e+00i & 4.277e+00+4.310e+00i & 9.677e-02-4.614e-01i & 9.490e-03-2.135e-02i \\
3 & 4.516e+01+7.100e+01i & 1.486e+02+2.038e+01i & 4.058e+00+4.203e+00i & 4.735e-01-3.770e-01i & 1.092e-02-5.956e-02i \\
3 & 7.735e+01+4.484e+01i & 1.696e+02+2.690e+01i & 3.589e+00+3.764e+00i & 8.196e-01-1.942e-01i & 5.654e-02-1.024e-01i \\
2 & 1.982e+01+9.373e+01i & 1.636e+02+9.516e+00i & 4.410e+00+4.472e+00i & 1.794e-01-4.563e-01i & 5.159e-03-2.629e-02i \\
2 & 5.817e+01+7.955e+01i & 1.758e+02+2.552e+01i & 4.164e+00+4.325e+00i & 5.254e-01-3.602e-01i & 1.232e-02-6.269e-02i \\
2 & 9.227e+01+4.940e+01i & 1.997e+02+3.074e+01i & 3.679e+00+3.857e+00i & 8.450e-01-1.823e-01i & 5.790e-02-1.014e-01i \\
3 & 3.182e+01+1.060e+02i & 1.903e+02+1.504e+01i & 4.532e+00+4.620e+00i & 2.506e-01-4.477e-01i & 2.738e-03-3.095e-02i \\
3 & 7.247e+01+8.821e+01i & 2.054e+02+3.093e+01i & 4.264e+00+4.439e+00i & 5.700e-01-3.441e-01i & 1.402e-02-6.511e-02i \\
3 & 1.084e+02+5.402e+01i & 2.322e+02+3.470e+01i & 3.766e+00+3.945e+00i & 8.671e-01-1.717e-01i & 5.906e-02-1.004e-01i \\
2 & 0.000e+00+1.254e+02i & 2.134e+02+0.000e+00i & 4.777e+00+4.777e+00i & 0.000e-01-4.745e-01i & 1.034e-02-1.034e-02i \\
2 & 4.517e+01+1.185e+02i & 2.195e+02+2.101e+01i & 4.644e+00+4.755e+00i & 3.124e-01-4.370e-01i & 1.616e-03-3.514e-02i \\
2 & 8.806e+01+9.696e+01i & 2.376e+02+3.658e+01i & 4.359e+00+4.545e+00i & 6.086e-01-3.289e-01i & 1.585e-02-6.696e-02i \\
2 & 1.258e+02+5.868e+01i & 2.672e+02+3.876e+01i & 3.849e+00+4.029e+00i & 8.867e-01-1.622e-01i & 6.006e-02-9.934e-02i \\
3 & 1.210e+01+1.415e+02i & 2.420e+02+5.776e+00i & 4.907e+00+4.935e+00i & 7.420e-02-4.748e-01i & 5.618e-03-1.394e-02i \\
3 & 5.987e+01+1.310e+02i & 2.511e+02+2.737e+01i & 4.750e+00+4.881e+00i & 3.663e-01-4.253e-01i & 1.370e-03-3.881e-02i \\
3 & 1.049e+02+1.058e+02i & 2.722e+02+4.245e+01i & 4.449e+00+4.645e+00i & 6.424e-01-3.147e-01i & 1.769e-02-6.834e-02i \\
3 & 1.445e+02+6.338e+01i & 3.046e+02+4.292e+01i & 3.928e+00+4.109e+00i & 9.043e-01-1.537e-01i & 6.090e-02-9.827e-02i \\
\end{tabular}
}
\end{table}
\end{document}